\documentclass[a4paper, fleqn, usenatbib]{mnras}

\usepackage{newtxtext,newtxmath}

\usepackage[T1]{fontenc}
\usepackage{ae,aecompl}
\usepackage{booktabs}  

\usepackage{graphicx}   
\usepackage{amsmath}    
\usepackage{amssymb}    
\usepackage{gensymb}
\usepackage{threeparttable}
\usepackage{pdflscape}
\usepackage{CJK}

\newcommand{\angstrom}{\textup{\AA}}

\newcommand{\RomanNumeralCaps}[1] {\MakeUppercase{\romannumeral #1}}

\def\lsim{\mathrel{\rlap{\lower4pt\hbox{\hskip1pt$\sim$}}
    \raise1pt\hbox{$<$}}}                
\def\gsim{\mathrel{\rlap{\lower4pt\hbox{\hskip1pt$\sim$}}
    \raise1pt\hbox{$>$}}}                

\title[Corona-disk-jet connection in RLQs]{The $L_\mathrm{x}$--$L_\mathrm{uv}$--$L_\mathrm{radio}$ relation and corona-disk-jet connection in optically selected radio-loud quasars}

\author[S. F. Zhu et al.]{S. F. Zhu,$^{1,2}$\thanks{E-mail: SFZAstro@gmail.com (PSU)}
W. N. Brandt,$^{1,2,3}$
B. Luo,$^4$
Jianfeng Wu,$^5$
\newauthor
Y. Q. Xue,$^{6,7}$
and G. Yang$^{8,9}$
\\
$^1$Department of Astronomy \& Astrophysics, The Pennsylvania State University, University Park, PA 16802, USA\\
$^2$Institute for Gravitation and the Cosmos, The Pennsylvania State University, University Park, PA 16802, USA\\
$^3$Department of Physics, 104 Davey Lab, The Pennsylvania State University, University Park, PA 16802, USA\\
$^4$School of Astronomy and Space Science, Nanjing University, Nanjing, Jiangsu 210046, China\\
$^5$Department of Astronomy, Xiamen University, Xiamen, Fujian 361005, China\\
$^6$CAS Key Laboratory for Research in Galaxies and Cosmology, Department of Astronomy, University of Science and Technology of China, Hefei 230026, China\\
$^7$School of Astronomy and Space Sciences, University of Science and Technology of China, Hefei 230026, China \\
$^8$Department of Physics and Astronomy, Texas A\&M University, College Station, TX 77843-4242, USA \\
$^9$George P. and Cynthia Woods Mitchell Institute for Fundamental Physics and Astronomy, Texas A\&M University, College Station, TX 77843-4242, USA}

\date{Accepted XXX. Received YYY; in original form ZZZ}

\pubyear{2020}

\begin{document}
\label{firstpage}
\pagerange{\pageref{firstpage}--\pageref{lastpage}}
\maketitle
\begin{abstract}
Radio-loud quasars (RLQs) are more X-ray luminous than predicted by the \mbox{X-ray}--optical/UV relation 
(i.e. $L_\mathrm{x}\propto L_\mathrm{uv}^\gamma$) for radio-quiet quasars (RQQs).
The excess \mbox{X-ray} emission depends on the radio-loudness parameter ($R$) and radio spectral slope ($\alpha_\mathrm{r}$).
We construct a uniform sample of 729 optically selected RLQs with high fractions of X-ray detections and $\alpha_\mathrm{r}$ measurements.
We find that steep-spectrum radio quasars (SSRQs; $\alpha_\mathrm{r}\le -0.5$) follow a quantitatively similar $L_\mathrm{x}\propto L_\mathrm{uv}^{\gamma}$ relation as that for RQQs,
suggesting a common coronal origin for the \mbox{X-ray} emission of both SSRQs and RQQs.
However, the corresponding intercept of SSRQs is larger than that for RQQs and increases with $R$, 
suggesting a connection between the radio jets and the configuration of the accretion flow.
Flat-spectrum radio quasars (FSRQs; $\alpha_\mathrm{r}>-0.5$) are generally more X-ray luminous than SSRQs at given $L_\mathrm{uv}$ and $R$,
likely involving more physical processes.
The emergent picture is different from that commonly assumed where the excess X-ray emission of RLQs is attributed to the jets.
We thus perform model selection to compare critically these different interpretations,
which prefers the coronal scenario with a corona-jet connection.
A distinct jet component is likely important for only a small portion of FSRQs.
The corona-jet, disk-corona, and disk-jet connections of RLQs 
are likely driven by independent physical processes.
Furthermore, the corona-jet connection implies that small-scale processes 
    in the vicinity of SMBHs, probably associated with the magnetic flux/topology instead of black-hole spin, are controlling the radio-loudness of quasars.

\end{abstract}




\section{Introduction}
\label{sec:intro}

Quasars are luminous active galactic nuclei (AGNs) whose
central engines are supermassive black holes (SMBHs) that are actively feeding.
A significant minority of quasars ($\approx10$--20\%; e.g. \citealt{padovani2017}) harbor a pair of powerful relativistic jets that launch from a region close to the SMBH.
Because relativistic jets are strong radio emitters,
the quasars with powerful jets (termed radio-loud quasars, RLQs)
are observationally distinguished from other
quasars (termed radio-quiet quasars, RQQs)
by requiring a radio-loudness parameter $R\equiv L_\mathrm{5GHz}/L_{4400\angstrom}\ge10$,
where $L_\mathrm{5GHz}$ and $L_{4400\angstrom}$ are monochromatic luminosities at
rest-frame 5~GHz and 4400~\angstrom, respectively (\citealt{kellermann1989}).
The jets may be powered by the rotational energy of the SMBH and/or the inner accretion flow
that is extracted by large-scale magnetic fields threading them (e.g. \citealt{bz1977, bp1982}; see the review paper of \citealt{blandford2018}).
However, the mechanism that triggers the production of powerful relativistic jets
in only a minority of quasars is not clear.

X-ray emission is nearly ubiquitous for quasars (\citealt{brandt2015}, and references therein).
The primary X-ray emission ($\sim$1--100 keV) from RQQs is radiated from a ``coronal'' structure containing hot plasma (e.g. \citealt{haardt1993}),
which might be powered by magnetic reconnection (e.g. \citealt{beloborodov2017}).
UV photons from the inner accretion disk are up-scattered
by electrons in the plasma to produce X-rays.
This thermal Compton-scattering process leaves an
imprint on the \mbox{X-ray} spectrum as a high-energy cutoff at $\sim$100--200 keV,
which has been observed in local AGNs (e.g. \citealt{fabian2015}) and
a few high-redshift quasars (e.g. \citealt{lanzuisi2019}).
Interestingly, a non-linear correlation between the luminosities of
the coronal structure and accretion disk has been established,
$L_\mathrm{2keV}\propto L_\mathrm{2500\angstrom}^{\gamma}$,
which is referred to as the $L_\mathrm{2keV}$--$L_\mathrm{2500\angstrom}$ relation
($\gamma=0.6$--0.8; e.g. \citealt{avni1986, just2007, lusso2016}).\footnote{Throughout the paper, $\gamma$ with no subscript refers to the slope of the $L_\mathrm{x}$--$L_\mathrm{uv}$ relation for RQQs.}
The $\alpha_\mathrm{ox}$--$L_{2500\angstrom}$ relation describes the same non-linear correlation as a dependence of the
shape of the optical/UV--\mbox{X-ray} spectral energy distribution (SED) on disk luminosity.
Here, $\alpha_\mathrm{ox}\equiv \log(L_\mathrm{2keV}/L_\mathrm{2500\angstrom})/\log(\nu_\mathrm{2keV}/\nu_\mathrm{2500\angstrom})$
is the two-point spectral index between rest-frame 2~keV and 2500~\angstrom\ (\citealt{tananbaum1979}).

The X-ray properties of RLQs are different from those of RQQs.
RLQs are generally more X-ray luminous than RQQs of matched optical/UV luminosity (e.g. \citealt{zamorani1981, worrall1987, miller2011, ballo2012}).
Their \mbox{X-ray} spectra are systematically flatter than those of RQQs
(e.g. \citealt{wilkes1987, reeves1997, page2005}),
particularly for the flat-spectrum radio quasars (FSRQs; e.g. \citealt{grandi2006}).\footnote{The flat-spectrum and steep-spectrum objects have $\alpha_\mathrm{r}\ge-0.5$ and $\alpha_\mathrm{r}<-0.5$, respectively. Here, $\alpha_\mathrm{r}$ is the power-law spectral index
(i.e. $f_\nu\propto v^{\alpha_\mathrm{r}}$)
in the radio band.}
Furthermore, Compton-reflection features are weaker in RLQs (e.g. \citealt{reeves1997, reeves2000}).
Among low-redshift AGNs, radio galaxies are more \mbox{X-ray}
luminous than their radio-quiet counterparts (e.g. \citealt{gupta2018}); broad-line radio galaxies (BLRGs)
are found to have weaker reflection features than type-1 Seyfert galaxies (e.g. \citealt{wozniak1998, eracleous2000}).
However, no strong evidence supports the X-ray spectra of BLRGs
being flatter than those of radio-quiet Seyfert galaxies (e.g. \citealt{sambruna1999, grandi2006, gupta2018}).
The different X-ray properties of radio-loud and radio-quiet AGNs
could be explained if the radio ``core''
of the jets contributes significantly
in the X-rays a broadband component with a flat spectrum,
in addition to the typical disk/corona emission being present
(e.g. \citealt{lawson1997, grandi2004}).\footnote{\label{fn:core} The radio ``core'' here refers to the
sub-arcsec component of the radio image
that spatially coincides with the
optical and X-ray (point-source) position of the quasar. It is presumably related to the base of the jet.}
However, except for a few FSRQs (e.g. \citealt{grandi2006, madsen2015}),
the X-ray spectra of BLRGs (e.g. \citealt{wozniak1998, sambruna2009, ronchini2019})
and steep-spectrum radio quasars (SSRQs; e.g. \citealt{lohfink2017})
generally do not reveal a jet-linked flat continuum, suggesting that
the orientation of the jets with respect to our line of sight might play an important role.

The coupling between the coronal structure and the accretion disk
that is revealed by the $L_\mathrm{2keV}$--$L_\mathrm{2500\angstrom}$ relation of RQQs
probably exists for RLQs as well.
Both the jet-launching region and corona are in the immediate vicinity of the SMBH,
and connections (e.g. through a joint dependence on the magnetic field)
between relativistic jets and the corona might be expected.
The disks/coronae of quasars dissipate most of their radiated energy in the optical/UV and X-ray bands.
Powerful relativistic jets have characteristic synchrotron radio emission,
and might have contributions in the X-rays.
Therefore, X-ray, optical/UV, and radio are the key observational windows
through which to peer at the nature of RLQs.
An empirical $L_\mathrm{2keV}$--$L_\mathrm{2500\angstrom}$--$L_\mathrm{5GHz}$
relation has long been sought for RLQs (e.g. \citealt{tananbaum1983}),
in analogy to the $L_\mathrm{2keV}$--$L_\mathrm{2500\angstrom}$ relation for 
RQQs.\footnote{Such empirical relations not only advance our understanding of AGN physics but 
also have broad practical applications in, e.g., SED fitting (e.g. \citealt{yang2020}).}
Indeed, the amount of the X-ray excess of RLQs over RQQs of
comparable optical/UV luminosities
increases with both radio-loudness parameter
and radio luminosity (e.g. \citealt{miller2011}),
supporting the idea that the X-ray luminosity is determined
by considering the power of both the disk and the jets,
which are represented by $L_\mathrm{2500\angstrom}$ and $L_\mathrm{5GHz}$, respectively.
Except for extreme objects at extreme redshifts (i.e. quasars with $\log R\ge2.5$ at $z\ge4$; \citealt{wu2013, zhu2019}),
the $L_\mathrm{2keV}$--$L_\mathrm{2500\angstrom}$--$L_\mathrm{5GHz}$ relation does not have an apparent redshift dependence (e.g. \citealt{worrall1987, miller2011}).

However, studies of the $L_\mathrm{2keV}$--$L_\mathrm{2500\angstrom}$--$L_\mathrm{5GHz}$
relation for RLQs have generally lagged behind those of the $L_\mathrm{2keV}$--$L_\mathrm{2500\angstrom}$ relation for RQQs (e.g. \citealt{worrall1987, miller2011}).
On one hand, the sample size of RLQs is about an order of magnitude smaller than that of RQQs from optical quasar surveys.
On the other hand, to constrain the relations to a comparable precision,
studies of the $L_\mathrm{2keV}$--$L_\mathrm{2500\angstrom}$--$L_\mathrm{5GHz}$ relation within at least a three-dimensional parameter space
generally require the sample size to be larger than that of RQQs, for which a two-dimensional parameter space is sufficient.
Furthermore, an extra dimension of RLQ properties (i.e. radio luminosity) compared to RQQs
also indicates an extra dimension of the model space.
There are more candidate models for the $L_\mathrm{2keV}$--$L_\mathrm{2500\angstrom}$--$L_\mathrm{5GHz}$
relation from which to choose. Beaming effects of the jet emission might add another layer of complexity.

The ambiguity in the functional form of the $L_\mathrm{2keV}$--$L_\mathrm{2500\angstrom}$--$L_\mathrm{5GHz}$
relation makes its physical interpretation and implications unclear.
It is possible that RLQs have a similar X-ray emitting disk/corona
structure as that of RQQs and a {\it distinct} jet-linked X-ray component.
The $L_\mathrm{2keV}$--$L_\mathrm{2500\angstrom}$--$L_\mathrm{5GHz}$ relation
would then simply describe the general positive correlations of
total X-ray luminosities with the optical/UV and radio luminosities.

Alternatively, the $L_\mathrm{2keV}$--$L_\mathrm{2500\angstrom}$--$L_\mathrm{5GHz}$ relation might indicate
a {\it connection} between the jets and disk/corona;
e.g. the disks/coronae of quasars that have stronger relativistic jets could be more X-ray
luminous than those of quasars that have weaker or no relativistic jets.
Such a connection might link AGNs to the phenomena of Galactic black-hole X-ray binaries (BHXRBs; or microquasars sometimes called)
that are also powered by the black-hole accretion process and show couplings between jets and the accretion flow
(e.g. \citealt{marscher2002, merloni2003}).
Most BHXRBs are transients and show outbursts that last months to years (e.g. \citealt{remillard2006}).
During a typical outburst, BHXRBs may cycle through
(a few) state transitions that are marked by
changes in spectral and timing properties (e.g. \citealt{homan2005}),
as well as jet activity (e.g. \citealt{fender2004}).
The physical scales of SMBHs at the centers of massive galaxies are $>10^5$
times those of the stellar-mass black holes of BHXRBs, making
it practically difficult for observations to spot
state transitions of individual AGNs directly (e.g. \citealt{schawinski2015}).
Instead, snapshot (as compared to the timescale of state transitions) observations
across different wavelengths discover a great variety of AGNs (e.g. \citealt{padovani2017}).
While some of the varieties are caused by inclination-dependent geometry
as we are able to observe only one aspect of
each AGN (e.g. \citealt{netzer2015}),
accretion states of the central engine might also play an important role (e.g. \citealt{best2012}).
Investigating the disk/corona-jet connection of RLQs and establishing
a phenomenological correspondence between AGN types and BHXRB states
can shed light on the physics of black-hole accretion and relativistic jets.

Previous works did not systematically compare these scenarios
(e.g. \citealt{tananbaum1983, zamorani1983, zamorani1984, worrall1987, miller2011}).
Specifically, they usually focus on one functional model and obtain several sets of parameters
that reflect different X-ray properties of different samples.
Those sample-dependent empirical relations can be used to
predict the \mbox{X-ray} luminosity for given optical/UV
and radio luminosities, within a restricted parameter space.
However, the driving mechanisms are hidden due to the lack of generality.
Here we instead apply various models to the same sample and seek the most probable explanation of the data (i.e. model selection).
Perhaps no single model is suitable for all RLQs.
For example, FSRQs and SSRQs might require separate mechanisms to explain their \mbox{X-ray} data.
Then, we compare the results across all samples, investigating their differences as well as similarities.

We construct a large ($>700$ objects) optically selected RLQ sample without regards to their radio/\mbox{X-ray} properties in \S~\ref{sec:sampleSelection}.
Those RLQs span a broad parameter (i.e. luminosity, radio slope, and radio-loudness) space
and have a high X-ray detection fraction ($\gtrsim90$\%).
Furthermore, almost all of them ($\gtrsim96$\%) have basic radio spectral information.
We perform model-independent, model-fitting, and model-selection analyses in \S~\ref{sec:fitting}.
We compare with literature results and discuss physical implications in \S~\ref{sec:discussion}.
A summary of this paper and future prospects are in \S~\ref{sec:summary}.
In this paper, the quoted error bars represent $1\sigma$ uncertainties,
and the upper limits are at a 95\% confidence level,  unless otherwise stated.
The spectral index ($\alpha_\nu$) follows the convention that $f_\nu\propto \nu^{\alpha_\nu}$.
We use $L_\mathrm{x}$, $L_\mathrm{uv}$, and $L_\mathrm{radio}$
interchangeably with $L_\mathrm{2keV}$, $L_\mathrm{2500\angstrom}$, and $L_\mathrm{5GHz}$.
The median statistic is widely used throughout the paper.
We calculate medians using the Kaplan-Meier estimator (e.g. \citealt{kaplan1958}) in cases where the data contain non-detections.
We use the bootstrapping method if the uncertainties of medians are quoted.
We adopt a flat-$\Lambda$CDM cosmology
with $H_0=70$ km s$^{-1}$ Mpc$^{-1}$ and $\Omega_\mathrm{m}=0.3$.

\section{Sample Selection}
\label{sec:sampleSelection}

We select new RLQs utilizing the Sloan Digital Sky Survey \citep[SDSS;][]{york2000}.
The radio data are from the Faint Images of the Radio Sky at Twenty-Centimeters \citep[FIRST;][]{becker1995} and the NRAO VLA Sky Survey (NVSS; \citealt{condon1998}).
Archival {\it Chandra} and {\it XMM-Newton} observations are used to constrain the \mbox{X-ray} luminosities.
The newly selected RLQs are combined with those from \citet{miller2011} to form a final sample of 729 optically selected RLQs,
which is summarized in Table~\ref{tab:optsample}.
Compared with \citet{miller2011}, both the sample size and X-ray detection fraction are increased.
Furthermore, we double the numbers of spectroscopic redshifts and radio slopes,
the latter of which affects the X-ray properties of RLQs (see \S~\ref{sec:fitting}).
Importantly, the number of $\log R>3$ RLQs with reliable spectroscopic redshifts are significantly increased (by 70\%).
We will show in \S~\ref{sec:base} that such RLQs with the highest radio-loudness parameters have the
largest statistical power in discriminating between models.

\subsection{New RLQs from the SDSS DR14 Quasar catalog}
\label{sec:newRLQs}
\subsubsection{Initial selection}
\label{sec:init}
The SDSS DR14 Quasar catalog (DR14Q; \citealt{paris2018}) covers
a sky area of 9376 deg$^2$ and contains spectroscopically
identified quasars from the Legacy Survey of SDSS-\RomanNumeralCaps 1/\RomanNumeralCaps 2,
the Baryon Oscillation Spectroscopic Survey (BOSS) of SDSS-\RomanNumeralCaps 3,
and the extended Baryon Oscillation Spectroscopic Survey (eBOSS) of SDSS-\RomanNumeralCaps 4.
The size of DR14Q ($5.3\times10^5$ quasars) is a factor of $\approx7$ 
times that of DR5Q ($7.7\times 10^4$ quasars), which was utilized by \citet{miller2011}.
The sky coverage of the FIRST survey has an extent of 10575 deg$^2$ and largely coincides with that of the SDSS.
The NVSS covers the entire sky north of $\mathrm{Dec}=-40$ degrees but with a beam about 10 times larger than that of FIRST.
Since the FIRST images have a better resolution,
we select new RLQs from the matching results of DR14Q with the final catalog of the FIRST survey (\citealt{helfand2015}).
Considering the fact that the quasars in DR14Q are generally fainter than those in DR5Q,
we only consider RLQs with $\log R\ge2$, which ensures $m_i\le21$ quasars
can be detected in the radio band given the ($5\sigma$) flux limit of about 1~mJy of the FIRST survey.
The matching is performed as follows.
We refer to each row in the FIRST catalog as a radio component.
We adopt the method of \citet{banfield2015} to distinguish resolved and unresolved components (cf. their Eq.~1 and Fig.~2).
The radio flux of a resolved (unresolved) radio component refers to its integrated (peak) flux.
We add the radio fluxes of all radio components within a radius of 90 arcsec around the optical
position of a quasar and calculate a first corresponding radio-loudness parameter,
which results in 24772 candidate $\log R\ge2$ quasars in the redshift range of $0.5< z\le4$.
Here, we have assumed $\alpha_{\rm r}=-0.5$ to calculate $L_{5\rm\ GHz}$ from the observed 1.4 GHz flux.
The $i$-band apparent magnitude ($m_i$) is utilized to calculate $L_{4400\rm\ \angstrom}$,
where the K-correction of \cite{richards2006} and an optical spectral index of $\alpha_{\rm o}=-0.5$ are assumed.

In the above, a very large matching radius (90 arcsec) is adopted to ensure that
the extended radio emission (e.g. from jets and lobes) associated with each quasar is recovered.
However, in many cases, the $R$ value calculated here is merely an upper limit,
because background radio sources that are not associated with the quasar are also included.
Visual inspection is required to eliminate such contamination from background radio sources \citep[e.g.][]{lu2007}.
To minimize the work of visual inspection,
we first match the list of candidate $\log R\ge2$ quasars
with the observation catalogs of {\it Chandra} and {\it XMM-Newton} and apply unbiased empirical quality cuts.
For the {\it Chandra}/ACIS observations, we require
\begin{equation}
    T_{\rm exp}>1000+35\times10^{(\theta/8)^2},
\end{equation}
where $T_{\rm exp}$ is the exposure time in seconds and $\theta$ is the off-axis angle of
the quasar on the X-ray image, in units of arcmin.
This criterion requires the exposure time to be at least 1 ks, and it requires additional exposure time
at large off-axis angles to compensate for the loss of sensitivity due to the larger point spread function.
Similarly, the quality cuts for {\it XMM-Newton}/EPIC-pn and {\it XMM-Newton}/EPIC-MOS are
\begin{equation}
    T_{\rm exp}>1000+20\times10^{(\theta/8)^2}
\end{equation}
and
\begin{equation}
    T_{\rm exp}>2000+20\times10^{(\theta/8)^2},
\end{equation}
respectively.
Note that $\theta$ is bounded by the field of view of the telescopes ($\theta<15^\prime$).
In addition to the cut on the exposure time, we also
require each quasar not to fall onto the edge of the detector or in CCD gaps.
The requirements for sensitive X-ray coverage and quality cut here reduce the sample size to 1090.

For these quasars, if there is no radio component in the annulus of $2\le r\le 90$ arcsec,
no visual inspection is performed (330 quasars) and the association is assigned automatically.
We visually inspected the 4 $\times$ 4 arcmin$^2$ FIRST images of the remaining 760 quasars,
where the radio components with apparent optical counterparts are labeled.
Radio components that are associated with other background sources in the field of view are discarded.
In cases where real association exists (see Appendix~A of \citealt{miller2011} for the utilized matching method),
we associate the quasar with one or multiple radio components.

Furthermore, even though FIRST has a deeper nominal flux limit than that of NVSS,
it is known to be less sensitive to extended radio components (e.g. \citealt{white2007}).
Sometimes, the faint extended radio emission is ``resolved out'' and completely missing in the FIRST catalog 
(e.g. \citealt{blundell2003}).\footnote{By matching the DR14Q with FIRST, we might already miss RLQs with only faint and diffuse radio emission.
However, those cases are rare ($\sim2.3$\%; e.g. \citealt{lu2007}) such that quasars with an extended morphology are almost always luminous in the radio band.}
The LOFAR Two-metre Sky Survey (LoTSS; \citealt{lofar2017}) aims to image the \mbox{120--168} MHz northern sky with a sensitivity of 100$\mu$Jy.
The LoTSS data release 1 (DR1; \citealt{lofar2019}) covers a sky area of about 400 square degrees, which contains 63 of our RLQs.
We compare the LoTSS and FIRST images of those RLQs and find three cases where FIRST resolves out extended radio components,
among which the FIRST fluxes are significantly lower than those of NVSS by 30\%--50\% for two RLQs.
Even though those are rare cases ($2/63\approx3\%$),
we preferentially use the radio fluxes from NVSS over those of FIRST to avoid underestimating total radio emission,
provided that the former is not contaminated by very nearby background radio sources.
Note that we regard the FIRST component $<2$ arcsec away from the quasar as the radio core.
Using the ratio of the peak flux of the radio core to quasar total flux, we assess the dominance of the core (see Footnote~\ref{footnote:ccs}).
Given the angular resolution of the FIRST survey ($\approx5$ arcsec),
the core dominance here might overestimate the true core dominance that is revealed in very-long-baseline interferometric data.

After the above procedures, the resulting sample has a size of 545.
We label quasars as serendipitously observed if their X-ray observations have $\theta\ge1$ arcmin,
while the rest are labeled as targeted.
We also label the quasars that are initially selected in SDSS color
space,\footnote{The target selection algorithms of the BOSS and eBOSS programs might also utilize infrared photometry information where available.}
regardless of their radio/X-ray properties; this allows us to construct an optically selected sample to match the methodology of \citet{miller2011}.
We consider those quasars spectroscopically confirmed in the Legacy Survey
that have LEGACY\_TARGET1 flags of ``QSO'',  ``HIZ'' and ``serendipitous'' as optically selected, following \citet{miller2011}.
For quasars targeted in BOSS, we consider the
``CORE'' and ``BONUS'' samples as indicated by the BOSS\_TARGET1 flags.
For quasars targeted in eBOSS, we include those belonging to
the ``CORE'' sample as indicated by the EBOSS\_TARGET1 flags. 
We refer readers to \citet{paris2018} and references therein for the target selection flags of DR14Q.

\subsubsection{X-ray data analysis}
\label{sec:xdata}

We first matched the quasar positions to the {\it Chandra} Source
Catalog Release 2.0 \citep[CSC 2.0;][]{evans2010}
and the latest {\it XMM-Newton} Serendipitous Source Catalog \citep[3XMM-DR8;][]{rosen2016} to
obtain their X-ray fluxes (0.5--7 keV for {\it Chandra} observations and 0.5--4.5 keV for {\it XMM-Newton} observations).
Galactic-absorption correction is performed subsequently \citep[][]{dickey1990}.

Then, for the observations that are not included in the source catalogs or where the quasar is not
detected by the catalog pipeline, we download the observations and analyze the data manually.
Specifically, data reduction, cleaning, and image extraction are performed using {\sc ciao} (v4.11) and {\sc sas} (v17.0.0) for {\it Chandra} and {\it XMM-Newton} observations, respectively.
Raw source counts are extracted from a circular region centered at the quasar position and enclosing $\approx$90\% of the total energy.
Background counts are extracted from a source-free concentric annulus or nearby circular region.
Using the source and background counts of each observation, we calculate the binomial no-source probability ($P_\mathrm{B}$; \citealt{weisskopf2007}),
which is the conditional probability of producing counts that are equal to or larger
than the observed counts in the source extraction region given the intensity of the background.
We take a source as detected if $P_\mathrm{B}<0.01$ and calculate its net count rate.
Instrumental response files (i.e. response matrix files and ancillary response files)
at the position of the source are created using the standard routines of {\sc ciao} and {\sc sas};
these include the appropriate aperture correction.
With these instrumental response files, the net count rate is converted to X-ray flux using {\sc sherpa}.
We have fixed the photon index to $\Gamma = 1.5$ for the power-law model, following \citet{miller2011};
the dependence of the X-ray flux on the choice of $\Gamma$ is mild for reasonable choices of $\Gamma$.
Note that the X-ray flux is corrected for Galactic absorption by
specifying the $N_\mathrm{H}$ value (\citealt{dickey1990}) of the model in {\sc sherpa}.
A source is treated as a non-detection if $P_\mathrm{B}\ge0.01$.
We calculate the upper limits on the net counts
at a 95\% confidence level using the algorithm of \citet{kraft1991}
for non-detected quasars,
and the upper limits of the X-ray fluxes are calculated using the same method as detections.

The Galactic-absorption corrected X-ray flux (or upper limit), from either the source catalogs or manual analysis,
is subsequently converted to the monochromatic luminosity at rest-frame 2 keV, $L_\mathrm{2keV}$.

\begin{table*}
\centering
\caption{Summary of optically selected RLQs utilized in the paper.}
\label{tab:optsample}
\begin{threeparttable}[b]
\begin{tabular}{lccccc}
\hline
\hline
{Sample} & No. of Sources & X-ray Detections & Serendipitous & Spectroscopic $z$& Radio Slope \\
\hline
All RLQs & 729& 657 (90.1\%) & 622 (85.3\%) & 587 (80.5\%) & 704 (96.6\%) \\
FSRQs & 394 & 363 (92.1\%) & 340 (86.3\%) & 319 (81.0\%) & 394 (100\%) \\
SSRQs & 310 & 275 (88.7\%) & 258 (83.2\%) & 253 (81.6\%) & 310 (100\%)\\
\hline
\end{tabular}
\end{threeparttable}
\end{table*}

\begin{table*}
\centering
\caption{Properties of optically selected RLQs utilized in the paper, in ascending order of RA. Only the top 5 objects are listed.}
\label{tab:list}
\begin{threeparttable}[b]
\begin{tabular}{lccccccccccc}
\hline
\hline
    SDSS Name & $z$ & $m_i$ & $\log L_\mathrm{2500\angstrom}$ & $\log L_\mathrm{5GHz}$ & $\log L_\mathrm{2keV}$ & $\log R$ & XFlag\tnote{a} & sd\tnote{b} & spec\tnote{c} & $\alpha_\mathrm{r}$ & cD\tnote{d}\\
\hline
    000442.18$+$000023.3& 1.008& 18.98   & 30.22&32.09&    26.91&  1.74& 1&    1 & 1 &  -       &  0.59\\
    000622.60$-$000424.4& 1.038& 19.58   & 30.05&34.94&    27.32&  4.77& 1&    1 & 1 &  $-0.55$ &  0.80\\
    001646.54$-$005151.7& 2.243& 21.00   & 30.20&32.66&    26.36&  2.34& 1&    1 & 1 &  $-0.21$ &  1.00\\
    001910.95$+$034844.6& 2.022& 20.30   & 30.35&32.91&    26.54&  2.44& 1&    1 & 1 &  $-0.43$ &  0.56\\
    003054.63$+$045908.4& 2.201& 20.95   & 30.19&33.81&    26.59&  3.50& 1&    1 & 1 &  $-0.41$ &  0.76\\
\hline
\end{tabular}
\begin{tablenotes}
\item[a] If the quasar is detected in X-rays, $\mathrm{XFlag}=1$, while $\mathrm{XFlag}=0$ if otherwise.
\item[b] $\mathrm{sd}=1$ if the X-ray observation is labeled serendipitous and $\mathrm{sd}=0$ if the quasar was targeted.
\item[c] If the quasar is spectroscopically confirmed, $\mathrm{spec}=1$. If the redshift is based on photometric data, $\mathrm{spec}=0$.
\item[d] The ratio of the flux of the radio core over the total radio flux (see \S~\ref{sec:init}).
\end{tablenotes}
\end{threeparttable}
\end{table*}

\subsubsection{Radio spectral indices}
\label{sec:radioSlope}
We matched the quasar list with sky surveys in the radio band to obtain radio slopes, using fluxes at 1.4 GHz and at another wavelength.
Those radio surveys include the Green Bank 6-cm (GB6) Radio Source Catalog \citep[4.85 GHz;][]{gregory1996},
Westerbork Northern Sky Survey \citep[325 MHz;][]{rengelink1997}, 
TGSS Alternative Data Release \citep[150 MHz;][]{intema2017},
and LoTSS DR1 \citep[144 MHz;][]{lofar2019}.
For quasars that are not matched with those radio surveys,
we gather their multi-wavelength radio fluxes in the NED\footnote{https://ned.ipac.caltech.edu/} and VizieR.\footnote{http://vizier.u-strasbg.fr/vizier/sed/}
For the 78 quasars that do not have multi-wavelength radio data in the NED or VizieR,
60 of them are covered by the ongoing VLASS (\citealt{vlass2019}),
which provides the 3~GHz flux.
We downloaded VLASS quick-look images\footnote{https://archive-new.nrao.edu/vlass/quicklook/}
of these 60 quasars and measured their 3 GHz fluxes using {\sc Aegean} \citep[][]{hancock2012, hancock2018}.
Note that the peak and integrated fluxes from these quick-look images are corrected by factors of 1.15 and 1.10, respectively (\citealt{lacy2019}).
We inspect the FIRST/NVSS images while matching RLQs with the above radio surveys to ensure true association of radio components
and to screen background contamination, which is important for the GB6 data since it has a large beam
size ($>3$ arcmin).\footnote{We measure FIRST/NVSS-VLASS radio slopes for those quasars having FIRST/NVSS-GB6 measurements to ensure that 
the inhomogeneous beam sizes do not strongly bias the results.}
In total, 263 quasars have high-frequency ($>1.4$ GHz) radio data,
while 346 quasars have low-frequency ($<1.4$ GHz) radio data.
Even though the radio slope measurements are not from homogeneous data sets,
they are sufficient to distinguish flat-spectrum and steep-spectrum objects.
Furthermore, for objects having fluxes at $\ge3$ different frequencies, the radio spectral shape 
is generally consistent with a single power-law,
which argues against a strong effect of redshift.
However, we label those ``GHz peaked'' and related quasars (28 objects) that are inconsistent 
with a power-law spectrum but seem to be peaked at
GHz to 100 MHz frequencies (e.g. \citealt{odea1998}). 
Note that these objects are eventually excluded from the final sample 
since their X-ray properties are often different from typical RLQs (e.g. \citealt{siemiginowska2008}).

\subsubsection{Cleaning the sample using multi-wavelength colors and optical/UV spectra}
\label{sec:cc}
We matched the 545 quasars resulting from initial selection (\S~\ref{sec:init}) with infrared sky surveys, including
the {\it Wide-field Infrared Survey Explorer} \citep[{\it WISE};][]{wright2010}, the
UKIRT Infrared Deep Sky Survey \citep[UKIDSS;][]{lawrence2007},
the UKIRT Hemisphere Survey \citep[UHS;][]{dye2018}, and the VISTA Hemisphere Survey \citep[VHS;][]{mcmahon2013}.
We also matched our sample with the Pan-STARRS Stack Object Catalog \citep[][]{chamber2016} to include
$Y$-band photometry.
Including the SDSS, we have high-quality photometric coverage from the UV to mid-infrared.
For the mid-infrared data, we have utilized the forced photometry from
the unWISE coadd images \citep[][]{lang2016}, which
improves the number of detections in the W1--W4 bands.
Since not all the forced-photometry data are high-quality measurements, we
replace those fluxes with signal-noise-ratio (SNR) smaller than 2 with their 95\% confidence upper limits, and
flag them as non-detections.\footnote{http://wise2.ipac.caltech.edu/docs/release/allwise/expsup/sec3\_1a.html}
For those quasars that do not have forced photometry,
we still use the measurements from the AllWISE catalog.
The forced photometry increases the number of detections by $\approx$ 30--50 for each mid-infrared band.
These photometric data are corrected for Galactic extinction using the dust map of \citet{dustMap2011} before the following analysis.

We found that some of the RLQs have abnormal
multi-wavelength colors and SDSS spectra
that are not consistent with those of typical quasars featuring
big blue bump-dominated SEDs and strong optical/UV emission lines.
We here further screen the sample based on their SDSS spectra and multi-wavelength colors.
We matched the 545 RLQs with the SDSS-DR7 Quasar Catalog \citep[DR7Q;][]{schneider2010, shen2011} and
SDSS-DR12 Quasar Catalog \citep[DR12Q;][]{paris2017}.
For the 501 quasars that are in DR7Q or DR12Q, 31 are classified as broad absorption line quasars (BALs).
For the remaining 44 quasars, one is a BAL because its Balnicity index is larger than 0 \citep{paris2018}.
We have removed these BAL quasars from further consideration because their X-ray properties are likely
affected by complex absorption (e.g. \citealt{miller2009}).
We also flagged 10 quasars whose SDSS spectra are Type II-like.
They are relatively local ($z<0.76$) with strong [O~{\sc iii}], weak H$\beta$, and often red continua.
Four of them are in the Type II quasar catalog of \citet{reyes2008} or \citet{yuan2016}.
These quasars also likely have complex X-ray absorption, so are screened.
We screened out quasars that have red SDSS and/or infrared colors that satisfy the following criteria:
\begin{align}
    u-W4 >12.0\ {\rm or}\ u-W3 >9.4\ {\rm or}\ u-i>0.95,\ z&\le1.5\\
    g-W4 >11.9\ {\rm or}\ g-W3 >9.7\ {\rm or}\ g-i>1.0,\ 1.5<z&\le2.5\\
    r-W3>8.4\ {\rm or}\ r-z>0.6,\ 2.5<z&\le4
\end{align}
Note that these color cuts effectively remove red quasars (79 objects) that either suffer from severe dust extinction or have
prominent jet emission in the infrared through UV bands.
Furthermore, we exclude quasars that do not have strong emission lines,
since their rest-frame optical/UV emission might be contaminated by strongly boosted jet emission,
rendering the observed $L_{\rm 2500\ \angstrom}$ an overestimate of their disk power.
Specifically, we measure the rest-frame equivalent width (REW) of Mg {\sc ii} (when it is covered by SDSS spectra)
and exclude those quasars (3 objects) with ${\rm REW}<10$~\angstrom.
After these procedures, the resulting sample has a size of 421,
among which 327 quasars are selected using SDSS colors (see the last paragraph of \S~\ref{sec:init}).
Note that serendipitous X-ray coverage makes up about 79\% of this sample,
so the majority of these quasars have X-ray observations unbiased by source properties.
Those quasar jets with the most extremely boosted emission are under-represented in our sample, due to the color cuts and the constraint on emission-line strength.
However, their fraction is estimated to be only $\lesssim6\%$, considering a conservative Lorentz factor of $\Gamma=8$ for the jets
and a half-opening angle of 30 deg for the dusty torus, and thus the results of this paper will not be strongly biased.

\begin{figure}
\centering
\includegraphics[width=0.45\textwidth, clip]{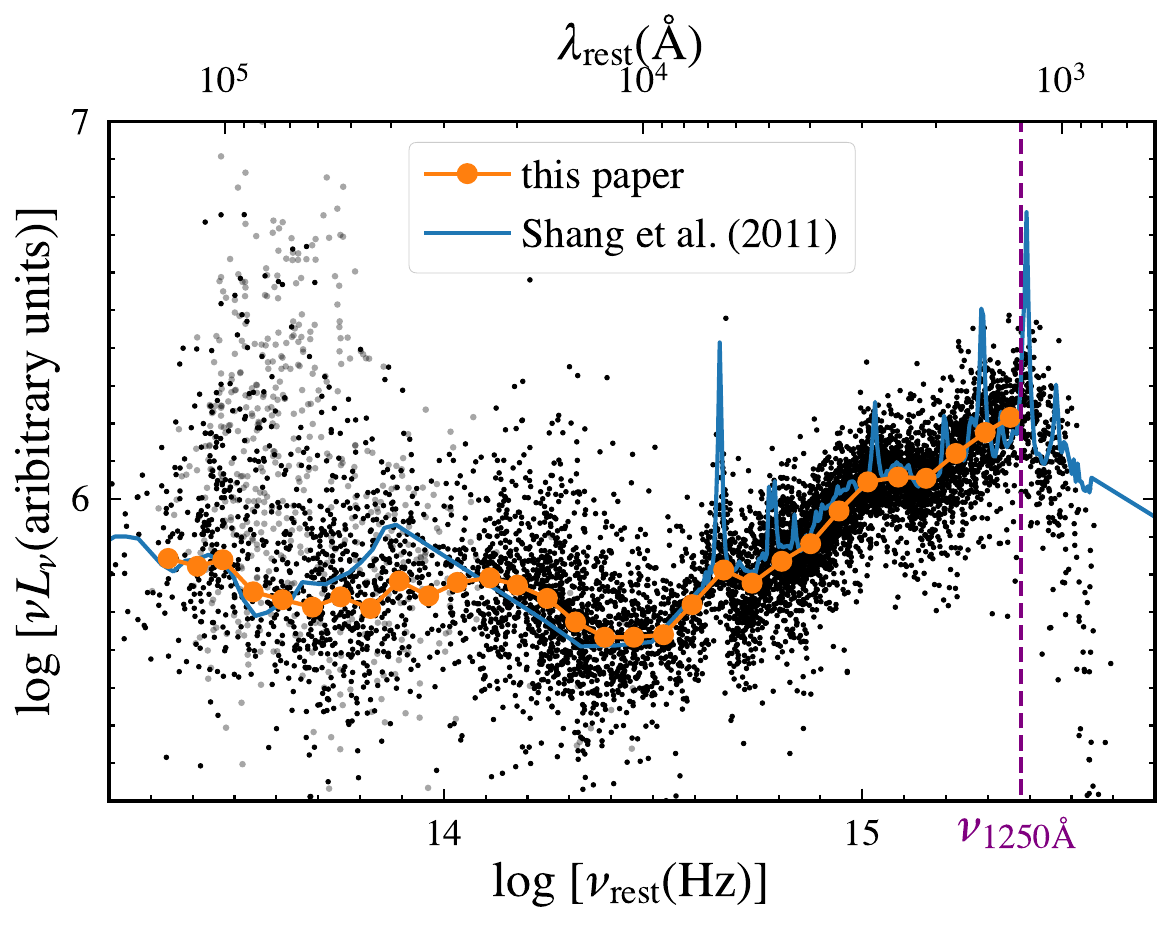}
\caption{The composite median SED (orange) of optically selected RLQs, from the mid-infrared to UV.
    The top and bottom ticks are in units of rest-frame frequency and wavelength, respectively.
The small dots are data points for individual quasars, where grey symbols are
non-detections in the {\it WISE} bands, distinguished from detections (black).
Any data points with rest-frame frequency $>\nu_{1250\angstrom}$ are not used in constructing the composite SED due to intergalactic absorption.
Also plotted is the composite SED of the local RLQs from \citet[blue]{shang2011} for comparison.}
\label{fig:compoSED}
\end{figure}

\begin{figure}
\centering
\includegraphics[width=0.48\textwidth, clip]{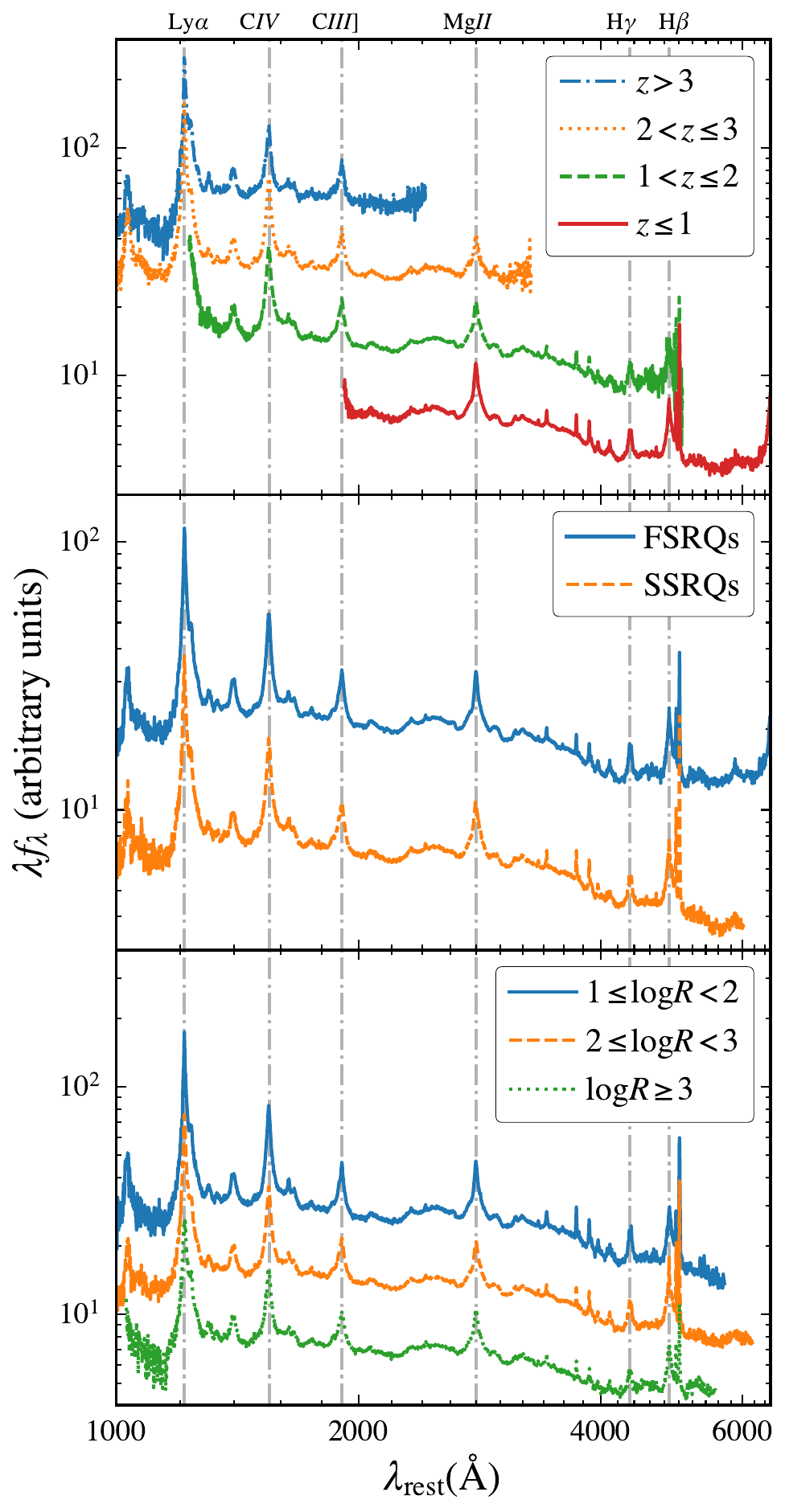}
\caption{The composite median SDSS spectra (in $\lambda f_\lambda$ representation) of optically selected RLQs.
From top to bottom, RLQs are divided into bins of redshift, radio slope, and radio-loudness.
The dash-dotted vertical lines show the frequencies of six emission lines as labeled.
Note that the composite spectra are arbitrarily shifted vertically to avoid overlapping.
In all cases, prominent emission lines are apparent,
indicating that the optical/UV emission of the RLQs is not contaminated by strong boosted jet emission.}
\label{fig:compoSpec}
\end{figure}

\begin{figure*}
\centering
\includegraphics[width=0.95\textwidth, clip]{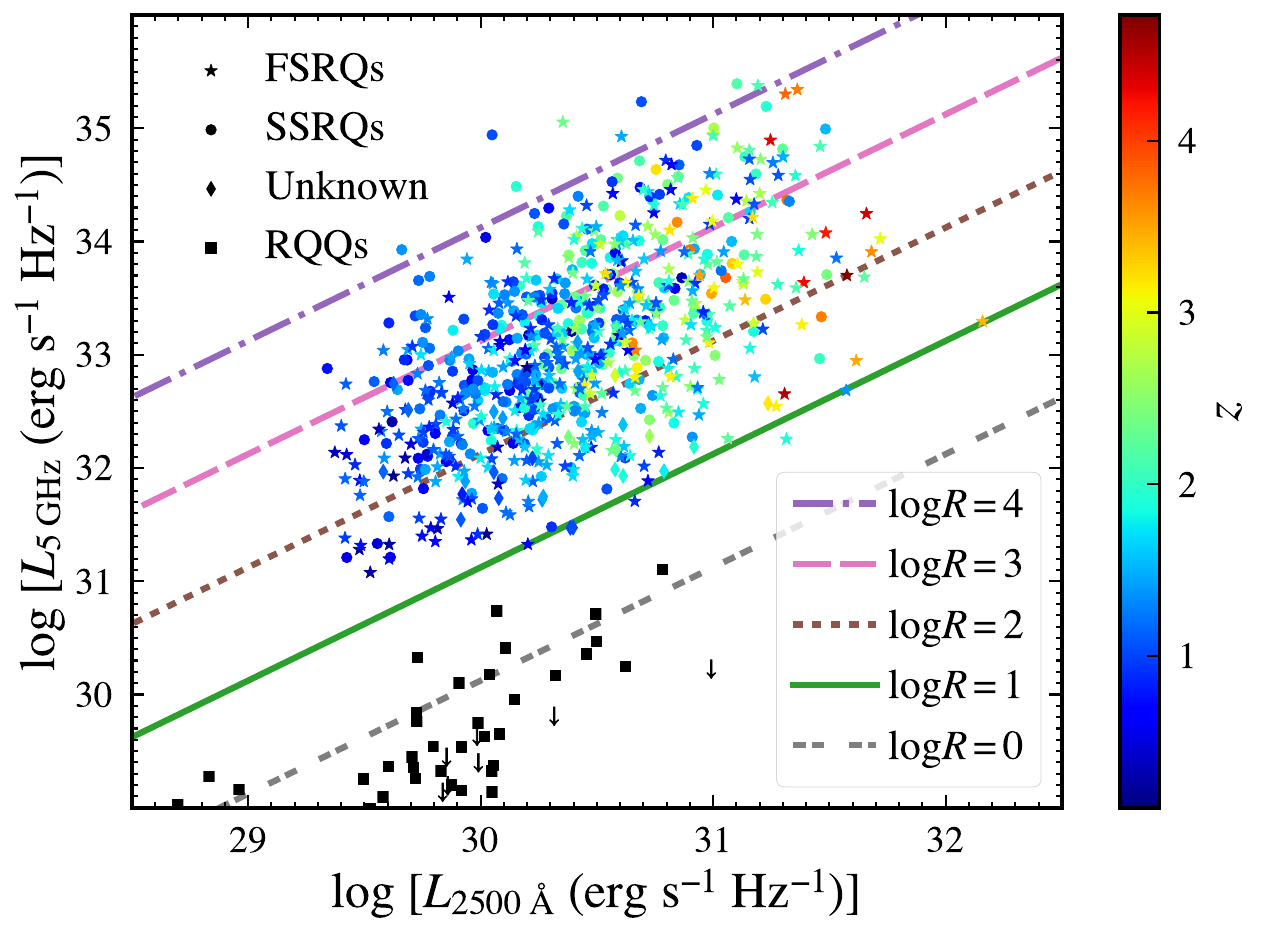}
\caption{The optically selected RLQ sample in the $L_\mathrm{radio}$-$L_\mathrm{uv}$ plane, where the data points are color-coded by their redshifts.
Different symbols have been used to indicate the radio spectral information.
Five lines defined by constant radio-loudness parameters are also shown.
We also show 59 RQQs from \citet{laor2008}, where squares and downward arrows are detections (50 quasars) and non-detections (9 quasars) in the radio band, respectively.}
\label{fig:Luv-Lr}
\end{figure*}

\begin{figure}
\centering
\includegraphics[width=0.48\textwidth, clip]{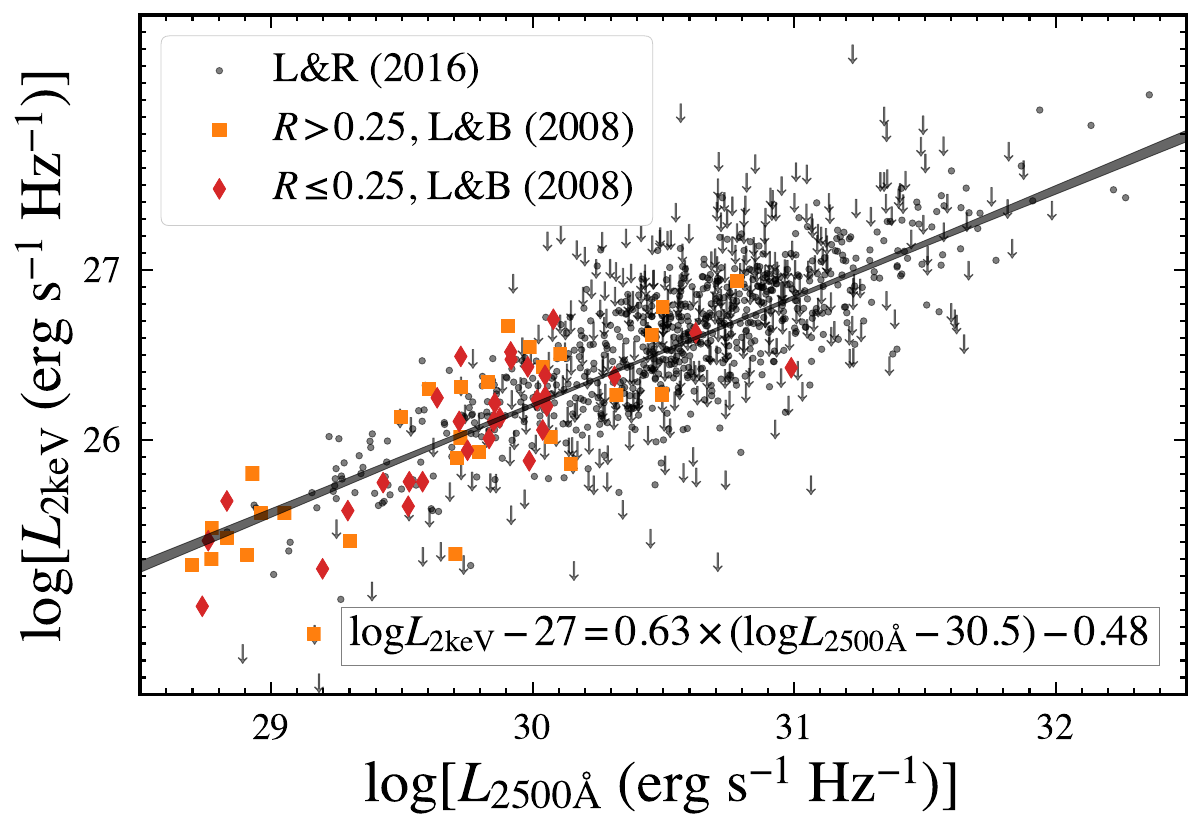}
\caption{The RQQ samples in the $L_\mathrm{x}$--$L_\mathrm{uv}$ plane.
The black symbols are (SDSS) RQQs from \citet{lusso2016}, while the (PG) RQQs from \citet{laor2008} are further divided into two bins (orange squares and red diamonds) by their median radio-loudness parameter ($R=0.25$).
Fitting a line to the black data points results in the equation at the lower-right corner (see Table~\ref{tab:pars}).
The uncertainties of this relation are indicated by the black-shaded region.
The PG quasars are consistent with the fitted $L_\mathrm{x}$--$L_\mathrm{uv}$ relation.
No apparent radio dependence for PG RQQs is found.}
\label{fig:rqq}
\end{figure}

\subsection{Improving the Miller et al. (2011) RLQ sample}
\label{sec:vlassM11}
\citet{miller2011} construct a sample of 791 quasars with $\log R^*\ge 1$.\footnote{\citet{miller2011} use $R^*\equiv L_\mathrm{5GHz}/L_\mathrm{2500\angstrom}$
to quantify radio-loudness, which can be converted to $R$ using $\log R=\log R^*-0.123$.
In \citet{miller2011}, quasars with $1\le\log R^*<2$ and $\log R^*\ge2$ are referred to as radio-intermediate quasars (RIQs) and RLQs, respectively.
We call all their quasars RLQs, even though some of them have $\log R$ values slightly smaller than unity.}
The overall X-ray detection fraction of this sample is high (85\%).
Their full sample consists of 654 optically selected primary RLQs,
and the rest are supplementary RLQs that are generally not optically selected.
The X-ray data for the primary sample are from archival {\it Chandra}, {\it XMM-Newton}, and {\it ROSAT} observations,
most (86\%) of which are serendipitous and thus unbiased with respect to RLQ properties.

However, about half of their primary RLQs lack radio spectral information.
Furthermore, the primary sample of \citet{miller2011} includes 312 spectroscopically confirmed quasars culled from the SDSS DR5 Quasar catalog (DR5Q; \citealt{schneider2007}),
while the remaining 342 quasars are photometrically selected but not spectroscopically confirmed (\citealt{richards2009}).
The improvement upon RLQs from \citet{miller2011} thus mainly comes from higher fractions of radio-slope measurements and spectroscopic redshifts.

We measure the radio slopes using the same method as for newly selected RLQs
by matching with radio sky surveys, NED and VizieR, and the quick-look images of VLASS.
The spectral indexes are measured for 719 quasars, making up 91.1\% of all quasars from \citet{miller2011}.
As to the optically selected RLQs, their radio-slope measurements are 91.4\% complete.
Note that to obtain a final data set with uniform measurements in the radio bands,
we also update the total radio fluxes for RLQs from \citet{miller2011} using NVSS as in \S~\ref{sec:init}.

For the 342 photometrically selected quasars in the primary sample of \citet{miller2011}, SDSS/BOSS/eBOSS spectra
from the SDSS/BOSS spectrograph can now be found for 183 of them.
We exclude 9 quasars whose rest-frame optical/UV spectra are BL Lac-like with a featureless continuum.
Another 9 quasars are excluded because they are likely to be BAL quasars.
For the rest (165) of the quasars with new spectroscopic measurements, we replace their photometric redshifts with spectroscopic
redshifts and update their luminosities accordingly. Note that catastrophic failures in the
photometric redshifts where $|z_\mathrm{photo}-z_\mathrm{spec}|/(1+z_\mathrm{spec})>0.15$ make up $\approx$5\% of these quasars,
suggesting that the photometric redshifts of the remaining (159) quasars that still lack spectroscopic measurements are largely reliable.

As for the new RLQs, we also gather multi-wavelength data (mid-infrared to UV) for the sample of \citet{miller2011}
and apply color cuts described in \S~\ref{sec:cc}, which eliminate 93 objects.
Another 9 objects are excluded since their X-ray observations do not satisfy the criteria in \S~\ref{sec:init}.

\subsection{The final optically selected RLQ sample}
We merge the newly selected sample with the refined sample of \citet{miller2011}, resulting in 940 unique RLQs.
Note that some quasars might be present in both samples,
mainly because many photometric quasars of \citet{miller2011} now have optical/UV spectra.
For the physical quantities describing these quasars, we use their newly measured values in this paper.
Since some quasars in the SDSS QSO catalog are selected by their radio or X-ray properties,
we utilize only the optically selected subset (789 objects) that does not suffer from apparent radio/X-ray selection biases
in the following sections.
Note that 49 quasars are not considered since they either have peaked radio spectra or are compact objects with steep 
spectra.\footnote{\label{footnote:ccs} Compact steep spectrum objects generally show a radio spectrum that is peaked around 100 MHz.
We exclude those quasars with core dominance $>0.9$ and radio slope $<-0.8$, even if a peak is not revealed.}
Another 11 quasars are excluded as rare cases that are known to be X-ray weak (2 quasars; e.g. \citealt{risaliti2005}),
are in the vicinity of luminous clusters (1 quasar; e.g. \citealt{rykoff2009}),
or have exceedingly flat X-ray spectra signifying strong absorption (8 quasars; e.g. \citealt{miyaji2006}).
The properties of the resulting 729 RLQs are listed in Table~\ref{tab:list}.

We show in Fig.~\ref{fig:compoSED} the infrared-to-UV composite SED of the final optically selected RLQs,
which has the blue color of typical Type~I quasars that are dominated by the emission from the accretion disk.
Note that the composite SED (orange) in Fig.~\ref{fig:compoSED} utilizes $\approx10^4$ photometric data points of $>700$ quasars.
In comparison, we show the composite SED of RLQs from \citet{shang2011} in the same plot.
Fig.~\ref{fig:compoSpec} shows the composite median spectra for various subsets of optically selected RLQs.
They are divided into redshift, radio slope, and $R$ bins in the three panels of Fig.~\ref{fig:compoSpec}.
Prominent emission lines are present in all composite median spectra, indicating that
the optical/UV continuum of the RLQs is generally not contaminated by boosted jet emission.
Indeed, from the catalogs of \citet{shen2011} and \citet{paris2017}, 
573/729 quasars of the final sample have REW measurements for at least one emission line among (C~{\sc iv}, Mg~{\sc ii}, H$\beta$), 
and the median REWs are $(42_{-2}^{+3}$, $37_{-2}^{+2}$, $77_{-8}^{+6})\;\angstrom$ and $(48_{-5}^{+3}, 40_{-2}^{+1}, 83_{-6}^{+8})\;\angstrom$ for FSRQs and SSRQs, respectively.
This indicates that even the FSRQs in our sample do not have a substantial optical/UV continuum contribution from a jet.

We show the final sample in the $L_\mathrm{5GHz}$-$L_\mathrm{2500\angstrom}$ plane in Fig.~\ref{fig:Luv-Lr}.
The color of each RLQ indicates its redshift, according to the color bar on the right-hand side.
Furthermore, we show four lines on the plane that are defined by constant radio-loudness parameters.

\subsection{Comparison optically selected RQQ samples}
\label{sec:rqq}
To compare the X-ray properties of RLQs with those of RQQs, we also utilize
a large sample of RQQs from \citet{lusso2016}.
These RQQs were selected from the SDSS quasar catalog,
and the X-ray data are exclusively from {\it XMM-Newton} observations.
Following \citet{lusso2016}, we select RQQs from their Table~1 and Table~2 that satisfy $E(B-V)\le0.1$, $S/N>5$,
and $1.9\le\Gamma_\mathrm{x}\le2.8$ from the main sample (cf. Table~3 of \citealt{lusso2016}).
We further exclude a small number of RQQs that are not optically selected.
The resulting sample has a size of 1074, among which 699 have detected X-ray emission.
We use this sample to constrain the $L_\mathrm{x}$--$L_\mathrm{uv}$ relation,
which is established for RQQs that span $>4$ decades in luminosity (e.g. \citealt{just2007, lusso2016})
and show no redshift dependence up to $z>6$ (e.g. \citealt{vito2019}).

Furthermore, we utilize $z<0.5$ PG quasars from \hbox{\citet{boroson1992}},
which are optically selected and have relatively deep radio constraints (e.g. \citealt{kellermann1989, kellermann1994}).
We only consider a subsample of 59 RQQs that do not have strong C{\sc iv} absorption from \citet{laor2008},
50 of which have detected radio emission while the remaining 9 have upper limits.
All 59 quasars are detected by {\it ROSAT} in X-rays (\citealt{brandt2000, steffen2006})
except for one (i.e. PG 1259$+$593), which is instead detected by {\it XMM-Newton} (\citealt{ballo2012}).

The two RQQ samples are shown in Fig.~\ref{fig:rqq}, where the RQQs from \citet{lusso2016} are shown as black dots (detections) and downward arrows (non-detections).
We fit the $L_\mathrm{x}$--$L_\mathrm{uv}$ relation of this sample using the method described in \S~\ref{sec:method}.
The results are given in both Fig.~\ref{fig:rqq} and Table~\ref{tab:pars}, which are consistent with those of \citet{lusso2016}.
The 59 PG quasars are consistent with the $L_\mathrm{x}$--$L_\mathrm{uv}$ relation in Fig.~\ref{fig:rqq}.
The 20th and 80th percentiles of their radio-loudness parameters are $R=0.10$ and $R=0.64$, respectively.
We divide these PG quasars into two bins separated by their median radio-loudness parameter ($R=0.25$) to assess potential radio dependence,
which is not found in Fig.~\ref{fig:rqq}.
However, we do not treat this statement as conclusive since the sample has a small size and contains only low-luminosity quasars.

\begin{figure}
\centering
\includegraphics[width=0.47\textwidth, clip]{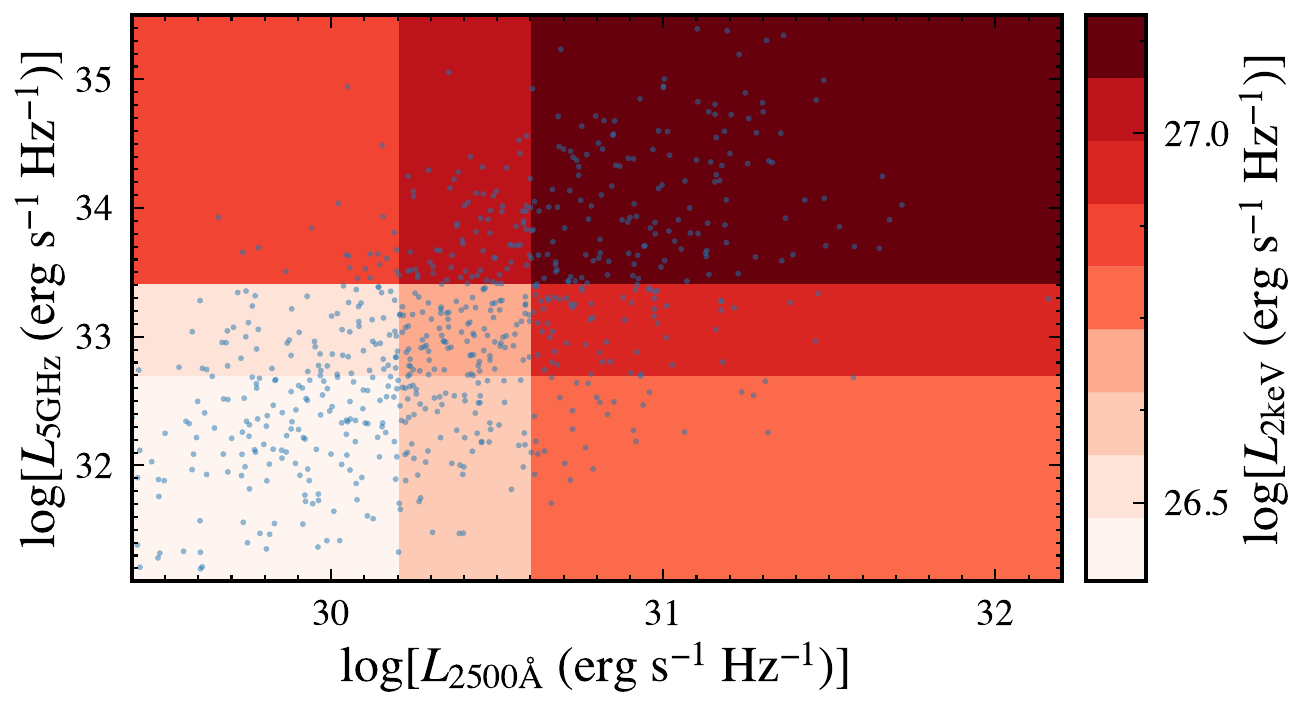}\\
\includegraphics[width=0.47\textwidth, clip]{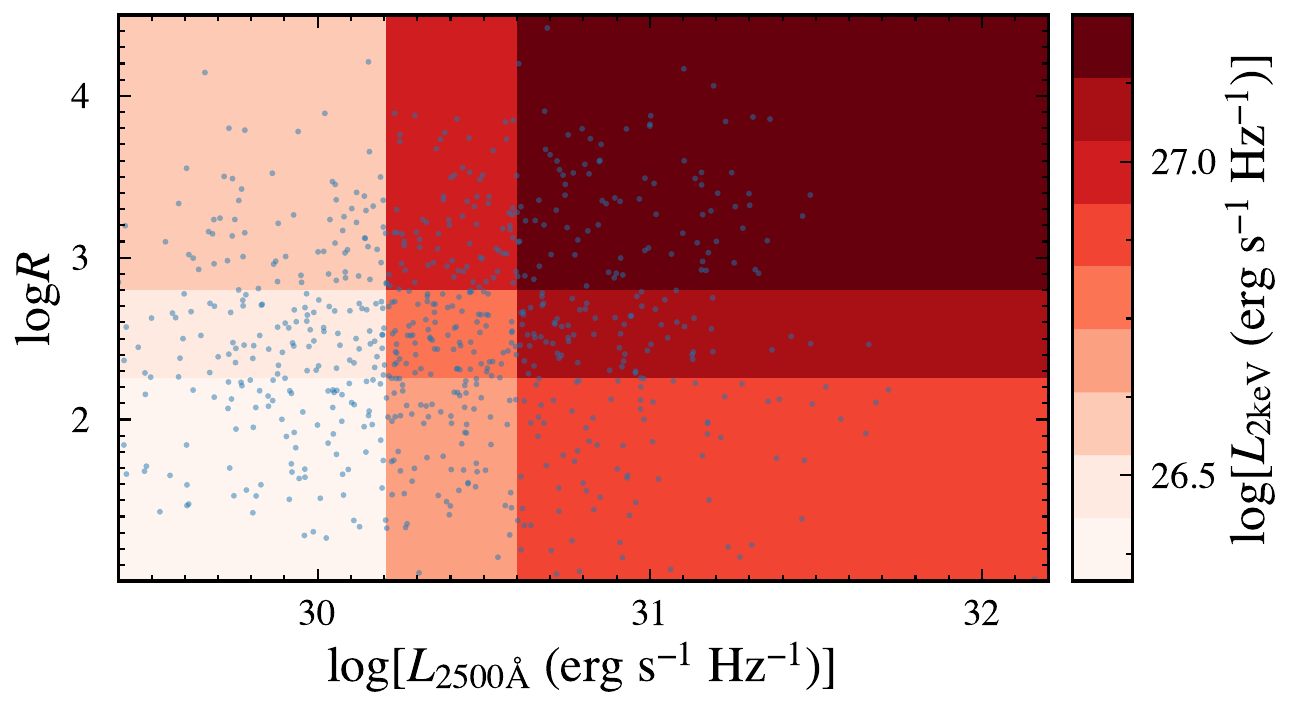}
\caption{Top: The dependence of X-ray luminosity on $\log L_\mathrm{5GHz}$ ($y$-axis) and $\log L_\mathrm{2500\angstrom}$ ($x$-axis).
The scattered points are all optically selected RLQs, which are further grouped into nine bins.
Those bins are separated by the 33rd and 66th percentiles along each axis.
We calculate the median X-ray luminosity of each bin as indicated by the color bar on the right-hand side.
    Clearly, $L_\mathrm{x}$ increases with both increasing $L_\mathrm{5GHz}$ and $L_\mathrm{2500\angstrom}$.
Bottom: Same as the top panel for the $R$--$L_\mathrm{2500\angstrom}$ plane.}
\label{fig:2dbinALT}
\end{figure}

\begin{figure}
\centering
\includegraphics[width=0.45\textwidth, clip]{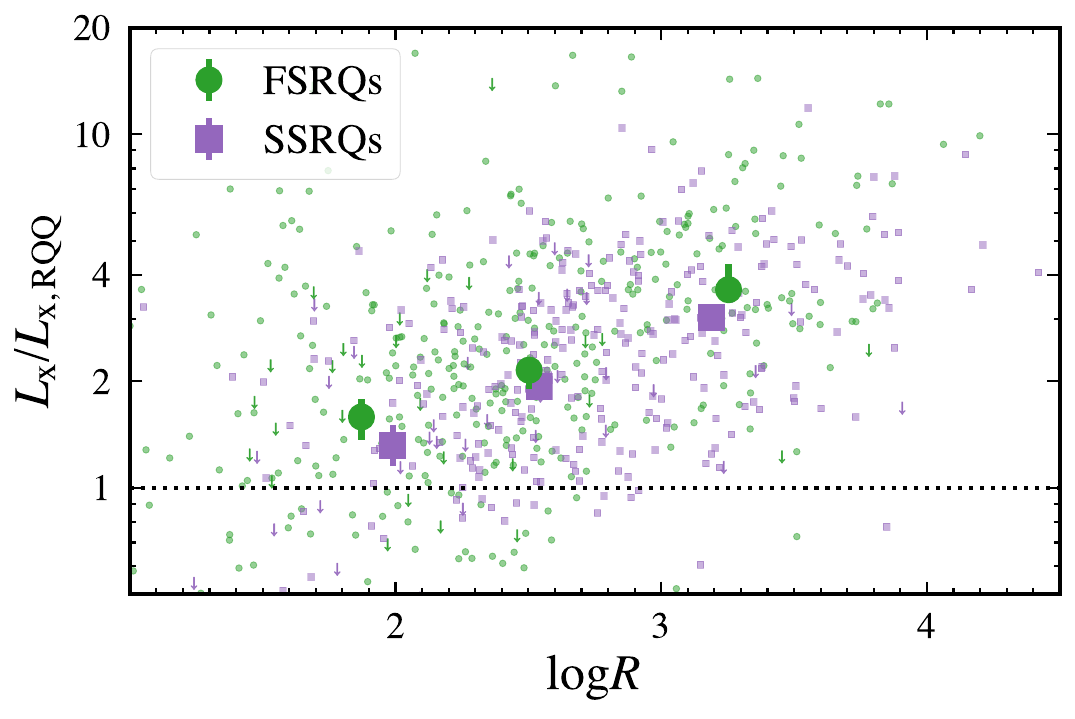}
\caption{The X-ray luminosities of FSRQs (green dots) and SSRQs (purple squares) over
those of RQQs at given $L_\mathrm{2500\angstrom}$ are plotted against $\log R$.
We group each sample into three $R$ bins and calculate their medians and uncertainties.}
\label{fig:lxOverlxRqq}
\end{figure}

\begin{figure}
\centering
\includegraphics[width=0.47\textwidth, clip]{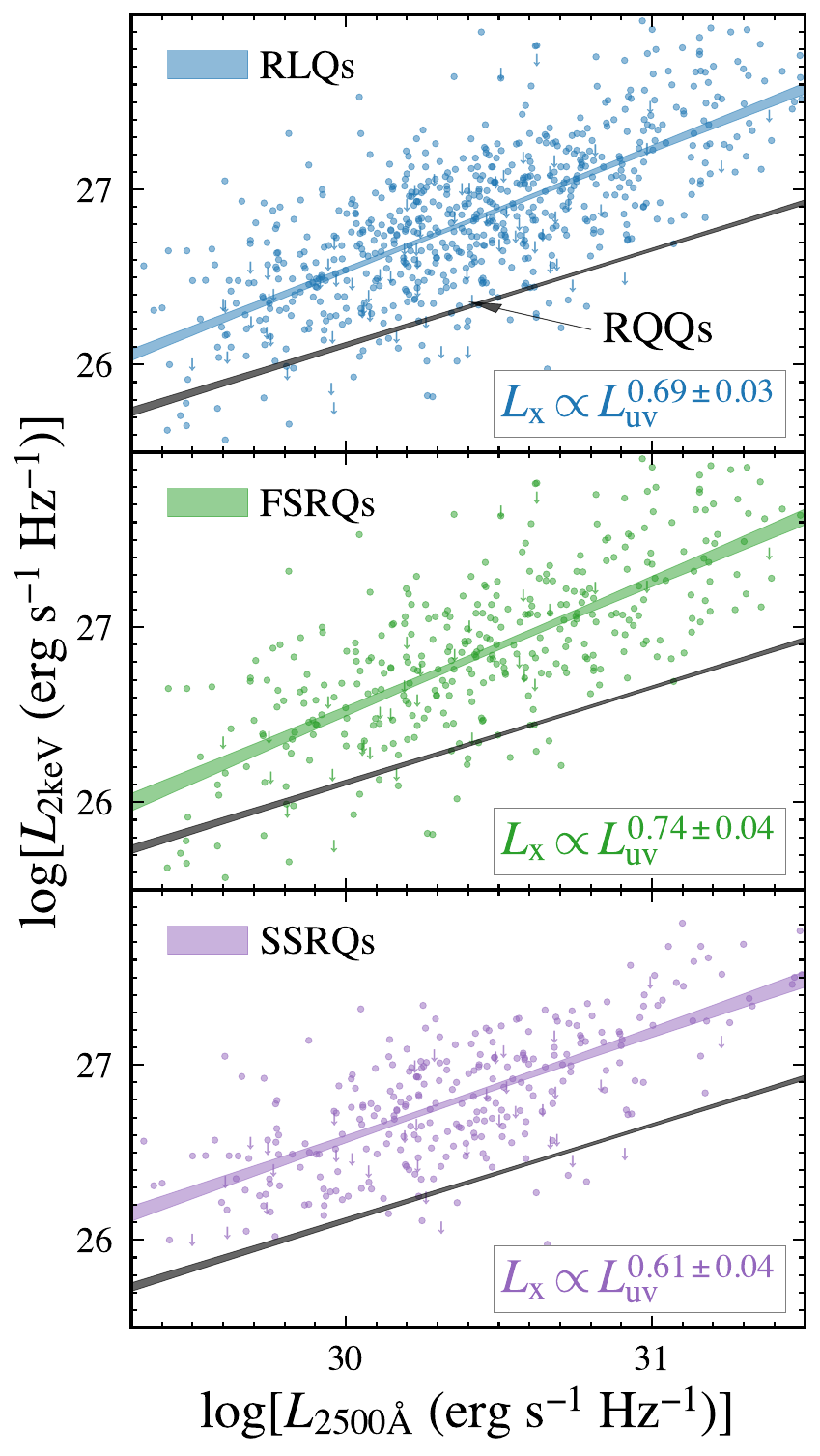}
\caption{The $L_\mathrm{x}$--$L_\mathrm{uv}$ relations for all RLQs, FSRQs, and SSRQs, from top to bottom.
The resulting slope with uncertainty estimation is at the lower-right corner of each panel.
The $L_\mathrm{x}$--$L_\mathrm{uv}$ relation for RQQs (black) is also plotted for comparison.
At given optical/UV luminosity, RLQs are more X-ray luminous than RQQs in all panels.
The slope of SSRQs is consistent with that of RQQs ($\gamma=0.63_{-0.02}^{+0.02}$), 
    while the slope of FSRQs is steeper at a $\approx2.5 \sigma$ significance level.}
\label{fig:lx_luv_tri}
\end{figure}

\begin{figure*}
\centering
\includegraphics[width=0.8\textwidth, clip]{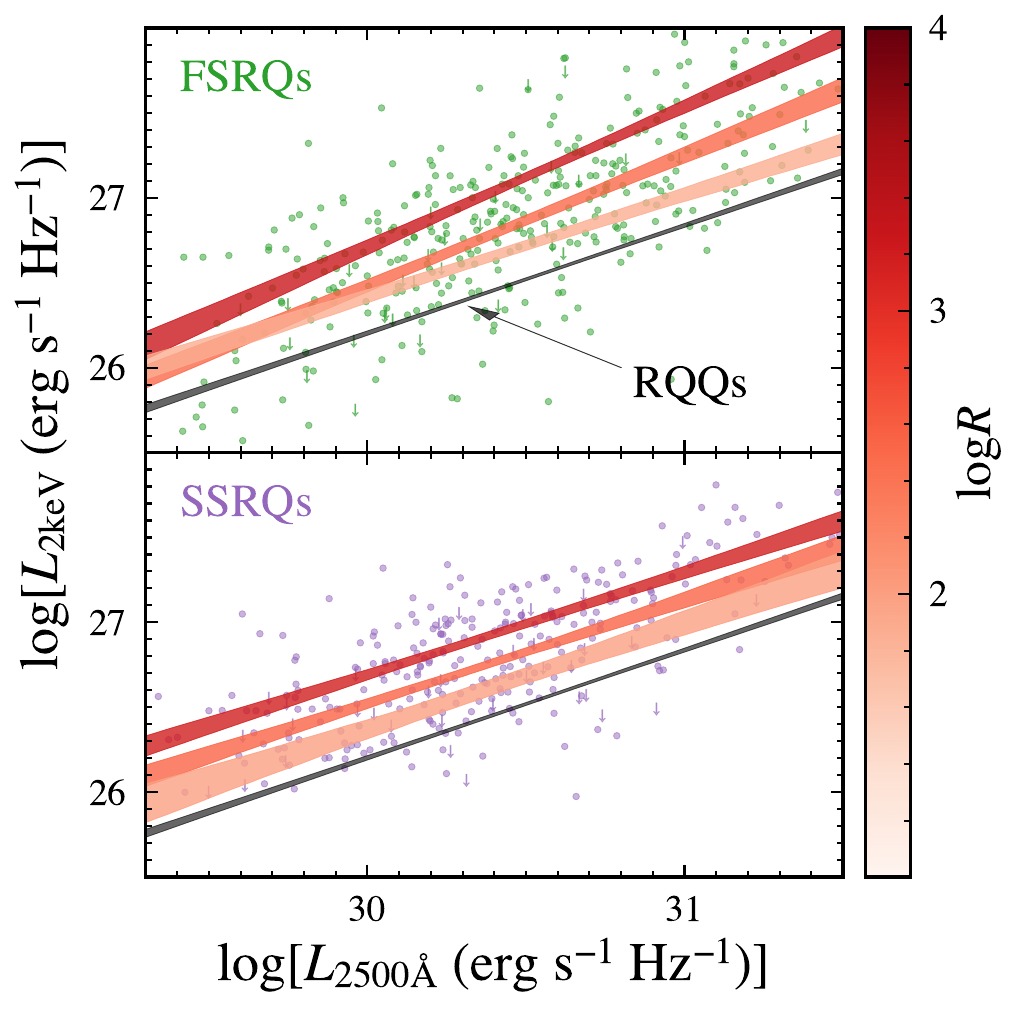}
\caption{The $L_\mathrm{x}$--$L_\mathrm{uv}$ relation for FSRQs (top) and SSRQs (bottom) divided into three $R$ bins as indicated by the color-bar on the right-hand side.
The slopes and intercepts for the relations are given in Fig.~\ref{fig:contour}.
The $L_\mathrm{x}$--$L_\mathrm{uv}$ relation for RQQs (black) is plotted for comparison.
For SSRQs, the slope is always consistent with that of RQQs while the intercept increases with $R$.
FSRQs follow another pattern where both the intercept and slope increase with $R$.}
\label{fig:lx_luv_bi}
\end{figure*}

\begin{figure}
\centering
\includegraphics[width=0.48\textwidth, clip]{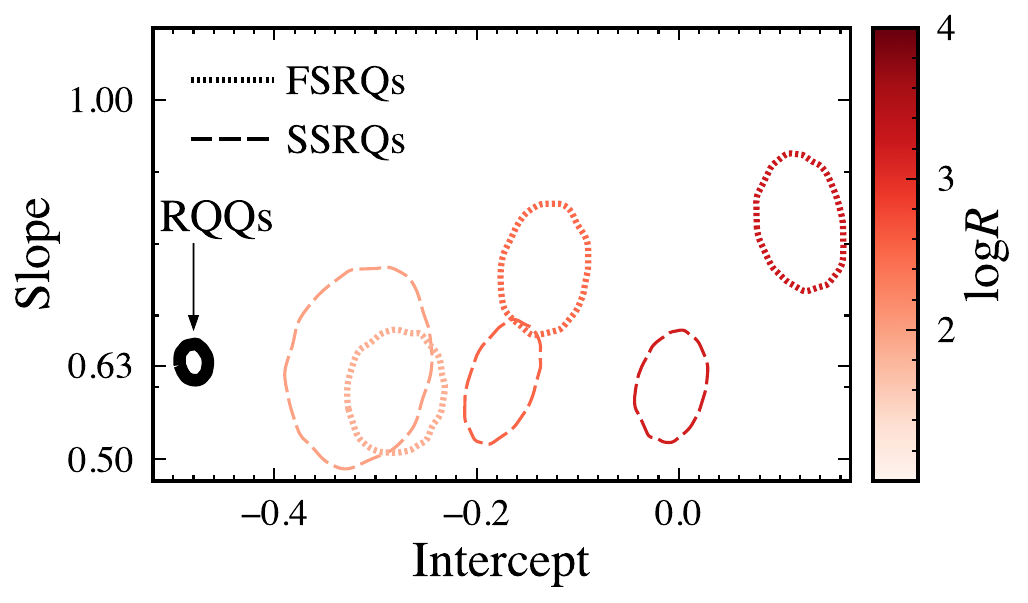}
\caption{The $1\sigma$ confidence regions for slope and intercept of the $L_\mathrm{x}$--$L_\mathrm{uv}$ relations in Fig.~\ref{fig:lx_luv_bi}.}
\label{fig:contour}
\end{figure}

\section{The relation between X-ray, Optical/UV, and Radio luminosities}
\label{sec:fitting}

\subsection{Insights from scatter plots}
\label{sec:scatter}
In the top panel of Fig.~\ref{fig:2dbinALT}, the $L_\mathrm{radio}$--$L_\mathrm{uv}$ plane is divided into $3\times3$ sub-regions,
the boundaries between which are the 33rd and 66th percentiles along each axis.
For each bin, the median $L_\mathrm{x}$ is calculated and indicated by the depth of the color according to the color bar on the right-hand side.
The median $L_\mathrm{x}$ increases in the positive direction for both $L_\mathrm{radio}$ and $L_\mathrm{uv}$.
We repeat the analysis for the $R$--$L_\mathrm{uv}$ plane in Fig.~\ref{fig:2dbinALT} (bottom),
where the distribution of data points is more homogeneous than that in the upper panel of Fig.~\ref{fig:2dbinALT}.
Both figures indicate that there are strong radio and optical/UV dependences
for the X-ray luminosities of RLQs.

We investigate the radio dependence for FSRQs and SSRQ separately in Fig.~\ref{fig:lxOverlxRqq},
where their X-ray luminosities divided by those of RQQs at given $L_\mathrm{uv}$ (using the $L_\mathrm{x}$--$L_\mathrm{uv}$ relation in Fig.~\ref{fig:rqq}) are shown as a function of $\log R$.
FSRQs and SSRQs are each divided into three $R$ bins by the 33rd and 66th percentiles of all RLQs ($\log R=2.25$ and 2.78).
The medians with estimated uncertainties for each bin are calculated.
In Fig.~\ref{fig:lxOverlxRqq}, the \mbox{X-ray} luminosities of both RLQ types are larger than those of RQQs and show strong dependence on $\log R$.
The most X-ray luminous objects at each $\log R$ are almost always FSRQs, probably due to their seemingly larger scatter than that of SSRQs.
Furthermore, the medians of FSRQs are larger than those of SSRQs by 10\%--20\%.

The $L_\mathrm{x}$--$L_\mathrm{uv}$ relations for RLQs and RQQs are compared in Fig.~\ref{fig:lx_luv_tri} (top).
The slope of the regression line for RLQs is slightly steeper than that for RQQs (i.e. $\gamma=0.63_{-0.02}^{+0.02}$).
Note that the fitting method here is the same as for Fig.~\ref{fig:rqq} and is described in \S~\ref{sec:method}.
In the middle and bottom panels of Fig.~\ref{fig:lx_luv_tri}, FSRQs and SSRQs are compared with RQQs, respectively.
The $L_\mathrm{x}$--$L_\mathrm{uv}$ relation for FSRQs (slope$=0.74_{-0.04}^{+0.04}$)  appears steeper (at about $2.5\sigma$ significance) than that for RQQs,
while the slope for SSRQs (slope$=0.61_{-0.04}^{+0.04}$) is consistent with $\gamma$ (to within $\approx3\%$).
Therefore, the X-ray luminosities of FSRQs and SSRQs probably have different radio and optical/UV dependences (at about $2.3\sigma$ significance) as indicated by Fig.~\ref{fig:lxOverlxRqq} and Fig.~\ref{fig:lx_luv_tri}.

In short, the X-ray luminosity of RLQs depends on $L_\mathrm{uv}$, $L_\mathrm{radio}$ (or $R$), and
radio slope,\footnote{The radio slope reflects, roughly, the direction of the jet relative to our line of sight.}
which has also been concluded by many previous works (e.g. \citealt{worrall1987, brinkmann1997, grandi2006, miller2011}).
In addition to the disk-corona interplay that is revealed by the $L_\mathrm{x}$--$L_\mathrm{uv}$ relation for RQQs,
we probably need at least two more mechanisms, or one mechanism that is affected by two key parameters, for those dependences in RLQs.
We focus on the $L_\mathrm{x}$--$L_\mathrm{uv}$ relation and consider $R$ and radio slope as controlled additional factors.
Specifically, we fit the $L_\mathrm{x}$--$L_\mathrm{uv}$ relations for the
three $R$ bins of FSRQs and SSRQs (as in Fig.~\ref{fig:lxOverlxRqq}), separately, and show the results in Fig.~\ref{fig:lx_luv_bi}.
The corresponding $1\sigma$ confidence regions are in Fig.~\ref{fig:contour}.

A notable pattern for SSRQs emerges that the slope of the $L_\mathrm{x}$--$L_\mathrm{uv}$ relation
is always consistent with that for RQQs, while the intercept increases monotonically with $R$.
If the excess \mbox{X-ray} luminosity of SSRQs relative to RQQs is caused by emission from the core region of the jets,
the contribution of this component increases with $R$, as expected.
However, two properties are required further by Fig.~\ref{fig:lx_luv_bi} (bottom)
that (a) the jet component depends strongly on $L_\mathrm{uv}$ and (b) the slope of this dependence is consistent with $\gamma$.
For example, the SSRQs in the most radio-loud bin are on average a factor of about 3 more X-ray luminous than RQQs (see Fig.~\ref{fig:lxOverlxRqq}), which
means that the jet component (if it exists) dominates their X-ray luminosities;
the dependence of their $L_\mathrm{x}$ on $L_\mathrm{uv}$ is still strong with a slope $=0.60_{-0.05}^{+0.05}$ (see Fig.~\ref{fig:contour}),
which is only different from $\gamma$ by $\approx5\%$.
Indeed, perhaps we do not need two distinct X-ray components (i.e. from the corona and jet core)
that are indistinguishable with regard to their correlations with $L_\mathrm{uv}$ and
their spatial properties.\footnote{The current X-ray telescope with superb sub-arcsec resolution (i.e. {\it Chandra}) cannot resolve the nuclear X-ray emission of RLQs into different spatial components.}
A more natural and compact explanation is that the X-ray emission of SSRQs reveals only one component and is produced in the same way as for RQQs.;
this component is attributed to the hot corona 
that is coupled with the disk (i.e. the $L_\mathrm{x}$--$L_\mathrm{uv}$ relation).
Furthermore, there is a connection between the activity of the corona (i.e. the intercept of the $L_\mathrm{x}$--$L_\mathrm{uv}$ relation) in the immediate vicinity of the SMBH
and that of the large-scale jets (i.e. $R$), such that those RLQs harboring more powerful relativistic jets also have more powerful coronae.\footnote{In  principle, another possible
explanation for Fig.~\ref{fig:lx_luv_bi} (bottom)
could be that even RQQs have a jet-linked X-ray continuum (as many RQQs do have small jets, and some envision the X-ray
corona in radio-quiet objects to be the base of a jet). Following this explanation, we should expect a corona-jet connection for RQQs as well, which is, however, not supported (see \S~\ref{sec:ssrqs} and Fig.~\ref{fig:norm}).
More generally, luminosity correlations and X-ray spectral properties support a thermal Compton-scattering origin for the 
X-ray continuum of RQQs (see \S~\ref{sec:intro}).}

The results for FSRQs are not as clear in Fig.~\ref{fig:lx_luv_bi} and Fig.~\ref{fig:contour}.
For the lowest $R$ bin, the slope is consistent with $\gamma$, supporting the idea that the observed X-rays are also dominated by the corona,
and the intercept is consistent with that of SSRQs within error bars.
However, the slope increases with $R$ to 0.7--0.9 for FSRQs in the other two bins,
and the intercepts are also larger than for the corresponding SSRQs (although only suggestively for the second $R$ bin).
Therefore, the $L_\mathrm{x}$--$L_\mathrm{uv}$ relation for the moderate-to-most radio-loud FSRQs probably requires additional mechanisms.
However, we stress that the medians of FSRQs and SSRQs of similar radio-loudness in Fig.~\ref{fig:lxOverlxRqq} are only different at a 10\%--20\% level.
Therefore, the additional mechanisms affecting FSRQs are likely secondary
such that their X-ray luminosities are generally controlled by the enhancement of the putative coronal
emission as for SSRQs.\footnote{In consequence, this statement is probably true for almost all RLQs.}
We test in the next section whether a distinct component from the jet core might play an important role or not.

\begin{table*}
\centering
\caption{Models used to explain the X-ray luminosity of RLQs.}
\label{tab:models}
\begin{threeparttable}[b]
\begin{tabular}{lcccc}
\hline
\hline
    Functional Model & \multicolumn{4}{c}{Decomposition\tnote{a}\ \ : $L_\mathrm{x}=\mathcal AL_\mathrm{uv}^{\Gamma_\mathrm{uv}}+\mathcal BL_\mathrm{radio}^{\Gamma_\mathrm{radio}}$}\\
    \cmidrule(lr){2-5}
    & $\mathcal A$ & $\Gamma_\mathrm{uv}$ & $\mathcal B$ & $\Gamma_\mathrm{radio}$ \\
\hline
\multicolumn{5}{l}{Radio-loud quasars}\\
\hline
\RomanNumeralCaps 1: $\log L_\mathrm{2\;keV}=\alpha+\gamma_\mathrm{uv}\log L_\mathrm{2500\;\angstrom} + \gamma_\mathrm{radio}^\prime\log L_\mathrm{5\;GHz}$&$10^\alpha\Big(\frac{2500}{4400}\Big)^{\gamma_\mathrm{radio}^\prime-0.5}R^{\gamma_\mathrm{radio}^\prime}$ &$\gamma_\mathrm{uv}+\gamma_\mathrm{radio}^\prime$ &- &- \\
\RomanNumeralCaps 2: $\log L_\mathrm{2\;keV}=\log\Big(AL_\mathrm{2500\;\angstrom}^{\gamma_\mathrm{uv}}+BL_\mathrm{5\;GHz}^{\gamma_\mathrm{radio}}\Big)$&$A$&$\gamma_\mathrm{uv}$&$B$ & $\gamma_\mathrm{radio}$\\
\RomanNumeralCaps 3: $\log L_\mathrm{2\;keV}=\log\Big(AL_\mathrm{5\;GHz}^{\gamma_\mathrm{radio}^\prime}L_\mathrm{2500\;\angstrom}^{\gamma_\mathrm{uv}}+BL_\mathrm{5\;GHz}^{\gamma_\mathrm{radio}}\Big)$& $A\Big(\frac{2500}{4400}\Big)^{\gamma_\mathrm{radio}^\prime-0.5}R^{\gamma_\mathrm{radio}^\prime}$&$\gamma_\mathrm{uv}+\gamma_\mathrm{radio}^\prime$&$B$&$\gamma_\mathrm{radio}$ \\
\hline
\multicolumn{5}{l}{Radio-quiet quasars}\\
\hline
$\log L_\mathrm{2keV}=\beta + \gamma\log L_\mathrm{2500\angstrom}$ & $10^\beta$ & $\gamma$ & - & - \\
\hline
\end{tabular}
\begin{tablenotes}
\item[a] The X-ray emission from RLQs is divided into a component that is from the
    disk/corona ($\mathcal AL_\mathrm{uv}^{\Gamma_\mathrm{uv}}$) and
    another component that is from the jet core ($\mathcal BL_\mathrm{radio}^{\Gamma_\mathrm{radio}}$),
    the latter of which may not be explicitly present in all the models (e.g. Model~\RomanNumeralCaps 1).
    The quantities that are supposed to have the same physical meaning and
    thus be comparable between different functional models are also listed.
\end{tablenotes}
\end{threeparttable}
\end{table*}

\subsection{Parameterized modeling}
\label{sec:reason}
Qualitatively similar plots to those in \S~\ref{sec:scatter} can also be found in previous works (e.g. \citealt{worrall1987, miller2011}),
although our quantitative constraints here are considerably tighter owing to our improved data quality.
For example, the results in Fig.~\ref{fig:contour} are in line with those in Fig.~1 of \citet{worrall1987}.
However, despite such consistent plots, this paper and the literature reach different conclusions.
Specifically, we suggest that the disk/corona of RLQs are systematically more X-ray luminous than those of RQQs,
while both \citet{worrall1987} and \citet{miller2011} attribute the excess X-ray emission of RLQs to the jet core.
These two interpretations need to be further compared.
In this section we use formal model selection to show that our interpretation in \S~\ref{sec:scatter} is indeed preferred by the data.
Furthermore, the varying contributions of the two additive X-ray components (jets and corona) 
are probably more easily revealed by direct model fitting than by
finding the proportionality of total X-ray luminosity with luminosities at other wavelengths (e.g. \S~\ref{sec:scatter}).

\subsubsection{Models for the X-ray-optical/UV-radio relation of RLQs}
\label{sec:model}

We first set up a decomposition for the \mbox{X-ray} emission from RLQs, and then discuss three functional models within this framework.
We write the X-ray luminosity of RLQs as
\begin{equation}
    L_\mathrm{x} = \mathcal{A} L_\mathrm{uv}^{\Gamma_\mathrm{uv}} + \mathcal{B}L_\mathrm{radio}^{\Gamma_\mathrm{radio}},
\label{eq:phy}
\end{equation}
where $\mathcal{A}L_\mathrm{uv}^{\Gamma_\mathrm{uv}}$ and $\mathcal BL_\mathrm{radio}^{\Gamma_\mathrm{radio}}$
represent the parts of the emission from the corona and jets, respectively.
Here, the \mbox{X-ray} luminosity from the disk/corona depends on the disk luminosity ($L_\mathrm{uv}$) to the power of $\Gamma_\mathrm{uv}$
and is subject to a normalization factor $\mathcal A$.
This $L_\mathrm{x}$--$L_\mathrm{uv}$ relation for the disk/corona emission of RLQs has
thus been assumed to be of similar functional form to that of RQQs.
Whether the specific parameters of this relation are universal to both RLQs and RQQs is not assumed.
Given that the X-ray and radio luminosities of the sample span 4--6 decades,
we adopt a power-law dependence for the case of the jet component as well (i.e. $L_\mathrm{x,jet}=\mathcal{B}L_\mathrm{radio}^{\Gamma_\mathrm{radio}}$).
This functional form is reasonably flexible and can accommodate many potential underlying physical emission mechanisms.
In practice, the data might not require both components to be present,
in cases where one of the normalization factors is consistent with zero (most likely $\mathcal B$).

In Table~\ref{tab:models}, we list three functional models that describe different
$L_\mathrm{2keV}$--$L_\mathrm{2500\angstrom}$--$L_\mathrm{5GHz}$ relations,
all of which are interpreted in the context of Eq.~\ref{eq:phy}.
For example, when we assess the dependence of the X-ray luminosity on the optical/UV luminosity of the disk/corona, i.e. $\Gamma_\mathrm{uv}$,
we will use $\gamma_\mathrm{uv}+\gamma_\mathrm{radio}^\prime$ of Model~\RomanNumeralCaps 1 and Model~\RomanNumeralCaps 3,
while we will use $\gamma_\mathrm{uv}$ of Model~\RomanNumeralCaps 2.

Model~\RomanNumeralCaps 1 was utilized by previous works because
it describes a joint dependence of the \mbox{X-ray} luminosity on both the optical/UV and radio luminosities (see Fig.~\ref{fig:2dbinALT})
with the smallest number of parameters (e.g. \citealt{tananbaum1983, worrall1987, miller2011}).
To preserve consistency with past work, we still use the form $L_\mathrm{2keV}\propto L_\mathrm{2500\angstrom}^{\gamma_\mathrm{uv}}L_\mathrm{5GHz}^{\gamma_\mathrm{radio}^\prime}$
in model fitting, while we interpret the results using another form that
$L_\mathrm{2keV}\propto R^{\gamma_\mathrm{radio}^\prime}L_\mathrm{2500\angstrom}^{\gamma_\mathrm{uv}+\gamma_\mathrm{radio}^\prime}$.\footnote{Another advantage of this preference in model fitting is
that there is non-zero covariance between the measurement errors of $R$ and $L_\mathrm{2500\angstrom}$, while $L_\mathrm{5GHz}$ and $L_\mathrm{2500\angstrom}$ are independent measurements.}
We note from Fig.~\ref{fig:lx_luv_bi} that, for fixed $R$, the $L_\mathrm{x}$--$L_\mathrm{uv}$ relation of RLQs has a slope
such that $\Gamma_\mathrm{uv}=\gamma_\mathrm{uv}+\gamma_\mathrm{radio}^\prime\approx\gamma$ for the majority of bins.
Previous model fitting results using Model~\RomanNumeralCaps 1 (e.g. Table~7 of \citealt{miller2011}) are roughly consistent with this suggestion.
Here, a corona-jet connection is parameterized
so that the normalization factor (intercept) correlates with the radio-loudness parameter to the power of $\gamma_\mathrm{radio}^\prime$ 
(i.e. $\mathcal{A}\propto R^{\gamma_\mathrm{radio}^\prime}$).
An X-ray emitting region that is associated with the relativistic jets is not
explicitly included (i.e. $\mathcal{B}$ is set to zero).

Similar to Eq.~\ref{eq:phy}, Model~\RomanNumeralCaps 2 explicitly divides the \mbox{X-ray} emission of RLQs into
corona ($AL_\mathrm{2500\angstrom}^{\gamma_\mathrm{uv}}$) and jet core
($BL_\mathrm{5GHz}^{\gamma_\mathrm{radio}}$) components.
However, it stands for a special but commonly assumed scenario where $AL_\mathrm{2500\angstrom}^{\gamma_\mathrm{uv}}$ might be consistent with the X-ray luminosity of RQQs;
therefore, RLQs are treated as a basic combination of a radio-quiet quasar engine with additional powerful jets from the perspective of the X-rays (e.g. \citealt{worrall1987}).
For the jet component, the X-rays might correlate linearly (i.e. $\gamma_\mathrm{radio}=1$) with
the radio emission from the same region as suggested by previous works (e.g. \citealt{zamorani1984, browne1987, worrall1987}).
No apparent connection between the jets and the corona is present in this model;
each contribution is mathematically independent.
Model~\RomanNumeralCaps 2 has been applied to FSRQs (e.g. \citealt{zamorani1984, worrall1987}) and SSRQs (e.g. \citealt{zamorani1984}) in previous works.
With small-size samples, their results actually hinted at a difference between the coronae of RLQs and those of RQQs,
though no solid conclusion was reached.\footnote{\citet{zamorani1984} does not provide error bars.
\citet{worrall1987} have a slightly larger sample than that of \citet{zamorani1984}, but the constraint is still statistically insignificant.}

It is also possible that both Model~\RomanNumeralCaps 1 and Model~\RomanNumeralCaps 2 are describing part of reality.
In \S~\ref{sec:scatter}, we require a corona-jet connection for the differences between RQQs and SSRQs
and cannot rule out a jet component in FSRQs.
We thus propose Model~\RomanNumeralCaps 3 (see Table~\ref{tab:models}) that combines those features.
Model~\RomanNumeralCaps 1 and Model~\RomanNumeralCaps 2 are special cases of Model~\RomanNumeralCaps 3, where
either $B$ or $\gamma_\mathrm{radio}^\prime$ is set to zero,
respectively. From this point of view, we utilize only Model~\RomanNumeralCaps 3 and discuss whether its certain parameters are zero or not below.

\subsubsection{Methods}
\label{sec:method}
Following \citet{miller2011}, we first normalize the quasar luminosities to $\log L_\mathrm{2500\;\angstrom}=30.5$, $\log L_\mathrm{5\;GHz}=33.3$,
and $\log L_\mathrm{2\;keV}=27.0$, which are near to the median luminosities of our sample.
The maximum-likelihood estimates are then obtained for the models in Table~\ref{tab:models}.
The model fitting is performed using {\sc lmfit} (\citealt{lmfit}),
and the likelihood function takes into account the upper limits on the X-ray luminosities (e.g. \citealt{isobe1986}).
The maximum-likelihood and uncertainty estimations of model parameters
are calculated using a Markov chain Monte Carlo (MCMC) algorithm ({\sc emcee}; \citealt{emcee2013}).
This method is thus equivalent to a Bayesian approach with flat priors.
We initially leave all parameters free to vary to reveal the most general results.
However, we fix the parameter values for cases where the parameters are either unconstrained or unphysical (see details in Appendix~\ref{sec:approx}),
which also prevents the model performance from being imprecisely characterized by the model-selection methods we use below (e.g. \citealt{nelson2020}).

Second, as to model selection, the likelihood-ratio test (e.g. $F$-test)
is not applicable here since Model~\RomanNumeralCaps 1 and
Model~\RomanNumeralCaps 2 are not nested to each other (e.g. \citealt{protassov2002}).
We thus use standard information criteria (ICs) to compare the performance of models.
We adopt the widely used Akaike Information Criterion (AIC; \citealt{akaike1974})
and Bayesian Information Criterion (BIC; \citealt{schwarz1978}), which are defined as
\begin{align}
    \mathrm{AIC}&\equiv -2\ln\mathcal{L}_{\max}+2k, \label{eq:aic}\\
    \mathrm{BIC}&\equiv -2\ln\mathcal{L}_{\max}+k\ln N. \label{eq:bic}
\end{align}
Here, $\mathcal{L}_{\max}$ is the maximum likelihood, $k$ the number of free parameters of the model, and $N$ the number of data points.
Under such definitions, the model with the smallest information criterion is selected.
ICs thus favor high model likelihood ($\mathcal{L}_{\max}$) and penalize high model complexity ($k$).
BIC penalizes model complexity more severely than AIC when $N>7$.
The significance of the selection result (i.e. the evidence for the selected model) is indicated by $\Delta$AIC
(e.g. \S~2.6 of \citealt{burnham2002}) and $\Delta$BIC (e.g. \S~3.2 of \citealt{kass1995}) between models.
In this paper, AIC and BIC always agree on the best model,
and in most cases, the best models are strongly favored with $|\Delta\mathrm{AIC}|\gtrsim 5$ and $|\Delta\mathrm{BIC}|\gtrsim5$.
It is probably not safe to rely solely on statistical methods and to draw conclusions
without considering physical plausibility.
The results of the following two subsections are secured by the fact that the ICs-selected model
fits also depict the most reasonable physical pictures.

\subsubsection{Results of all RLQs without distinguishing radio slope}
\label{sec:base}

\begin{figure}
\centering
\includegraphics[width=0.42\textwidth, clip]{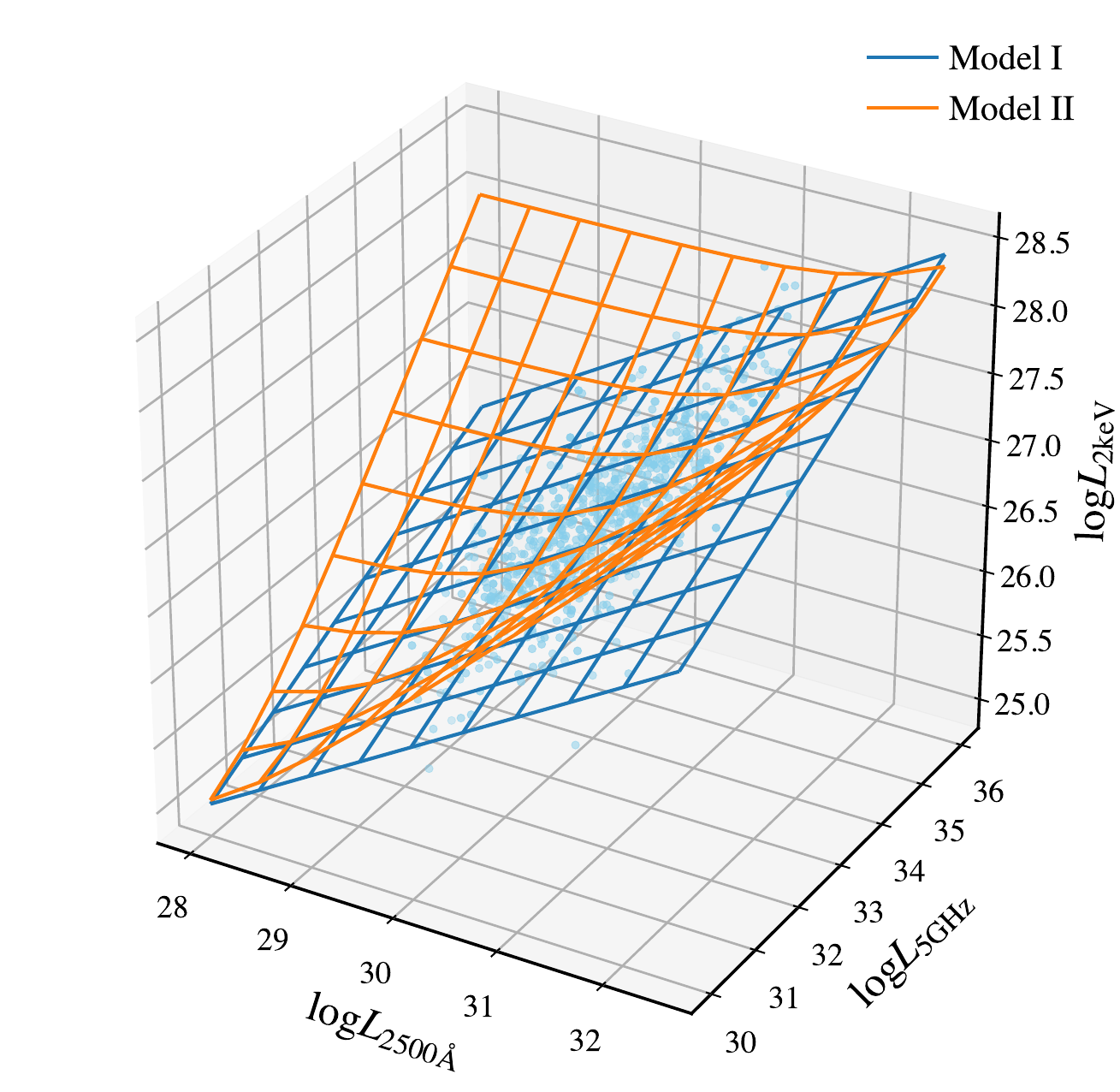}
\caption{The best-fit results of all optically selected RLQs using Model~\RomanNumeralCaps 1 (blue) and Model~\RomanNumeralCaps 2 (orange).
Model~\RomanNumeralCaps 1 defines a plane while Model~\RomanNumeralCaps 2 defines a curved surface.}
\label{fig:3d}
\end{figure}

\begin{figure*}
\centering
\includegraphics[width=0.42\textwidth, clip]{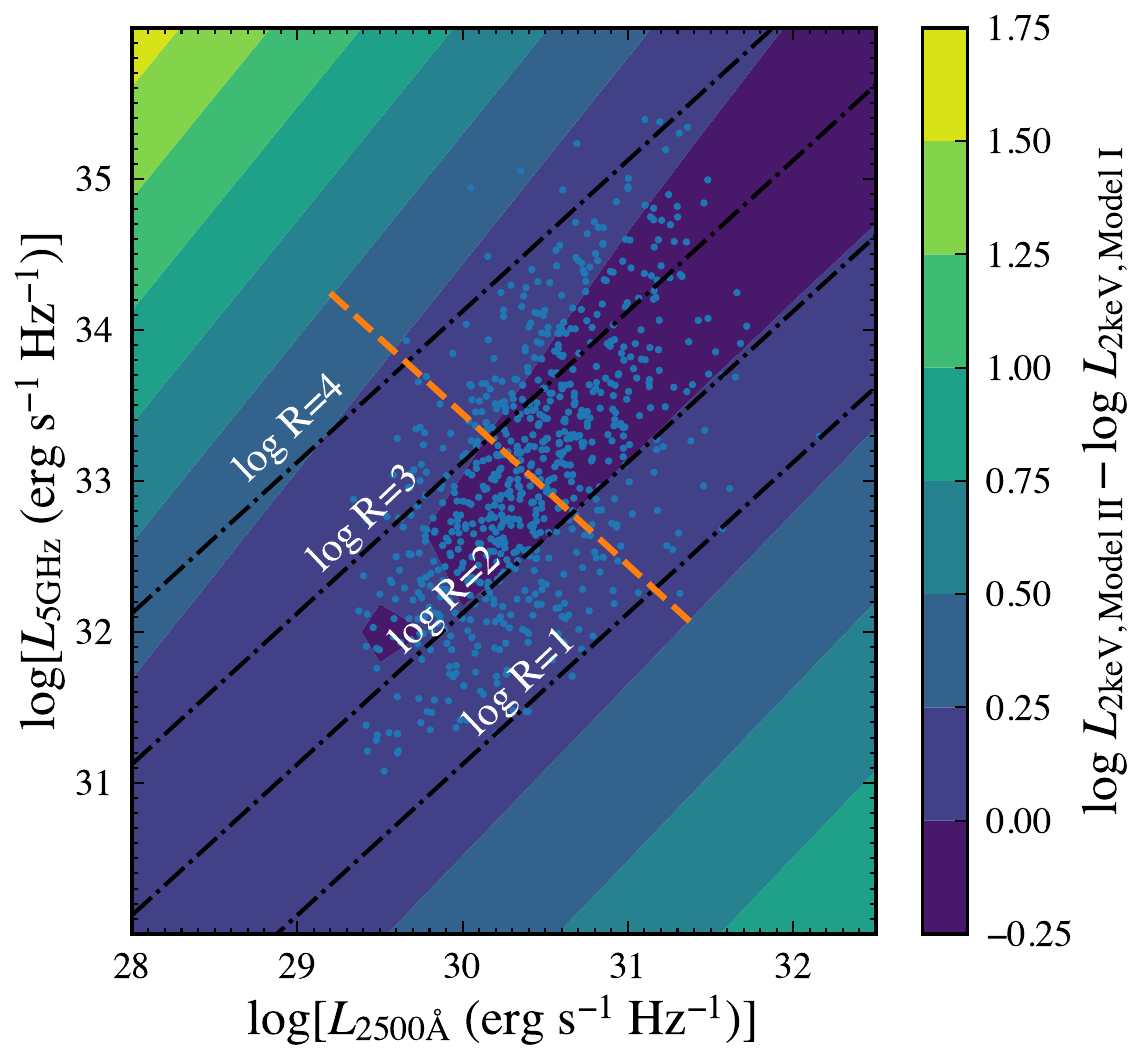}
\hspace{0.4cm}
\includegraphics[width=0.45\textwidth, clip]{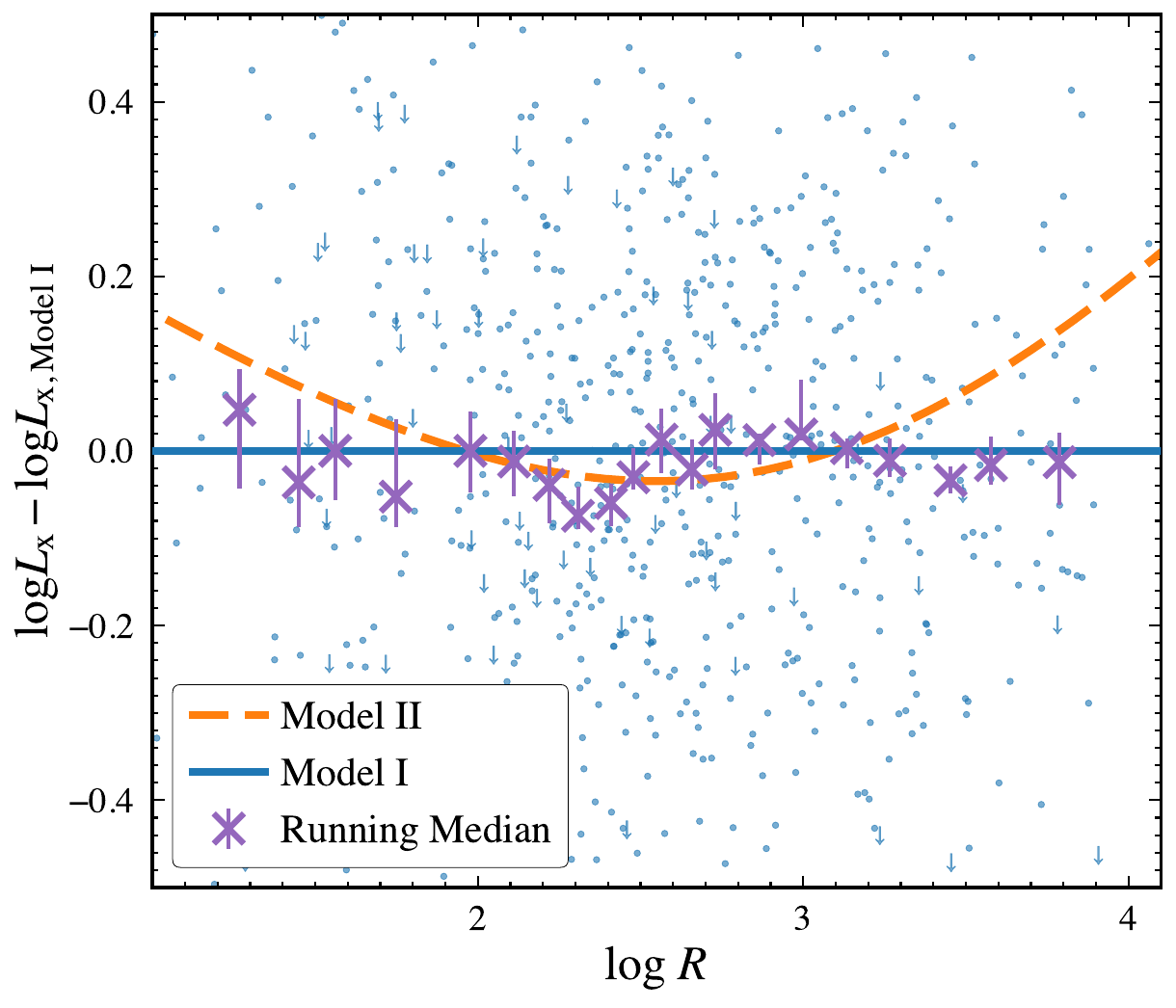}
\caption{Left: A ``from above'' view of Fig.~\ref{fig:3d},
    where the plane is filled by colors that represent the difference between best-fit Model~\RomanNumeralCaps 1 and Model~\RomanNumeralCaps 2.
The distribution of the optically selected RLQs on the $L_\mathrm{5GHz}$--$L_\mathrm{2500\angstrom}$ plane with overlaid constant-$\log R$ lines (dash-dotted) is also shown.
The dashed line (orange) is orthogonal to the constant-$\log R$ lines and passes through the median values of $L_\mathrm{2500\angstrom}$ and $L_\mathrm{5GHz}$ of the sample.
The profile (side view) of Model~\RomanNumeralCaps 2 at the position of the dashed line is shown in the right panel.
Right: A ``side view'' of Fig.~\ref{fig:3d}, where the $x$-axis is $\log R$, and the $y$-axis is $\log L_\mathrm{2keV}$ normalized to the prediction of Model~\RomanNumeralCaps 1.
Therefore, Model~\RomanNumeralCaps 1 is the horizontal line at zero.
The dashed curve (orange) is the profile of Model~\RomanNumeralCaps 2, the top view of which is shown in the left panel.
The blue dots and downward arrows are the detected and non-detected RLQs in X-rays.
We calculate the running medians (purple) to smooth the raw data points.
The running medians are apparently more consistent with Model~\RomanNumeralCaps 1 than Model~\RomanNumeralCaps 2.}
\label{fig:compareModels}
\end{figure*}

The fitting results for all RLQs, FSRQs, and SSRQs are listed in Table~\ref{tab:pars},
where the bottom row is the $L_\mathrm{x}$--$L_\mathrm{uv}$ relation for the comparison RQQ sample from \citet{lusso2016}.
We first check whether the fitting results for all RLQs are consistent with the hypotheses of the models.
Model~\RomanNumeralCaps 1 results in $\Gamma_\mathrm{uv}=\gamma_\mathrm{uv}+\gamma_\mathrm{radio}^\prime=0.69_{-0.02}^{+0.03}$,\footnote{To take account of the covariance of the errors of $\gamma_\mathrm{uv}$ and $\gamma_\mathrm{radio}^\prime$,
Markov chains resulting from model fitting are used to estimate the uncertainty of $\Gamma_\mathrm{uv}$, which is not necessarily larger than those of $\gamma_\mathrm{uv}$ and $\gamma_\mathrm{radio}^\prime$.}
which is $1.6\sigma$ ($p=0.05$) away from $\gamma=0.63_{-0.02}^{+0.02}$.
Due to the fact that the resulting error bar is relatively large for Model~\RomanNumeralCaps 2,
comparing $\Gamma_\mathrm{uv}=\gamma_\mathrm{uv}=0.72_{-0.05}^{+0.08}$ and $\gamma$ results in a similar statistical significance (i.e., $1.5\sigma$; $p=0.06$).
The normalization factor $A=0.44_{-0.05}^{+0.05}$ is 2.2$\sigma$ ($p=0.02$) away from the $0.33_{-0.01}^{+0.01}$ value applicable for RQQs.
However, these two parameters together indicate that the corona component of RLQs is inconsistent with that of RQQs at a $3.1\sigma$ ($p=9\times10^{-4}$) significance level.
Consequently, RLQs cannot be treated as radio-quiet central engines plus jet cores in X-rays.\footnote{In fact,
the fitting results of Model~\RomanNumeralCaps 2 indicate that the coronae of RLQs have elevated X-ray activity compared with those of RQQs at given disk optical/UV luminosity, which motivates the proposal of Model~\RomanNumeralCaps 3.}
The value of $\Gamma_\mathrm{uv}=\gamma_\mathrm{uv}+\gamma_\mathrm{radio}^\prime=0.66_{-0.03}^{+0.02}$ resulting from Model~\RomanNumeralCaps 3 is consistent with $\gamma$.
Therefore, the physical picture of Model~\RomanNumeralCaps 2 is not supported, while the results of Model~\RomanNumeralCaps 3 are favored.
We note that the only difference between the fitting results of Model~\RomanNumeralCaps 1 and Model~\RomanNumeralCaps 3 is that in Model~\RomanNumeralCaps 3 there
is a minute amount of jet-linked X-ray emission ($B=0.03_{-0.01}^{+0.01}$).
This component only deviates from zero at a $1.9\sigma$ ($p=0.03$) significance level and
makes up $B/(A+B)\lesssim3\%$ of the total X-ray luminosity.
Note that we can use $B/(A+B)$ to assess the typical jet contribution because the luminosities
are normalized to values that are near to their medians before model fitting.
We thus omit the difference between Model~\RomanNumeralCaps 1 and Model~\RomanNumeralCaps 3 and focus on comparing them with Model~\RomanNumeralCaps 2.

Secondly, the corresponding information criteria are listed in Table~\ref{tab:ic}.
For the results for all RLQs, Model~\RomanNumeralCaps 3 has smaller ICs
and is preferred over Model~\RomanNumeralCaps 2 with $\Delta\mathrm{AIC}=-7.61$
and $\Delta\mathrm{BIC}=-7.61$.\footnote{Model~\RomanNumeralCaps 2 and Model~\RomanNumeralCaps 3 (with fixed $\gamma_\mathrm{radio}$) have the same number of free parameters.
In this case, $\Delta\mathrm{AIC}=\Delta\mathrm{BIC}$ and they indicate that Model~\RomanNumeralCaps 3 has a larger $\mathcal{L}_\mathrm{max}$ than that of Model~\RomanNumeralCaps 2.}
Note that we performed another fitting test in which the corona term of Model~\RomanNumeralCaps 2 is fixed to the X-ray luminosity of RQQs, explicitly.
The ICs of Model~\RomanNumeralCaps 2 become worse, and Model~\RomanNumeralCaps 2 is defeated by Model~\RomanNumeralCaps 3 with $\Delta\mathrm{AIC}=-22.38$ and $\Delta\mathrm{BIC}=-13.19$.
Therefore, it is highly unlikely that the corona component of RLQs is identical to that of RQQs.
Model~\RomanNumeralCaps 2 is as strongly disfavored if we compare it with Model~\RomanNumeralCaps 1 as well.

We plot in Fig.~\ref{fig:3d} the best-fit Model~\RomanNumeralCaps 1 (blue),
Model~\RomanNumeralCaps 2 (orange), and observational data for RLQs in the
$L_\mathrm{2keV}$--$L_\mathrm{2500\angstrom}$--$L_\mathrm{5GHz}$ space, in order to provide an overall geometrical impression of the best-fit models.
Model~\RomanNumeralCaps 3 is not shown as its fitting results are similar to those of Model~\RomanNumeralCaps 1.
With axes in logarithmic scales, Model~\RomanNumeralCaps 1 is a plane while the surface of
Model~\RomanNumeralCaps 2 shows upward curvature.
In Fig.~\ref{fig:compareModels} (left) we show a ``from above'' view (i.e. looking down along the $\log L_\mathrm{2keV}$ axis) of Fig.~\ref{fig:3d},
where the $L_\mathrm{2500\angstrom}$--$L_\mathrm{5GHz}$ plane 
is filled by colors that represent the difference between the two models.
The distribution of RLQs overlaid with constant-$R$ lines are also shown.
It is easily noticed that there is a rough consistency
between the direction of the color boundaries and the constant-$R$ lines.
In other words, the difference between Model~\RomanNumeralCaps 1 and Model~\RomanNumeralCaps 2 has a strong dependence on $R$.

Fig.~\ref{fig:compareModels} (left) cannot tell us which model is better. However, it
hints at an appropriate azimuth angle for a useful ``side view'' that can---we should look in the direction of constant $R$.
Such a side view is plotted in the right panel of Fig.~\ref{fig:compareModels}, where the $y$-axis is 
normalized to the prediction of Model~\RomanNumeralCaps 1.
Therefore, Model~\RomanNumeralCaps 1 is a horizontal line at zero.
We choose a position that is indicated by the dashed line (orange) in the left panel of Fig.~\ref{fig:compareModels} for the profile of Model~\RomanNumeralCaps 2.
This dashed line is perpendicular to constant-$\log R$ lines and passes through the median values of $\log L_\mathrm{2500\angstrom}$ and $\log L_\mathrm{5GHz}$.
This profile of Model~\RomanNumeralCaps 2 is plotted as the dashed curve (orange) in the right panel of Fig.~\ref{fig:compareModels}.
The small dots and downward arrows in Fig.~\ref{fig:compareModels} (right) represent detections and non-detections in \mbox{X-rays}, respectively.
We use a running-box filter with a minimum of 40 and a maximum of 120 data points (sliding from left to right)
to smooth the considerable scatter of the points.
The step size of the running-box filter is 40 data points.
We calculate the median of the data points within the box after each step.
The resulting running medians with uncertainty estimates are shown as cross symbols with error bars (purple).
The running medians are consistent with zero, supporting the superiority of Model~\RomanNumeralCaps 1 in describing the data.
There are systematic discrepancies between Model~\RomanNumeralCaps 2 and the smoothed data (running median) at the smallest and largest radio-loudness parameters.
The latter case has the largest discriminatory leverage since the running medians here have relatively small error bars.
Therefore, the superiority of Model~\RomanNumeralCaps 1 is apparently revealed by the median properties of RLQs with similar
radio-loudness.\footnote{The same conclusion is revealed if the $y$-axis is instead normalized to the best-fit Model~\RomanNumeralCaps 2.}
The dispersion (0.29~dex at a $1\sigma$ level) of individual RLQs in Fig.~\ref{fig:compareModels} (right) has multiple sources.
We assume typical uncertainties of 20\%/30\%/40\% for the radio/UV/X-ray luminosities (e.g. \citealt{miller2011}),
including observational uncertainties and quasar variability, the latter of which probably dominates (e.g. \citealt{gibson2008, gibson2012}).
Subtracting those effects, about $0.23$~dex of residual dispersion might be attributed to unaccounted for physical processes,
which is similar to the magnitude estimated for RQQs (e.g. \citealt{salvestrini2019}).

\begin{landscape}
\begin{table}
\centering
\caption{Model fitting results for RLQs and RQQs, both of which are optically selected.}
\label{tab:pars}
\begin{threeparttable}[b]
\begin{tabular}{lcccccccccccc}
\hline
\hline
    Sample\tnote{a} & \multicolumn{12}{c}{Model Parameters\tnote{b}}\\
\cmidrule(lr){2-13}
    & \multicolumn{3}{c}{\RomanNumeralCaps 1: $\log L_\mathrm{2keV}= \alpha + \gamma_\mathrm{uv}\log L_\mathrm{2500\angstrom}+ \gamma_\mathrm{radio}^\prime\log L_\mathrm{5GHz}$} & \multicolumn{4}{c}{\RomanNumeralCaps 2: $\log L_\mathrm{2keV}=\log \Big(AL_\mathrm{2500\angstrom}^{\gamma_\mathrm{uv}}+BL_\mathrm{5GHz}^{\gamma_\mathrm{radio}}\Big) $} & \multicolumn{5}{c}{\RomanNumeralCaps 3: $\log L_\mathrm{2keV}=\log \Big(AL_\mathrm{5GHz}^{\gamma_\mathrm{radio}^\prime}L_\mathrm{2500\angstrom}^{\gamma_\mathrm{uv}}+BL_\mathrm{5GHz}^{\gamma_\mathrm{radio}}\Big)$} \\
\cmidrule(lr){2-4}\cmidrule(lr){5-8}\cmidrule(lr){9-13}
     &$\left. \alpha\right/10^\alpha$  & $\gamma_\mathrm{uv}$& $\gamma_\mathrm{radio}^\prime$& $A$ &$B$& $\gamma_\mathrm{uv}$\tnote{c}& $\gamma_\mathrm{radio}$& $A$& $B$\tnote{d}& $\gamma_\mathrm{uv}$ & $\gamma_\mathrm{radio}$\tnote{c} & $\gamma_\mathrm{radio}^\prime$\\
\hline
    All RLQs& $\left. -0.09_{-0.01}^{+0.01}\right/ 0.81_{-0.02}^{+0.02}$& $0.47_{-0.03}^{+0.03}$&$0.22_{-0.02}^{+0.02}$ &$0.44_{-0.05}^{+0.05}$&$0.31_{-0.05}^{+0.05}$&$0.72_{-0.05}^{+0.07}$&$0.55_{-0.06}^{+0.07}$ &$0.76_{-0.04}^{+0.03}$&$0.03_{-0.01}^{+0.01}$&$0.49_{-0.04}^{+0.03}$&\underline{1.00}&$0.18_{-0.03}^{+0.03}$\\
    FSRQs& $\left. -0.06_{-0.02}^{+0.02}\right/ 0.88_{-0.04}^{+0.03}$& $0.45_{-0.04}^{+0.05}$&$0.26_{-0.03}^{+0.02}$ &$0.45_{-0.06}^{+0.05}$&$0.34_{-0.06}^{+0.09}$&$0.67_{-0.05}^{+0.08}$&$0.66_{-0.08}^{+0.08}$ &$0.76_{-0.06}^{+0.06}$&$0.06_{-0.02}^{+0.03}$&$0.49_{-0.05}^{+0.05}$&\underline{1.00}&$0.18_{-0.04}^{+0.04}$\\
    SSRQs\tnote{e}& $\left. -0.14_{-0.01}^{+0.02}\right/ 0.72_{-0.03}^{+0.03}$& $0.39_{-0.04}^{+0.04}$&$0.22_{-0.03}^{+0.03}$ &$0.42_{-0.05}^{+0.06}$&$0.26_{-0.05}^{+0.04}$&\underline{0.63}&$0.48_{-0.04}^{+0.05}$ &$0.72_{-0.04}^{+0.02}$&$<0.02$&$0.39_{-0.03}^{+0.06}$&\underline{1.00}&$0.21_{-0.05}^{+0.01}$\\
\hline
    & \multicolumn{2}{c}{$\log L_\mathrm{2keV}=\beta + \gamma\log L_\mathrm{2500\angstrom}=\log \Big(10^\beta L_\mathrm{2500\angstrom}^\gamma\Big)$}& \\
\cmidrule(lr){2-3}
    &$\left.\beta\right/10^\beta$&$\gamma$ \\
\hline
    RQQs & $\left. -0.48_{-0.01}^{+0.01}\right/0.33_{-0.01}^{+0.01}$& $0.63_{-0.02}^{+0.02}$  \\
\hline
\end{tabular}
\begin{tablenotes}
\item[a] Note that the luminosities of the quasars are normalized to $\log L_\mathrm{2keV}=27$, $\log L_\mathrm{2500\angstrom}=30.5$, and $\log L_\mathrm{5GHz}=33.3$ before model fitting (see \S~\ref{sec:method}). Effectively, radio-loudness parameters are normalized to $\log R=2.677$.
    The resulting normalization factors cannot be compared directly with those of other works using different normalization.
\item[b] For the results of RLQs using Model~\RomanNumeralCaps 1 and the results of RQQs,
    the normalization factors are given in both logarithmic ($\alpha$) and physical ($10^\alpha$) units.
\item[c] We have fixed $\gamma_\mathrm{uv}=0.63$ of Model~\RomanNumeralCaps 2 for SSRQs and fixed $\gamma_\mathrm{radio}=1$ of Model~\RomanNumeralCaps 3 for all RLQ samples (i.e. the underlined cells). See Appendix~\ref{sec:approx} for details.
\item[d] If $B$ is consistent with zero, we only show the upper bound of the 90\% probability interval.
\item[e] The fitting results for SSRQs using Model~\RomanNumeralCaps 1 are consistent with the corresponding results listed in Table~7 of \citet{miller2011}, although our constraints here are quantitatively tighter. 
    Note that, however, our interpretation of the results is different from that of \citet{miller2011}. See \S~\ref{sec:reason}.
\end{tablenotes}
\end{threeparttable}
\end{table}
\end{landscape}

\begin{table*}
\centering
    \caption{Information criteria of different functional models for the RLQ samples. The corresponding model fitting results are listed in Table~\ref{tab:pars}.}
\label{tab:ic}
\begin{threeparttable}[b]
\begin{tabular}{lcccccccc}
\hline
\hline
    Sample  & \multicolumn{6}{c}{Information Criteria\tnote{a}} & \multicolumn{2}{c}{Significance\tnote{b}}\\
    \cmidrule(lr){2-7} \cmidrule(lr){8-9}
    & \multicolumn{2}{c}{Model \RomanNumeralCaps 1 ($k=3$)} & \multicolumn{2}{c}{Model \RomanNumeralCaps 2 ($k=4$ or 3)\tnote{c}} & \multicolumn{2}{c}{Model \RomanNumeralCaps 3 ($k=4$)\tnote{c}} & $\Delta$AIC & $\Delta$BIC\\
    \cmidrule(lr){2-3}\cmidrule(lr){4-5}\cmidrule(lr){6-7}
    & AIC & BIC & AIC & BIC & AIC & BIC \\
\hline
    All RLQs & -& - & $-994.56$& $-971.61$ & $\bf -1002.17$ & $\bf -979.22$ &  $-7.61$ & $-7.61$ \\
    FSRQs & -& -& $-480.78$& $-460.90$ & $\bf -485.79$ & $\bf -465.91$  & $-5.01$ & $-5.01$\\
    SSRQs & $\bf -534.62$& $\bf -519.67$ & $-533.41$& $-518.46$ & - & - & $-1.21$ & $-1.21$\\
\hline
\end{tabular}
\begin{tablenotes}
\item[a] We have bold-faced the smallest AIC and BIC of each row. The difference between the fitting results of Model~\RomanNumeralCaps 1 and Model~\RomanNumeralCaps 3 is always very small for all samples. 
    We omit Model~\RomanNumeralCaps 1 for all RLQs and FSRQs, in which a small amount of jet component is present (see \S~\ref{sec:base} and \S~\ref{sec:radioSpec}).
    We omit Model~\RomanNumeralCaps 3 for SSRQs where the jet component seems absent (see \S~\ref{sec:radioSpec}).
\item[b] $\Delta\mathrm{AIC}$ and $\Delta\mathrm{BIC}$ are calculated as the smallest values of AIC and BIC subtracted by the second smallest ones.
\item[c] The number of free parameter is reduced by one if a parameter of Model~\RomanNumeralCaps 2 or Model~\RomanNumeralCaps 3 is fixed. See Appendix~\ref{sec:approx} for details.
\end{tablenotes}
\end{threeparttable}
\end{table*}

\begin{table}
\centering
    \caption{Mean and percentiles of $f_\mathrm{jet}$ (in units of percent) for different RLQ samples. $f_\mathrm{jet}$ represents the fraction of the X-ray emission from jets.}
\label{tab:fjet}
\begin{threeparttable}[b]
\begin{tabular}{lcccc}
\hline
\hline
    Sample & Mean (\%)&  \multicolumn{3}{c}{Percentile (\%)} \\
    \cmidrule(lr){3-5}
    & & 10th & 50th (Median) & 90th \\
\hline
    All RLQs & 5.2 & 0.3 & 2.3 & 14.4 \\
    FSRQs& 9.5 & 0.6 & 4.2 & 27.5\\
    SSRQs\tnote{a}& -& - & - & 5.9\\
\hline
\end{tabular}
\begin{tablenotes}
\item[a] We only list the 90th percentile for SSRQs, the X-ray luminosities of which contain a negligible jet component.
\end{tablenotes}
\end{threeparttable}
\end{table}

\subsubsection{The results for FSRQs and SSRQs and the amount of X-ray emission from jets}
\label{sec:radioSpec}

\begin{figure}
\centering
\includegraphics[width=0.45\textwidth, clip]{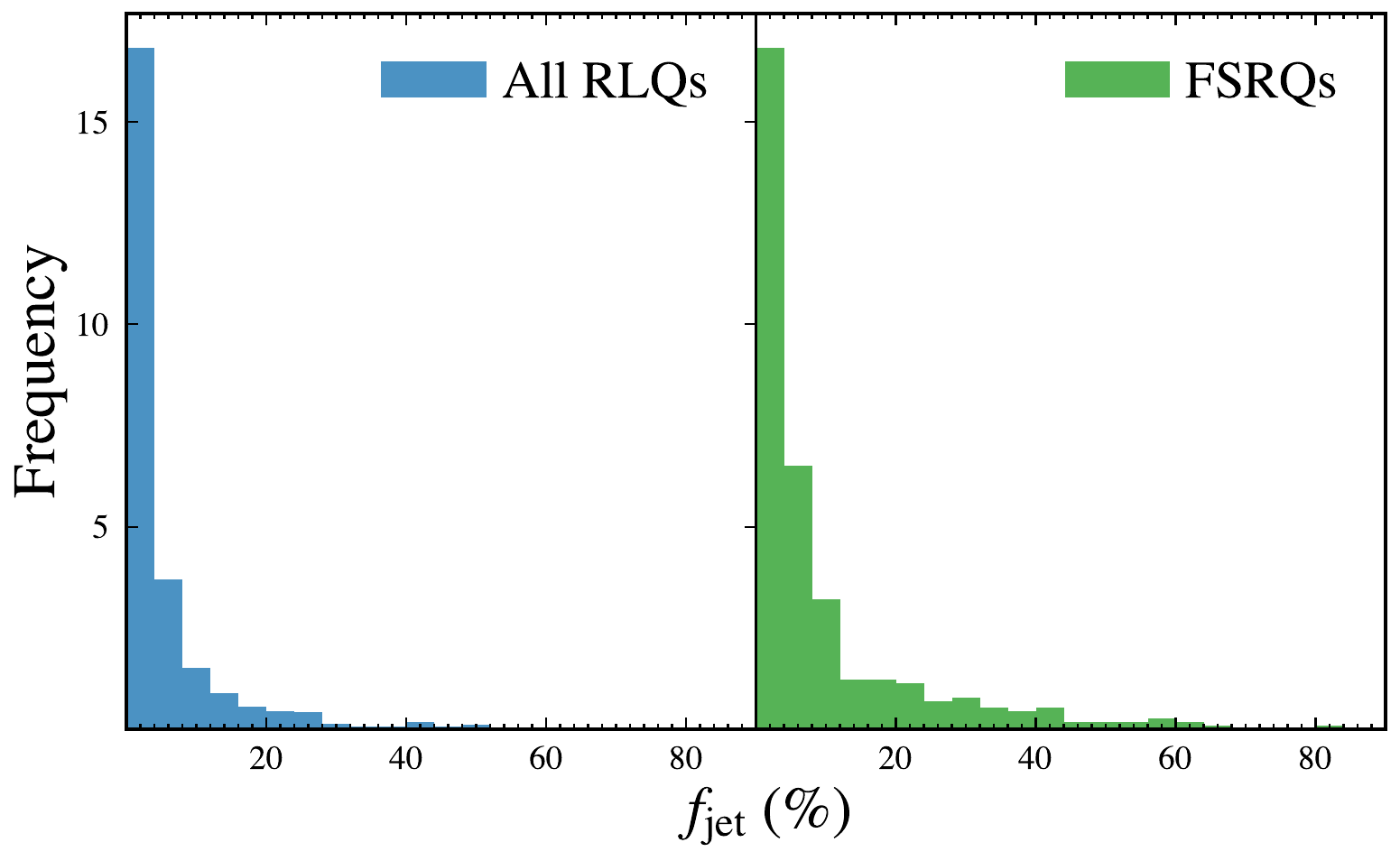}
\caption{The distributions of $f_\mathrm{jet}$ for all RLQs and FSRQs. The height of the highest bin is fixed to be identical in the two panels.
    In both cases, $f_\mathrm{jet}$ is clustered around very small values, while FSRQs have a long tail that extends to $f_\mathrm{jet}\approx 80\%$.}
\label{fig:fjet}
\end{figure}

\begin{figure*}
\centering
\includegraphics[width=0.45\textwidth, clip]{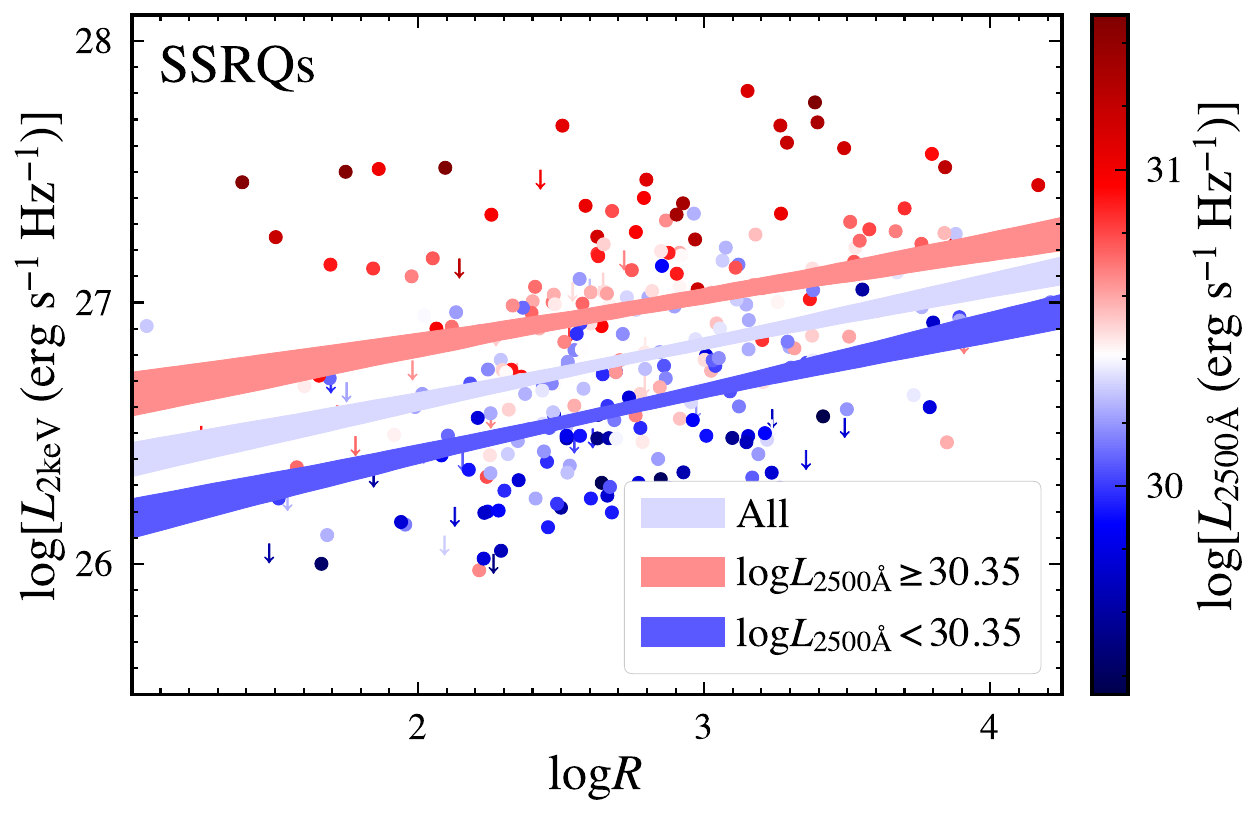}
\hspace{0.4cm}
\includegraphics[width=0.45\textwidth, clip]{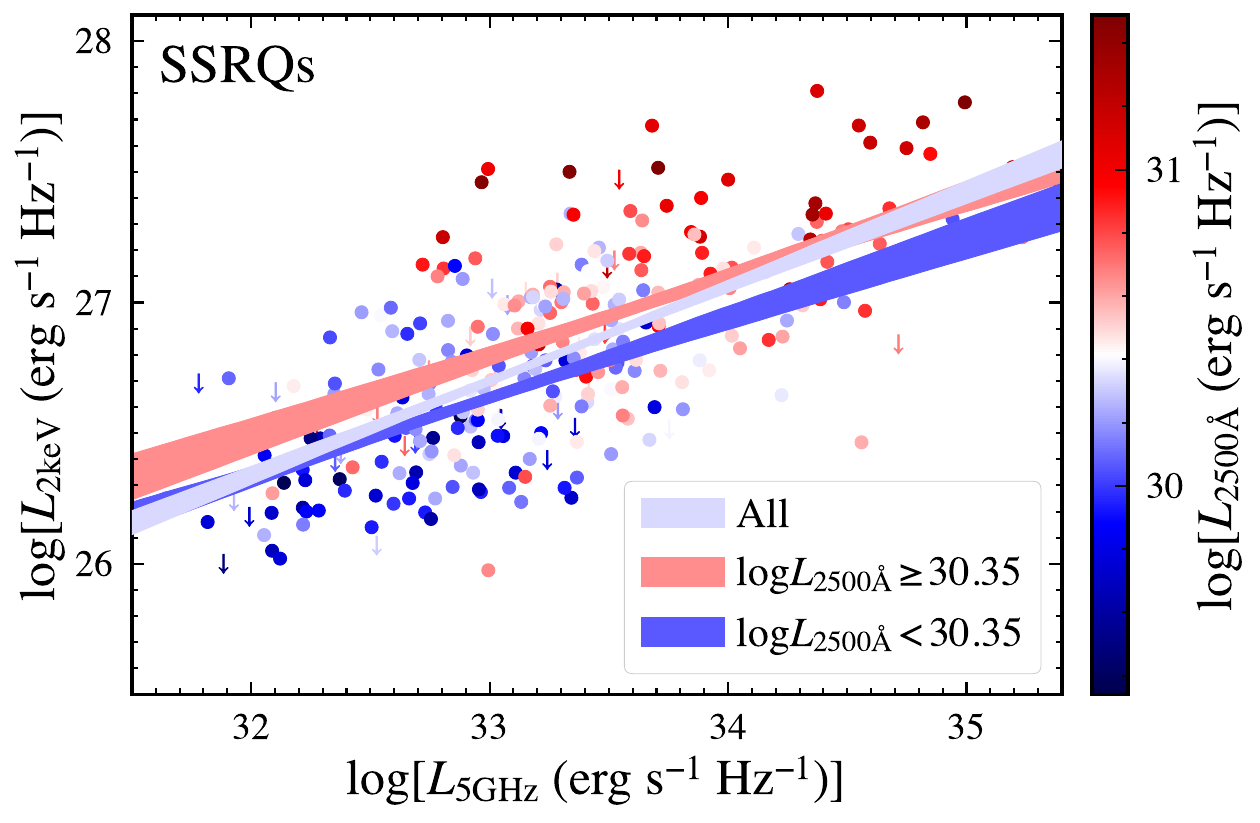}
\caption{Left: Fitting the data for SSRQs using $L_\mathrm{x}\propto R^{\gamma_\mathrm{radio}^\prime}$.
    The dots and downward arrows represent the detected and non-detected objects in X-rays.
    The color of each symbol represents the optical/UV luminosity,
    as indicated by the color bar on the right-hand side.
    We also divide the data into two bins of equal size at the median
    optical/UV luminosity ($\log L_\mathrm{2500\angstrom}$=30.35) and fit them separately.
    Right: The same as the left panel for $L_\mathrm{x}\propto L_\mathrm{radio}^{\gamma_\mathrm{radio}^\prime}$.}
\label{fig:ssrqs}
\end{figure*}

\begin{landscape}
\begin{figure}
\centering
\includegraphics[width=1.2\textwidth, clip]{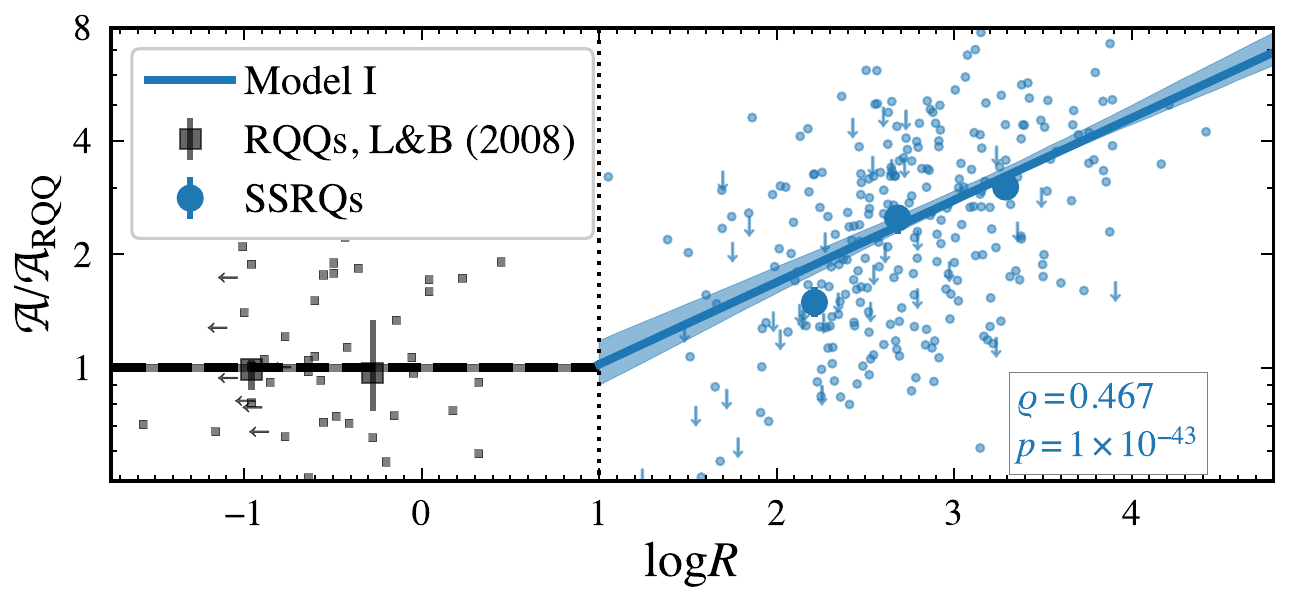}
\caption{The dependence of $\mathcal{A}$ on $\log R$ for RQQs and RLQs.
$\mathcal{A}$ is the normalization factor of the coronal component (Eq.~\ref{eq:phy}).
A vertical line (dotted black) has been used to separate RLQs and RQQs.
The small dots and downward arrows in blue are X-ray detected and non-detected SSRQs, respectively.
The generalized Spearman rank correlation coefficient ($\rho$) and corresponding $p$-value are given in the lower-right corner.
These SSRQs are further grouped into three $R$ bins of comparable sizes, where medians with uncertainty estimates are calculated (large blue dots with error bars).
The RQQs from \citet{laor2008} are small squares and leftward arrows in black, while binned data for those RQQs are large squares.
$\mathcal A_\mathrm{RQQ}$ is represented by the horizontal (dashed black) line at unity,
as no strong dependence on $\log R$ is reported in the literature or supported by the RQQ sample of \citet{laor2008}.
Note that the solid blue line is derived from the fitting results for SSRQs using Model~\RomanNumeralCaps 1 (see Table~\ref{tab:pars}).
The shaded regions represent $1\sigma$ uncertainties, which are small for the case of RQQs.
Note that the good match between dashed and solid lines at $\log R=1$ is not an enforced prerequisite but a direct result of the model fitting.}
\label{fig:norm}
\end{figure}
\end{landscape}

FSRQs and SSRQs are RLQs that are generally observed at relatively
small and large inclination angles, respectively (e.g. \citealt{urry1995}).
If part of the X-ray luminosity is from the beamed core region of jets,
the contribution of this component is expected to be larger in FSRQs than SSRQs.
Indeed, their properties seem to be different as suggested in \S~\ref{sec:scatter};
FSRQs are generally more X-ray luminous than SSRQs at fixed $L_\mathrm{uv}$ and $R$ (see Fig.~\ref{fig:lxOverlxRqq}).

For the fitting results for FSRQs in Table~\ref{tab:pars},
the parameters of Model~\RomanNumeralCaps 1 do not support $\Gamma_\mathrm{uv}=\gamma$,
while those of Model~\RomanNumeralCaps 2 do not support a component that is consistent with the coronae of RQQs.
Model~\RomanNumeralCaps 3 might point to the most plausible scenario with $\Gamma_\mathrm{uv}=\gamma_\mathrm{uv}+\gamma_\mathrm{radio}^\prime=0.67_{-0.04}^{+0.04}$.
Note that the jet component is set to zero in Model~\RomanNumeralCaps 1,
while its fraction in Model~\RomanNumeralCaps 3 is still only $B/(A+B)\approx7.0\%$.
Therefore, we do not distinguish between those two models, similar to the case of \S~\ref{sec:base}.
Model~\RomanNumeralCaps 2 is disfavored in comparison with Model~\RomanNumeralCaps 3 (Model~\RomanNumeralCaps 1)
with $\Delta\mathrm{AIC}=-5.01$ ($-2.27$) and $\Delta\mathrm{BIC}=-5.01$ ($-6.24$).
Note that the fitting results for FSRQs are largely consistent with
those for the sample of all RLQs in \S~\ref{sec:base},
except that the $BL_\mathrm{5GHz}^{\gamma_\mathrm{radio}}$ term is more significant.
Note that the residual scatter after subtracting the effects of measurement errors and variability
is about $0.17$ dex for FSRQs (and SSRQs below).

As for SSRQs, the fitting results of Model~\RomanNumeralCaps 3 are almost identical to those of Model~\RomanNumeralCaps 1 with $B$
consistent with zero (see Table~\ref{tab:pars}). In this case, $B$ is totally redundant in that
it increases the number of free parameters without improving the performance.
We thus omit Model~\RomanNumeralCaps 3 for SSRQs.
Model~\RomanNumeralCaps 1 results in $\Gamma_\mathrm{uv}=\gamma_\mathrm{uv}+\gamma_\mathrm{radio}^\prime=0.61_{-0.03}^{+0.03}$, consistent with $\gamma$.
The fitting results of Model~\RomanNumeralCaps 2 do not support a corona component that is identical to
those of RQQs ($A=0.42_{-0.05}^{+0.06}$ vs. $10^\beta=0.33_{-0.01}^{+0.01}$) and
$\gamma_\mathrm{radio}=0.48_{-0.04}^{+0.06}$ departs strongly from unity.
Furthermore, Model~\RomanNumeralCaps 2 is not favored in comparison with Model~\RomanNumeralCaps 1 with $\Delta\mathrm{AIC}=-1.21$ and $\Delta\mathrm{BIC}=-1.21$ (see Table~\ref{tab:ic}).
Note that the model selection results here are not as significant as for the cases of all RLQs and FSRQs,
due to a smaller sample size and lower X-ray detection fraction for SSRQs.

Benefiting from its flexibility, Model~\RomanNumeralCaps 3 either is selected as the best model or 
can be consistent with the best model for all samples.
Using the results of Model~\RomanNumeralCaps 3 in Table~\ref{tab:pars},
we now assess the importance of the jets in accounting for the X-ray luminosities of RLQs.
We input $L_\mathrm{2500\angstrom}$ and $L_\mathrm{5GHz}$ as well as estimates of parameters into Model~\RomanNumeralCaps 3.
We quantify the fraction of jet emission for each RLQ using $f_\mathrm{jet}=BL_\mathrm{5GHz}^{\gamma_\mathrm{radio}}/\big(AL_\mathrm{5GHz}^{\gamma_\mathrm{radio}^\prime}L_\mathrm{2500\angstrom}^{\gamma_\mathrm{uv}}+BL_\mathrm{5GHz}^{\gamma_\mathrm{radio}}\big)$.
The mean and three percentiles (10, 50, and 90th) of $f_\mathrm{jet}$ for samples that might have a jet component are listed
in Table~\ref{tab:fjet}, while the full distributions are plotted in Fig.~\ref{fig:fjet}.
For SSRQs, we only list their 90th percentile,
which indicates that for 90\% of SSRQs the jet component contributes $\lesssim6\%$ of the observed nuclear X-ray emission.
Note that $B/(A+B)$ is generally between the mean and median of $f_\mathrm{jet}$.

The distributions of $f_\mathrm{jet}$ are highly asymmetric in Fig.~\ref{fig:fjet}.
As a consequence, the median is always smaller than the mean in Table~\ref{tab:fjet}.
The median is probably a better statistic to assess the typical value of $f_\mathrm{jet}$, which is $<5\%$ for all samples.
FSRQs are the group of RLQs that have the largest jet contribution.
Notably, the 10th and 90th percentiles span several decades in Table~\ref{tab:fjet}, which indicates a large variance of $f_\mathrm{jet}$ from object to object.
The mean is strongly affected by a small number of RLQs that have relatively large $f_\mathrm{jet}$.
Less than $10\%$ of FSRQs have $f_\mathrm{jet}\gtrsim30\%$, which manifest themselves as a long tail in the right panel of Fig.~\ref{fig:fjet}.
Thus, only in a minority of FSRQs does the jet component become important.

At fixed $L_\mathrm{2500\angstrom}$, RLQs are generally a factor of 1.5--4 times more \mbox{X-ray}
luminous than RQQs, depending on the radio-loudness parameter (see Fig.~\ref{fig:lxOverlxRqq}).
Given our limits on $f_\mathrm{jet}$, this difference is mainly caused by the difference between
the coronae of RLQs and those of RQQs, not the emission from jets.
Furthermore, typical FSRQs are more X-ray luminous than SSRQs at fixed $L_\mathrm{uv}$ and $R$ by 10\%--20\% (Fig.~\ref{fig:lxOverlxRqq}),
which is only partially attributed to the jet component according to our estimation here, given that the median $f_\mathrm{jet}$ is only $4.2$\% for FSRQs.
Therefore, other processes are probably involved as well (e.g. a small level of jet contribution to the optical/UV emission for FSRQs) to explain the difference between FSRQs and SSRQs.

\subsection{The corona-jet connection in SSRQs}
\label{sec:ssrqs}
Owing to the apparent absence of significant X-ray emission from jets in general,
SSRQs form a useful sample to investigate the
corona-jet connection as implied by Model~\RomanNumeralCaps 1.

In \S~\ref{sec:model}, we interpret Model~\RomanNumeralCaps 1 as $L_\mathrm{x}\propto R^{\gamma_\mathrm{radio}^\prime}L_\mathrm{uv}^{\gamma_\mathrm{uv}+\gamma_\mathrm{radio}^\prime}$ 
instead of $L_\mathrm{x}\propto L_\mathrm{radio}^{\gamma_\mathrm{radio}^\prime}L_\mathrm{uv}^{\gamma_\mathrm{uv}}$.
The argument there was based on the fact that $\gamma_\mathrm{uv}+\gamma_\mathrm{radio}^\prime$ is close to $\gamma$ (the slope of the $L_\mathrm{x}$--$L_\mathrm{uv}$ relation for RQQs),
which might suggest an identical relation between $\Delta L_\mathrm{x}/{L_\mathrm{x}}$ and $\Delta L_\mathrm{uv}/{L_\mathrm{uv}}$ for both the RQQ and RLQ populations.
The results of \S~\ref{sec:base} and \S~\ref{sec:radioSpec} indeed confirm this approach.
We now show that the former expression is preferred even if the properties of RQQs are not considered,
and our interpretation of the corona-jet connection is thus independent of outside assumptions.
We fit the data for SSRQs using $L_\mathrm{2keV}\propto R^{\gamma^\prime_\mathrm{radio}}$ and $L_\mathrm{2keV}\propto L_\mathrm{5GHz}^{\gamma^\prime_\mathrm{radio}}$, 
which results in $\gamma_\mathrm{radio}^\prime=0.22_{-0.04}^{+0.04}$ and $\gamma_\mathrm{radio}^\prime=0.36_{-0.02}^{+0.02}$, respectively.
The former value is consistent with that from the three-parameter model in Table~\ref{tab:pars} (i.e. $\gamma_\mathrm{radio}^\prime=0.22_{-0.03}^{+0.03}$ in Model~\RomanNumeralCaps 1 for SSRQs).
The results are shown in Fig.~\ref{fig:ssrqs}, where the data points are color-coded by their optical/UV luminosities.
In the left panel of Fig.~\ref{fig:ssrqs}, the gradient of data-point colors appears perpendicular to the line of $L_\mathrm{2keV}\propto R^{\gamma_\mathrm{radio}^\prime}$,
while in the right panel $L_\mathrm{2keV}\propto L_\mathrm{5GHz}^{\gamma_\mathrm{radio}^\prime}$ is not able to separate high-luminosity and low-luminosity objects.
To confirm this, we divide SSRQs into high-luminosity and low-luminosity bins at the median $\log L_\mathrm{2500\angstrom}=30.35$ and fit them separately.
The results are shown in Fig.~\ref{fig:ssrqs} as well.
The lines in the left panel are largely parallel, with the high-luminosity and low-luminosity fits well straddling the original fit.
The right panel is complex, and no useful information can be easily extracted.
Furthermore, reading Fig.~\ref{fig:ssrqs} (left) and Fig.~\ref{fig:lx_luv_bi} (bottom) 
together reveals an attractive property:
$L_\mathrm{x}\propto R^{\gamma_\mathrm{radio}^\prime}L_\mathrm{uv}^{\gamma_\mathrm{uv}+\gamma_\mathrm{radio}^\prime}$ can be decomposed into two {\it statistically independent}
relations such that $L_\mathrm{x}\propto R^{\gamma_\mathrm{radio}^\prime}$ and
$L_\mathrm{x}\propto L_\mathrm{uv}^{\gamma_\mathrm{uv}+\gamma_\mathrm{radio}^\prime}$.\footnote{The commonly assumed situation where the properties of RLQ coronae are identical to those of
RQQ coronae probably needs finely tuned jets to produce such a property.}
Therefore, $L_\mathrm{x}$ depends primarily on $L_\mathrm{uv}$ and $R$, while its relation with $L_\mathrm{radio}$ is a \mbox{by-product}.
\citet{miller2011} notice that $L_\mathrm{x}/L_\mathrm{x, RQQ}$ has a stronger dependence on $R$ than $L_\mathrm{radio}$ (especially for SSRQs),
which is consistent with the idea that $R$ is the more meaningful parameter for describing the corona-jet connection of RLQs.
Consequently, the index of the X-ray-optical/UV relation for the corona is universal among RLQs and
RQQs (i.e. $\Gamma_\mathrm{uv}=\gamma_\mathrm{radio}^\prime+\gamma_\mathrm{uv}=\gamma$),
which is now a direct result of separated correlation analysis of two quasar populations.

More importantly, due to the above exposed statistical independence, another process is probably in control of the corona-jet connection in
RLQs aside from the disk-corona interaction that is at work for both RLQs and RQQs.
Following \S~\ref{sec:model}, the discussion is continued in the context of Eq.~\ref{eq:phy} where $\mathcal{A}$ is a function of $R$.
We plot $\mathcal A$ of SSRQs divided by that of RQQs against $R$ in Fig.~\ref{fig:norm}.
The scattered individual data points together with binned data points (separated by the 33rd and 66th $\log R$ percentiles of SSRQs) are shown.
We calculate the generalized Spearman's rank correlation coefficient of the individual points in Fig.~\ref{fig:norm},
using the Astronomy Survival Analysis package (\citealt{lavalley1992}).
The correlation is strongly supported with a $p$-value of $\approx1\times10^{-43}$.
The solid blue line (with shaded region showing estimated uncertainties) in Fig.~\ref{fig:norm} is a projection of
Model~\RomanNumeralCaps 1 in Table~\ref{tab:pars} for SSRQs, and not a new fitting of the individual data points.
The normalization factor of RQQs (i.e. $\mathcal{A}_\mathrm{RQQ}$) is at unity (dashed black line), where the shaded region represents its small uncertainty.
We are unaware of any established radio dependence of the $L_\mathrm{x}$--$L_\mathrm{uv}$ relation for RQQs (e.g. \citealt{steffen2006, just2007, lusso2010}).
More importantly, the radio luminosity of RQQs might not be dominated by the synchrotron emission from quasar jets (e.g. \citealt{kellermann2016, panessa2019}).
After subtracting the effects of measurement errors and quasar variability,
the physical intrinsic scatter of the $L_\mathrm{x}$--$L_\mathrm{uv}$ relation is about 0.2~dex (e.g. \citealt{lusso2016, chiaraluce2018, salvestrini2019}).
There is thus limited error budget for additional physical processes involving radio luminosity (or radio-loudness parameter)
to play an important role in RQQs as in RLQs.
We show the 59 RQQs from \citet{laor2008} as black symbols, where they are also grouped into two $R$ bins.
These PG quasars are consistent with the horizontal line and do not show an $R$ dependence.

The dashed and solid lines are satisfyingly consistent with each other at $R\approx10$; this agreement has not been enforced but arises entirely from the independent model fits.
Consequently, if we express the X-ray luminosity of RQQs as
\begin{equation}
    \label{eq:rqq}
    L_\mathrm{x}=\mathcal{A}_\mathrm{RQQ}L_\mathrm{uv}^\gamma,
\end{equation}
the X-ray luminosity of the corona of RLQs can be written as
\begin{equation}
    \label{eq:rlq}
    L_\mathrm{x}=\mathcal{A}_\mathrm{RQQ}\big (R/10\big )^{\gamma_\mathrm{radio}^\prime} L_\mathrm{uv}^\gamma.
\end{equation}
The good match between $\mathcal{A}_\mathrm{RQQ}$ and $\mathcal{A}_\mathrm{RLQ}$ around $R=10$ (instead of some much smaller radio-loudness parameter) in Fig.~\ref{fig:norm}
prevents extension of the corona-jet connection of RLQs into the radio-quiet regime.

It is also tempting to attempt to connect RQQs and RLQs with a single function that
contains a ``jet component'', which grows smoothly from being negligible at $R\le10$ to being significant at $R>10$.
However, this hypothetical component unavoidably has a strong dependence on optical/UV emission (e.g. Fig.~\ref{fig:lx_luv_bi}) and a relatively weak dependence on radio emission (e.g. Fig.~\ref{fig:ssrqs}),
in contrast with expectations for a ``jet component''.
We thus suggest that a break at $R\approx10$ is probably required, which points to an intrinsic difference between the innermost accretion flows of RLQs and RQQs.
Therefore, the radio-loudness parameter of quasars is more fundamental than simply empirical.
Future works focusing on the \mbox{X-ray} properties of quasars with $0\le \log R \le2$ might reveal the exact transition clearly.

\section{Discussion}
\label{sec:discussion}

\begin{figure*}
\centering
\includegraphics[width=0.85\textwidth, clip]{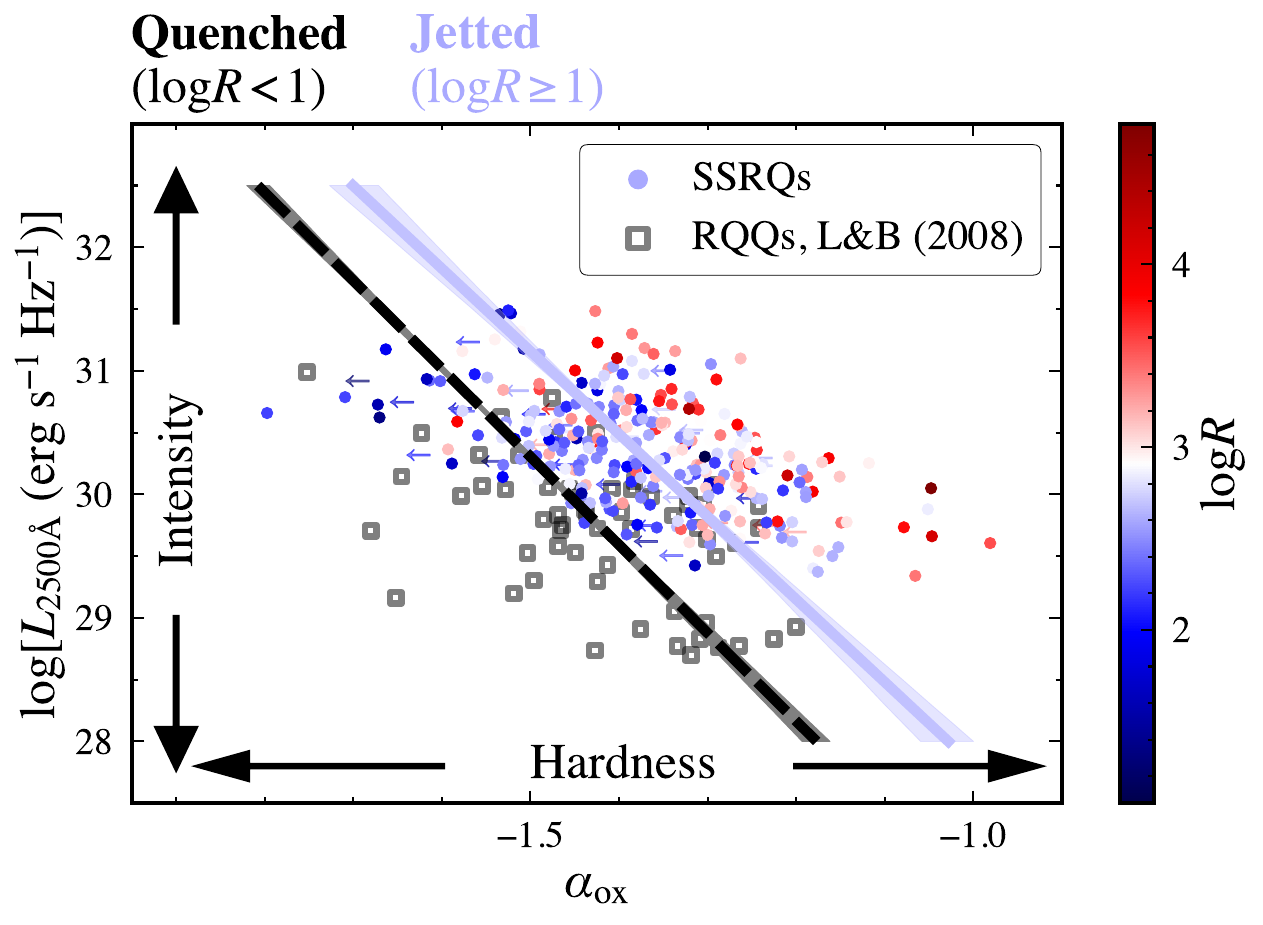}
\caption{The distributions of RQQs (open squares) and SSRQs (solid circles)
    in the $L_\mathrm{2500\angstrom}$--$\alpha_\mathrm{ox}$ plane.
   The RQQs are from \citet{laor2008}.
    SSRQs are color-coded according to their radio-loudness parameters (following the color-bar on the right-hand side).
    The dashed (black) line is the $\alpha_\mathrm{ox}$--$L_\mathrm{uv}$ relation for RQQs, derived from the $L_\mathrm{x}$--$L_\mathrm{uv}$ relation in Fig.~\ref{fig:rqq}.
    The solid (light blue) line is derived from the fitting results for SSRQs using Model~\RomanNumeralCaps 1 (see Table~\ref{tab:pars}), fixed at the median radio-loudness ($\log R=2.69$).
    We interpret this plot as a HID for quasars, in analogy to that for BHXRBs.
    We propose that the $\alpha_\mathrm{ox}$--$L_\mathrm{uv}$ relation for RQQs corresponds to an approximate
    ``jet line'' for quasars; the quasars lying well to the right of the jet
    line have powerful relativistic jets, which are quenched for quasars that are ``on'' or to the left of the jet line.}
\label{fig:jetline}
\end{figure*}

\begin{figure*}
\centering
\includegraphics[width=0.85\textwidth, clip]{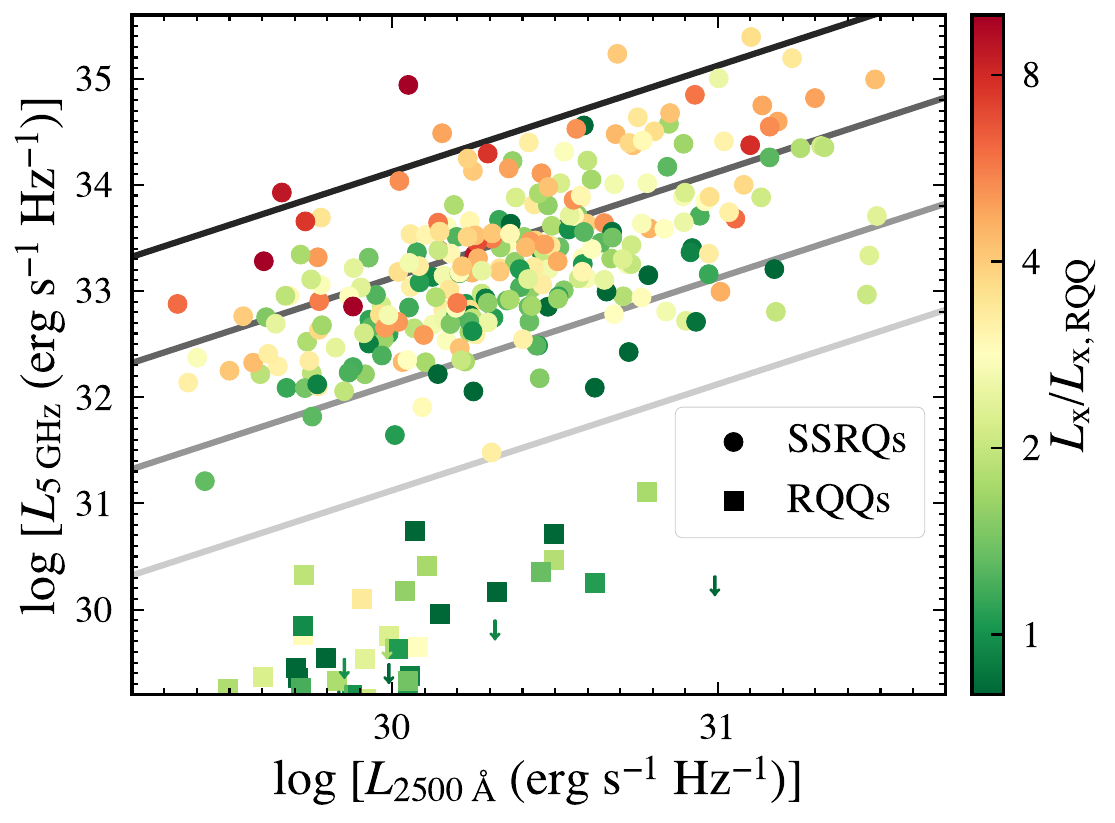}
\caption{The SSRQs of this paper and RQQs of \citet{laor2008} in the $L_\mathrm{radio}$--$L_\mathrm{uv}$ plane.
The symbol colors represent observed X-ray luminosities divided by those expected for RQQs at given matched optical/UV luminosity, $L_\mathrm{x}/L_\mathrm{x, RQQ}$, as indicated by the color-bar on the right-hand side.
X-ray non-detected SSRQs (11.3\%) are not shown, the distribution of which is consistent with the detections (see Fig.~\ref{fig:norm}).
Radio non-detected RQQs are shown as downward arrows.
Four lines indicate $\log R=1$ to 4, from light to dark.
There is an apparent correlation between $L_\mathrm{x}/L_\mathrm{x,RQQ}$ and $R$ for SSRQs (cf. Fig.~\ref{fig:lxOverlxRqq}).}
\label{fig:aboutR}
\end{figure*}

\subsection{Comparison with literature results}
\label{sec:literature}

Given that we find the corona is still responsible for most of the \mbox{X-ray} emission of RLQs (as for RQQs),
it is of value to compare with literature results on the spectral,
imaging, and timing properties of radio-loud AGNs in X-rays to assess overall consistency.

\subsubsection{X-ray spectral properties}
X-ray spectral analyses of radio-loud AGNs, except for some FSRQs,
using high-quality X-ray data rarely reveal a distinct jet component.
For example, the X-ray spectra of BLRGs like
3C 382 (e.g. \citealt{gliozzi2007, ballantyne2014, ursini2018}),
PKS 2251$+$11 (e.g. \citealt{ronchini2019}),
3C 120 (e.g. \citealt{rani2018}),
and 3C 390.3 (e.g. \citealt{sambruna2009, lohfink2015}) are consistent with being disk/corona-related.
Specifically, their observed primary power-law, reflected Fe K$\alpha$ line (and Compton hump),\footnote{The reflection features of BLRGs in statistical samples are known to be weaker
than those of radio-quiet Seyfert galaxies,
which can be attributed to mechanisms other than the dilution by continuum X-ray emission from jets (e.g. \citealt{eracleous2000}).}
and soft excess are all established features common to radio-quiet AGNs as well.
Also, importantly, those studies utilizing {\it NuSTAR} data in the hard X-rays generally find a high-energy
cutoff at $\sim100$ keV (e.g. \citealt{ballantyne2014, lohfink2015, rani2018}),
signifying a thermal Comptonization process of X-ray continuum production as for radio-quiet Seyfert galaxies.
\citet{lohfink2017} analyzed the {\it Swift}/{\it NuSTAR} X-ray spectra of an SSRQ (4C 74.26);
no sign of jet emission but a high-energy cutoff of $183_{-35}^{+51}$ keV was revealed.

Note that a jet-dominated spectrum usually shows an ``upturn'' (in SED representation) in hard X-rays, 
extending to MeV energies (and sometimes beyond), instead of a ``rollover''.
Indeed, the upturn is present in those FSRQs where the jet emission dominates
(e.g. \citealt{paliya2016, ghisellini2019}) or has a significant contribution (e.g. \citealt{grandi2004, madsen2015}) in X-rays.
However, those objects with strongly beamed jet X-ray emission are only a small portion of their parent population (see \S~\ref{sec:cc}).
Some sample-based archival X-ray spectral studies finding flat X-ray power-law continua for RLQs relative to those of RQQs (e.g. \citealt{wilkes1987, lawson1997, reeves1997, page2005})
utilize heterogeneous RLQ samples, dominated (at the 60--100\% level) by flat-spectrum RLQs, and potentially extreme objects.
Thus, these samples cannot be reliably used to constrain the X-ray spectral properties of general RLQs, the focus of this work.
Generally, once FSRQs are excluded, the X-ray spectra of the remaining BLRGs, narrow-line radio galaxies,
and SSRQs are similar to those of radio-quiet AGNs (e.g. \citealt{lawson1992, galbiati2005, grandi2006}).\footnote{Note that even if 
the \mbox{X-ray} spectra of general RLQs are somewhat flatter than those of RQQs, this would not necessarily demonstrate that the X-ray emission is jet linked (e.g. \citealt{laor1997}).}

\citet{gupta2018} investigated luminous radio galaxies and their radio-quiet counterparts with comparable Eddington ratios and black-hole masses.
The radio galaxies were found to be, on average, a factor of $\sim2$ more luminous in hard X-rays than the radio-quiet AGNs,
while their power-law spectral slopes and high-energy breaks were similar.
The authors concluded that the X-ray emission in both samples is produced by a common mechanism.
However, the X-ray production efficiency is apparently higher for the radio-loud group,
which is consistent with our suggestion of a corona-jet connection (e.g. \S~\ref{sec:ssrqs}).
\citet{gupta2019} extended the comparison and showed that Type~1 AGNs are not X-ray louder than Type~2 AGNs, within both radio-loud and radio-quiet groups.
Therefore, the dependence of observed X-ray luminosity on viewing angle seems weak, inconsistent with beamed jet X-ray emission.

\subsubsection{X-ray imaging properties}
\label{sec:xjet}

The {\it Chandra} observatory with sub-arcsec angular resolution has detected kpc-scale X-ray jets from many RLQs (e.g. \citealt{harris2006}),
which might contribute to the X-ray fluxes we use (see \S~\ref{sec:xdata}).
The observed X-ray luminosities of these extended jets are generally only a few percent those of the cores (e.g. \citealt{marshall2005, marshall2018}),
consistent with the amount of jet contribution we estimated in \S~\ref{sec:radioSpec}.

Imaging the nuclear region of quasars in X-rays is currently limited to indirect methods (e.g. gravitational lensing).
The sizes of the nuclear X-ray emission regions derived from gravitational-lensing studies
seem to be systematically larger in RLQs than in RQQs, for a given SMBH mass, 
although the source statistics are presently very limited (e.g. \citealt{xRayLensing2019}).
Therefore, we might not expect the coronae of RLQs and RQQs to have the same physical properties,
the former of which could be larger in units of black-hole radius.

\subsubsection{X-ray timing properties}
Long-term timing studies are generally more observationally intensive than single-epoch spectral studies of AGNs in X-rays,
and thus are sparse for general radio-loud AGNs.\footnote{Those rare radio-loud AGNs with one of their jets pointing close to our line of sight have generally gained more attention (e.g. \citealt{marscher2010, hayashida2015}).
However, their X-ray variability probably has a different origin and is strongly affected by Doppler boosting effects.}
\citet{leighly1997} investigated the 9-month X-ray variability of the BLRG 3C~390.3 and found its fractional amplitude of variability is about 33\%.
This amount of variability is not apparently different from those of radio-quiet Seyfert galaxies given the luminosity of 3C~390.3 and the timescales this study covers.
Furthermore, the hour-to-year power spectra of several BLRGs (e.g. 3C~120, \citealt{marshall2009}; 3C~390.3, \citealt{gliozzi2009}; 3C~111, \citealt{chatterjee2011}) are also similar to those of radio-quiet Seyfert galaxies.
The X-ray variability properties of RLQs are poorly constrained for systematically derived samples as well, let alone for subsamples with established radio slopes.
\citet{sambruna1997} investigated the soft X-ray variability of a few FSRQs on month-to-year timescales ($\ge6$ epochs)
and found a typical variability amplitude of 10--30\%, roughly consistent with that of RQQs (e.g. \citealt{gibson2012}).
\citet{gibson2012} studied the X-ray variability of about 20 non-BAL RLQs and found suggestive evidence that they are less variable than RQQs.
However, the observations of these RLQs generally have only 2--3 epochs.
Future X-ray variability studies utilizing RLQ samples with measured radio spectral slopes and observations with more epochs
might improve our understanding of the X-ray variability properties of RLQs and shed further light on
the origin of their X-ray emission.

\subsection{A jet line for AGNs}
\label{sec:jetline}
In this section, we investigate further the connection between the disk/corona and
jets from the perspective of the $\alpha_\mathrm{ox}$--$L_\mathrm{uv}$ relation.
Fig.~\ref{fig:jetline} shows the transposed $\alpha_\mathrm{ox}$-$L_\mathrm{uv}$ plot for RQQs and SSRQs,
which is no more than an alternate representation of Fig.~\ref{fig:lx_luv_bi} (bottom).
However, we gain new insights from it by analogy with BHXRBs.

We interpret Fig.~\ref{fig:jetline} as a quasar-version of the hardness-intensity diagram (HID).
The HID is frequently used in BHXRB studies. The $y$-axis serves as a proxy for the accretion rate of the black hole,
while the $x$-axis represents the relative contributions of the thermal (at 2500 \angstrom) and power-law (at 2 keV) components
that are radiated from the accretion disk and corona, respectively.
As discussed in detail earlier, a significant jet contribution in X-rays for SSRQs is highly unlikely, legitimizing our approach.
Note that the data points in an HID for BHXRBs often represent the evolutionary stages of a single black hole,
which in Fig.~\ref{fig:jetline} are represented by an ensemble of black holes that span ranges in accretion rate, ``hardness ratio'', and jet power.

In Fig.~\ref{fig:jetline}, the dashed line (black) represents the $\alpha_\mathrm{ox}$--$L_\mathrm{uv}$ relation for
RQQs, which was transformed from the $L_\mathrm{x}$--$L_\mathrm{uv}$ relation in Table~\ref{tab:pars} (bottom row) and Fig.~\ref{fig:rqq}.
For SSRQs, we utilize the fitting results of Model~\RomanNumeralCaps 1 in
Table~\ref{tab:pars} to derive their $\alpha_\mathrm{ox}$--$L_\mathrm{uv}$ relation (light blue)
at fixed $\log R=2.69$, which is the median radio-loudness parameter for SSRQs in our sample.
The dashed and solid lines are parallel to each other, as a consequence of $\Gamma_\mathrm{uv}\approx\gamma$.
Open squares and filled circles are RQQs from \citet{laor2008} and SSRQs, respectively.
SSRQs are further color-coded by their radio-loudness parameters.
Among RQQs, more luminous objects have steeper $\alpha_\mathrm{ox}$, and hence a smaller power-law fraction.
SSRQs to first approximation follow this trend as well.
However, the radio-loudness parameter also plays an important role
such that those more radio-loud quasars have flatter $\alpha_\mathrm{ox}$ at given $L_\mathrm{uv}$,
which is a consequence of $\mathcal{A}_\mathrm{RLQ}\approx\mathcal{A}_\mathrm{RQQ}(R/10)^{\gamma_\mathrm{radio}^\prime}$ (as per Eq.~\ref{eq:rqq} and Eq.~\ref{eq:rlq}).
To summarize, as shown in Fig.~\ref{fig:jetline}:
\begin{enumerate}
    \item The most radio-loud SSRQs (say, with $\log R=4$) lie on a line near the right edge of the data points
    that is parallel to the $\alpha_\mathrm{ox}$--$L_\mathrm{uv}$ relation for RQQs.
\item Horizontally shifting such a line leftward,
    it passes through SSRQs with generally decreasing $R$.
\item Finally, the line arrives at the region populated by RQQs.
\end{enumerate}
Radio observations of BHXRBs suggest the existence of a critical line in the HID such that
jets are active on the right-hand side of this line, but seem to be quenched
once a BHXRB moves leftward and passes the line (e.g. \citealt{fender2004}).
Fig.~\ref{fig:jetline} might reveal a similar jet line for AGNs,
which is approximately the $\alpha_\mathrm{ox}$--$L_\mathrm{uv}$ relation for RQQs.
The jet line for BHXRBs seems to be tilted in a way such that the
quenching happens at a larger hardness ratio when the luminosity is lower (e.g. \citealt{fender2009}),
similar to the alignment of the black-dashed line in Fig.~\ref{fig:jetline}.

The jet line adds to the mounting evidence for the idea that black-hole accretion/jet physics is largely scale-invariant (e.g. \citealt{merloni2003, mchardy2006, arcodia2020}).
RLQs and RQQs are likely scaled-up versions of BHXRBs in different accretion states (e.g. \citealt{nipoti2005, kording2006, sobolewska2009}).

\subsection{Origin of radio-loudness}
\label{sec:aboutR}

\subsubsection{The overall corona-disk-jet relation}
We show in Fig.~\ref{fig:aboutR} the SSRQs in the $L_\mathrm{radio}$--$L_\mathrm{uv}$ plane,
reorienting the discussion from, hitherto, the nature of \mbox{X-ray} emission to the origin of powerful radio jets.
The symbol colors represent the X-ray luminosities divided by those of RQQs at given matched $L_\mathrm{uv}$, i.e. $L_\mathrm{x}/L_\mathrm{x,RQQ}$,
utilizing the $L_\mathrm{x}$--$L_\mathrm{uv}$ relation in Fig.~\ref{fig:rqq}.
The \mbox{X-ray} non-detections (11.3\%) are omitted in Fig.~\ref{fig:aboutR},
the distribution of which is similar to that of the detections for SSRQs (cf. Fig.~\ref{fig:norm});
omitting non-detections will not bias the result.
The 59 RQQs of \citet{laor2008} are also shown for comparison, where radio non-detected objects are represented by downward arrows.
Apparently, RQQs follow a different track in Fig.~\ref{fig:aboutR} than SSRQs such that 
typical RQQs are about 3 orders of magnitude less radio-luminous than RLQs
at a given accretion rate (e.g. \citealt{rawlings1994}).
Whether the dichotomy between RLQs and RQQs is real or not has long been debated (e.g. \citealt{kellermann1989, kellermann2016, ivezic2002, cirasuolo2003, balokovic2012, condon2013}).
The results of this paper (e.g. Fig.~\ref{fig:norm}) support the idea that RLQs and RQQs are intrinsically different objects,
regardless of the bimodality of the distribution of $R$ (e.g. \citealt{padovani2016}).
The accretion flow for RQQs is probably in a state that is not compatible with launching powerful relativistic jets.

The most prominent relation for SSRQs in Fig.~\ref{fig:aboutR} is that between $L_\mathrm{radio}$ and $L_\mathrm{uv}$,
which is not one of the main topics of this paper but reveals probably no less important physics (e.g. \citealt{sergeant1998, vanVelzen2013}).
The $L_\mathrm{radio}$--$L_\mathrm{uv}$ relation
is expected from the near-linear correlation between the radiative powers
of the outflowing jets and the infalling accretion disk for RLQs
(e.g. \citealt{ghisellini2014}),
supporting a dependence of the jet power on accretion rate (e.g. \citealt{rawlings1991}).
Therefore, in addition to the corona-jet connection investigated in this work
and disk-corona interplay, RLQs also feature a disk-jet connection.
For the first time, not only has the overall corona-disk-jet relationship of RLQs been revealed,
but also the number of responsible physical processes has been determined.
In short, the corona-disk-jet relationship is three-fold
such that any two of these structures are physically linked.\footnote{RQQs share only a common disk-corona interplay with RLQs.}
Furthermore, the underlying drivers for these three relations are independent,
and none of these three relations is simply a side effect of the other two.
For example, we found that $R$ is a more useful property to consider than $L_\mathrm{radio}$ when the corona-jet connection is examined in \S~\ref{sec:ssrqs}, which now has a physical reason.
Specifically, the radio luminosity is affected by both the disk-jet and corona-jet connections;
using $R$ minimizes the intrusion of the former into studies of the latter.
Fig.~\ref{fig:aboutR} demonstrates more clearly the independence between the disk-jet and corona-jet connections.\footnote{We have already shown the independence between the corona-jet connection and disk-corona interplay in \S~\ref{sec:ssrqs}.}
There is an apparent correspondence between $R$ and $L_\mathrm{x}/L_\mathrm{x,RQQ}$ (cf. Fig.~\ref{fig:lxOverlxRqq})
such that the symbol-color gradient appears largely perpendicular to the constant-$\log R$ lines.
From Fig.~\ref{fig:aboutR}, it is strongly suggested that the spread of $R$ within the RLQ group is directly caused by small-scale processes in the vicinity of SMBHs (relating black-hole spin,
Eddington ratio, and magnetic field) that, at the same time, affect the corona.

\begin{figure}
\centering
\includegraphics[width=0.48\textwidth, clip]{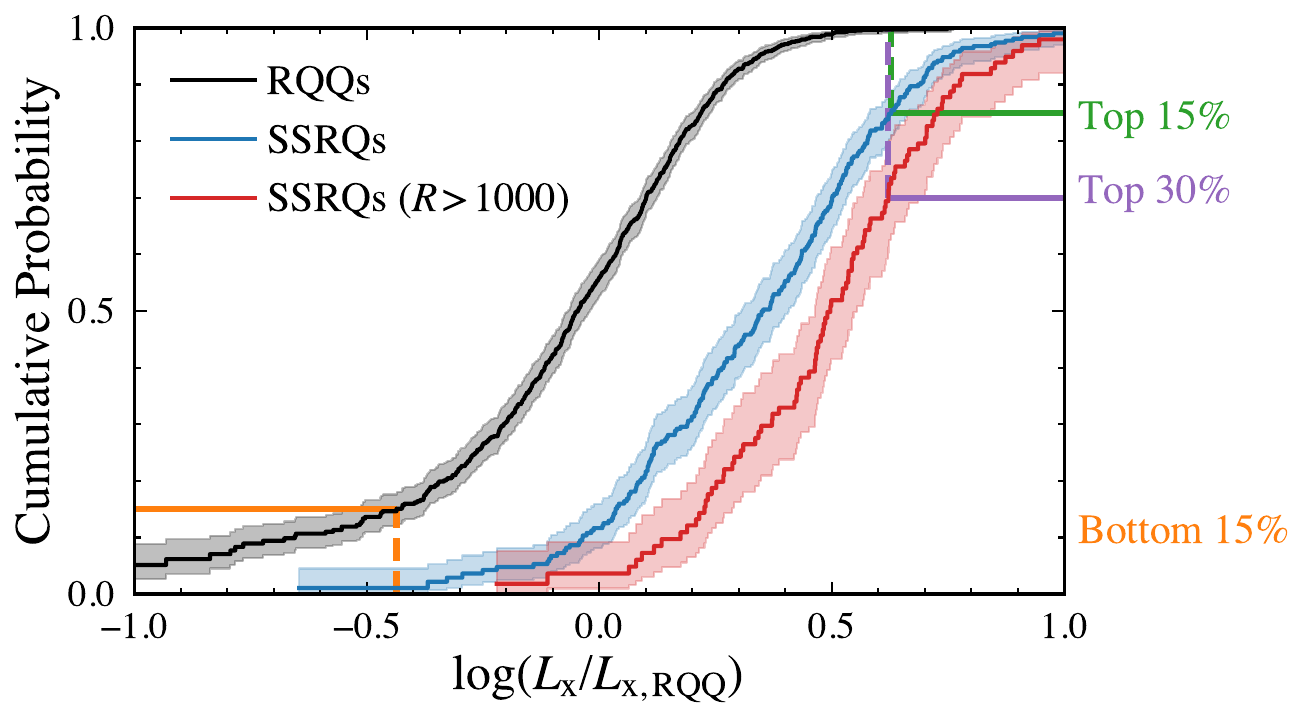}
    \caption{The cumulative distribution functions of $L_\mathrm{x}/L_\mathrm{x, RQQ}$ for RQQs (from \citealt{lusso2016}; black), SSRQs (blue), and a subset of SSRQs with $R>1000$ (red).
    Almost no RQQs are as X-ray luminous as the top $15\%$ of all SSRQs or the top $30\%$ of SSRQs with $R>1000$.
    Similarly, almost no SSRQs are less X-ray luminous than the bottom $15\%$ of RQQs.}
\label{fig:aboutSpin}
\end{figure}

\subsubsection{The role of black-hole spin}
Models for the launching of quasar jets need to explain our corona-jet connection in RLQs as well. 
The role of rapidly spinning black holes can thereby be constrained,
which are often taken to be an important ingredient for jet launching (e.g. \citealt{blandford2018}).
At first sight, it seems plausible that since higher prograde black-hole spins increase both the jet-production efficiency and radiative efficiency of the inner accretion flow,
a correlation between $R$ and $L_\mathrm{x}/L_\mathrm{x,RQQ}$ may be naturally produced (cf. Fig.~\ref{fig:lxOverlxRqq} and Fig.~\ref{fig:aboutR}).
However, the majority of spin parameters resulting from X-ray Fe~K$\alpha$ modeling of SMBHs are $\gtrsim0.9$ (e.g. \citealt{reynolds2019}, cf. \citealt{laor2019}), 
even for radio-quiet Seyfert galaxies.\footnote{Using the \citet{soltan1982} argument,
constraints on the accretion efficiency may also imply that most RQQs have rapidly spinning SMBHs (e.g. \citealt{elvis2002, yu2002, shankar2020}).
However, these So\l tan-argument constraints still have a sizeable variance owing to 
the uncertainties of several utilized parameters (e.g. \citealt{brandt2015, comastri2015}).}
Unless some mechanism can largely decouple black-hole spin and the corona for (only) radio-quiet AGNs,
those RQQs with the highest black-hole spins are then expected to be as \mbox{X-ray} luminous as their most radio-loud counterparts.
In Fig.~\ref{fig:aboutSpin}, we show the cumulative distribution functions of $L_\mathrm{x}/L_\mathrm{x,RQQ}$ for our SSRQs and the RQQ sample of \citet{lusso2016}.
Apparently, there are almost no corresponding RQQs to the top 15\% of SSRQs (i.e. the most radio-loud group) in $L_\mathrm{x}/L_\mathrm{x, RQQ}$.
If we instead use a subset of SSRQs with $R>1000$ (red curve in Fig.~\ref{fig:aboutSpin}), the value can be relaxed to the top 30\%.

Considering a worst-case scenario where the X-ray Fe~K$\alpha$ method significantly overestimates the spin parameter of radio-quiet AGNs 
and all RQQs have slowly rotating black holes,
a group of X-ray under luminous RLQs is expected since black holes with high retrograde spin parameters can likely launch relativistic jets as well (e.g. \citealt{tchekhovskoy2012mn}),
the radiative efficiency of which will be lower than those of RQQs.
However, no SSRQ extends beyond the 15\% of RQQs at the lower end in $L_\mathrm{x}/L_\mathrm{x, RQQ}$ (see Fig.~\ref{fig:aboutSpin}).

A decisive role of black-hole spin is also inconsistent with the jet line for AGNs in \S~\ref{sec:jetline}.
Indeed, a BHXRB can be in both jet-active and jet-inactive states without a significant change of the black-hole spin,
a conclusion which by analogy probably applies to SMBHs as well.

Therefore, the corona-jet connection supports the idea that rapid BH spin is a necessary but not sufficient condition for the development of jets, 
which likely requires a second, independent factor (e.g. \citealt{blandford2018}).

\subsubsection{The role of magnetic flux / topology}
We suggest that the magnetic flux threading the SMBH is likely the key physical factor controlling the radio-loudness of AGNs (e.g. \citealt{sikora2013}).
\citet{merloni2001, merloni2002} discussed the possibility of a connection between the coronae and relativistic jets in the most direct form such that
the former contains the magnetic fields required for the launching of the latter.
Therefore, only when these magnetically dominated coronae are sufficiently strong can jets be launched from them, and a possible corona-jet connection is expected.
Indeed, our results support the idea that the activity of the corona is at least a tracer for the magnetic flux responsible for the launching of relativistic jets of RLQs.
Note that the topology of the magnetic field might play a role in determining accretion states and jet launching as well (e.g. \citealt{livio2003, dexter2014}).

Furthermore, the inner accretion flows of RLQs may be a magnetically arrested disk (MAD; e.g. \citealt{igumenshchev2003, narayan2003}),
in contrast with the standard and normal evolution (SANE; e.g. \citealt{narayan2012grmdh}) scenario for RQQs.
In the MAD state, the magnetic flux threading the black hole is regulated by the accretion rate (e.g. \citealt{tchekhovskoy2012, zamaninasab2014}),
perhaps thereby naturally producing a $L_\mathrm{radio}$--$L_\mathrm{uv}$ relation for RLQs (see Fig.~\ref{fig:aboutR}).
As a consequence, to explain the corona-jet connection established in this work,
we need another factor that can independently affect the magnetic flux, in addition to the accretion rate.
We speculate that the thickness of the MAD is a good candidate since a thicker disk is able to support a stronger magnetic field (e.g. \citealt{tchekhovskoy2012mn}). 
The MAD disk contains optically thick accretion streams that are rapidly spiralling toward the black hole and
magnetically dominated low-density voids that can contribute to hard X-ray emission via the inverse-Compton scattering process (i.e. behaving like coronae; e.g. \citealt{igumenshchev2008}).
The magnetospheric radius (i.e. size of the MAD disk)\footnote{It is convenient to treat 
this radius as the inner truncation radius of the standard thin disk outside the MAD disk.}
also positively depends on the disk thickness (e.g. \citealt{xie2019}).
Therefore, a thicker disk can launch more powerful jets (e.g. \citealt{tchekhovskoy2012mn}) as well as produce an enlarged X-ray emitting corona,
producing the corona-jet connection. Note that the disk thickness might be affected by other more fundamental parameters (e.g. the Eddington ratio).

\section{Summary and Future Prospects}
\label{sec:summary}
\subsection{The main results of this paper}

In this paper, we construct a large uniform RLQ sample with high-quality multi-wavelength coverage (see \S~\ref{sec:sampleSelection}).
We investigate the \mbox{X-ray}-optical/UV-radio relation and the nature of the X-ray emission of RLQs.
The main results are summarized as follows:
\begin{enumerate}
\item The X-ray luminosities of RLQs are systematically larger than those of appropriately matched RQQs and depend on $L_\mathrm{uv}$, $L_\mathrm{radio}$ (or $R$), and radio spectral index.
    The slope of the $L_\mathrm{x}$--$L_\mathrm{uv}$ relation for SSRQs is consistent with that for RQQs to within $\approx3\%$ ($0.61_{-0.04}^{+0.04}$ vs. $0.63_{-0.02}^{+0.02}$),
        supporting the idea that SSRQ \mbox{X-ray} emission is mainly from the corona instead of the jets.
        The corresponding intercept for SSRQs is always larger than that for RQQs (by factors up to $\approx3$) and increases with $R$, suggesting a coordination between the activities of the jets and corona.
    Typical FSRQs are more X-ray luminous than SSRQs for given $L_\mathrm{uv}$ and $R$, likely involving more physical processes. See \S~\ref{sec:scatter}.
\item We perform model fitting and formal model selection in \S~\ref{sec:reason}. The models and methods are described in \S~\ref{sec:model} and \S~\ref{sec:method}, respectively.
In the model selection for all RLQs, a model with a strong jet X-ray contribution is ruled out.
    The preferred models either attribute all X-ray emission to the corona or contain only a small amount of jet X-ray emission. See \S~\ref{sec:base}.
\item For SSRQs and FSRQs considered as groups, a model with a strong jet contribution to the X-ray emission is not favored either. 
    The typical jet contribution for SSRQs is consistent with zero, again indicating that the corona is the dominant X-ray emitting structure.
    The observed X-rays from FSRQs might have a contribution from the jets, which is, however, only significant for $\lesssim10\%$ of FSRQs. See \S~\ref{sec:radioSpec}.
\item Given that the jets are apparently not contributing substantially to the \mbox{X-ray} luminosities of SSRQs, we discuss the corona-jet connection in detail in \S~\ref{sec:ssrqs}.
    Based on patterns apparent in our data,
        we suggest that the disk-corona interplay and corona-jet connection operate in largely independent ways.
    The former process is at work for both RQQs and RLQs, while the later process affects only RLQs.
    \item The literature results on the X-ray spectral, imaging, and timing properties of radio-loud AGNs are generally consistent with the idea that the disk/corona is responsible for the bulk of the X-ray emission of most RLQs.
        See \S~\ref{sec:literature}.
    \item We propose that the $\alpha_\mathrm{ox}$--$L_\mathrm{uv}$ relation for RQQs approximately defines a jet line for AGNs, drawing an analogy with BHXRBs. 
        See \S~\ref{sec:jetline}.
    \item RLQs feature corona-jet, disk-corona, and disk-jet connections, each of which seems to be driven by a distinct physical process. 
        The magnetic flux threading the SMBH instead of the black-hole spin is most likely controlling the jet-launching process of RLQs. See \S~\ref{sec:aboutR}.
\end{enumerate}

Perhaps the most important result of this paper is that the jets generally contribute much less to the X-ray emission than was previously thought.
The corona-jet connection and other results are then unavoidable corollaries.
Our modeling results do allow some modest X-ray emission to arise from the jets in general,
as is expected on basic physical grounds and also sometimes observed via X-ray imaging studies (see \S~\ref{sec:xjet}).
However, aside from a minority of FSRQs, this jet-linked X-ray emission generally appears secondary relative to that from the corona.

\subsection{Future work}

Many lines of study that extend our results can follow as future work.
First, as mentioned at the end of \S~\ref{sec:ssrqs},
the apparent break at $R\approx10$ needs further investigation, possibly with a large sample of $0\le\log R\le2$ quasars.
Secondly, X-ray spectral analyses for a sample of SSRQs
along a sequence of increasing $L_\mathrm{x}/L_\mathrm{x, RQQ}$ (or $R$)
might reveal how the corona changes, following the increasing production efficiency of jets.
Thirdly, in addition to the viewing-angle based unified model,
an evolutionary aspect of the AGN phenomenon is likely essential as well (e.g. \citealt{klindt2019}).
The identification of a jet line makes Fig.~\ref{fig:jetline} a useful diagnostic for accretion states of various AGN populations.
For example, at least some weak-line quasars and BAL quasars are intrinsically X-ray weak (e.g. \citealt{leighly2007, luo2014}),
thus perhaps representing different states than normal quasars.
Fourthly, one still-missing piece of the jigsaw for the X-ray properties of RLQs is systematic sample-based studies of their X-ray variability.
It would be valuable to compare RQQs and RLQs (with established $\alpha_\mathrm{r}$ values) as well as AGNs and BHXRBs in this regard.
Finally, the total radio luminosity we use traces the time-averaged jet power over perhaps up to $\sim10^7$ years,
while the X-ray luminosity is a more instantaneous tracer of the current activity of the corona.
Therefore, the corona-jet connection we found is probably subject to a delay 
and variability induced ``smearing'' already.
We might anticipate a closer correlation through long-term radio/X-ray variability monitoring of individual RLQs,
in particular the high-frequency radio emission that is from regions closer to (or even co-spatial with) the X-ray emitting corona and has shorter delays.

These future projects will benefit from on-going and scheduled multi-wavelength sky surveys,
especially in the radio bands that are witnessing a resurgence (e.g. \citealt{norris2017}).
The {\it eROSITA} telescope (\citealt{erositaScienceBook}) and Large Synoptic Survey Telescope (\citealt{ivezic2019})
will provide copious time-domain X-ray and IR/optical/UV data that can advance our understanding of the variability properties of AGNs.

\section*{Acknowledgements}
We appreciate the help of Mark Lacy in measuring VLASS fluxes.
We thank Tracy Clarke for her effort in checking the VLITE data of our RLQs.
We thank Ari Laor for his comments that greatly improved the clarity of this paper.
We also have benefitted from the help of Brendan Miller and discussions with Mike Eracleous.
We thank the referee, Giovanni Zamorani, for his constructive review.
SFZ and WNB acknowledge support from 
CXC grant AR8-19011X, NASA ADAP grant 80NSSC18K0878, and the Penn State ACIS
Instrument Team Contract SV4-74018 (issued by the {\it Chandra}
X-ray Center, which is operated by the Smithsonian Astrophysical
Observatory for and on behalf of NASA under contract
NAS8-03060).
The Chandra ACIS team Guaranteed Time Observations (GTO) utilized were
selected by the ACIS Instrument Principal Investigator, Gordon P.
Garmire, currently of the Huntingdon Institute for X-ray Astronomy, LLC,
which is under contract to the Smithsonian Astrophysical Observatory
via Contract SV2-82024.
BL acknowledges financial support from the National Natural Science
Foundation of China grants 11673010 and 11991053 and National Key R\&D
Program of China grant 2016YFA0400702.
YQX acknowledges support from NSFC-11890693, NSFC-11421303, the CAS Frontier Science Key Research Program (QYZDJ-SSW-SLH006),
and the K.C. Wong Education Foundation.
This research has made use of data obtained from the Chandra Data Archive and
the Chandra Source Catalog.
Based on observations obtained with {\it XMM-Newton},
an ESA science mission with instruments and contributions directly funded by
ESA Member States and NASA.

\bibliographystyle{mnras}
\bibliography{mn} 

\appendix
\section{The fixed parameters in model fitting}
\label{sec:approx}

We fix $\gamma_\mathrm{radio}=1$ for Model~\RomanNumeralCaps 3 in Table~\ref{tab:pars}
to have a better constraint on the X-ray contribution from the jets, which is one of our scientific goals.
This treatment is similar to that of \citet{browne1987}.
The $BL_\mathrm{5GHz}^{\gamma_\mathrm{radio}}$ term of Model~\RomanNumeralCaps 3 represents the X-ray luminosity from the core region of the quasar jets.
When it makes up a small portion of the total emission, the parameters $B$ and $\gamma_\mathrm{radio}$ can only be loosely constrained.
The fitting results in the case where no parameter of Model~\RomanNumeralCaps 3
is fixed are listed in Table~\ref{tab:free}.
The maximum-likelihood estimates of $\gamma_\mathrm{radio}$ are $0.92_{-0.32}^{+0.32}$ and $1.22_{-0.34}^{+0.34}$ for all RLQs and FSRQs,
which are close to unity.
The quoted error bars indicate $1\sigma$ uncertainties.
The fitting results of \citet{zamorani1984} and \citet{worrall1987} also suggest that there is a linear correlation between the radio and X-ray luminosities for the jet component.
\citet{miller2011} prefer a scenario where the excess X-ray emission of RLQs relative to RQQs is jet-linked, and
this jet-linked \mbox{X-ray} emission is beamed with a smaller bulk Lorentz factor than that of the radio emission,
which suggests $\gamma_\mathrm{radio}<1$. 
However, FSRQs are more X-ray luminous than SSRQs at given $L_\mathrm{uv}$ and $R$ (see Fig.~\ref{fig:lxOverlxRqq}),
which is inconsistent with the idea that the jet-linked \mbox{X-ray} emission has less anisotropy than the radio emission.
Note that fixing $\gamma_\mathrm{radio}=1$ does not affect the result that the jet component is a very minor term but allows for an estimate of $B$ with smaller uncertainty.

The fitting for SSRQs in Table~\ref{tab:free} is unphysical.
Firstly, $A\lesssim B$ would suggest that the jets are equal to or more
important than the corona in explaining the X-ray luminosities of SSRQs, which is in contrast with other samples.
In fact, we expect much less contribution from the jets in SSRQs than in FSRQs.
Secondly, the value of $\Gamma_\mathrm{uv}=\gamma_\mathrm{uv}+\gamma_\mathrm{radio}^\prime=1.01_{-0.20}^{+0.15}$ is not even close to $\gamma$.
Model~\RomanNumeralCaps 2 has similar issues when it is applied to SSRQs.
The fitting results using Model~\RomanNumeralCaps 2 with all four free parameters are listed in Table~\ref{tab:free2}.
They are non-physical for the same reasons as the case of Model~\RomanNumeralCaps 3.
These issues with Model~\RomanNumeralCaps 2 and Model~\RomanNumeralCaps 3 (without fixing parameters)
already suggest that they are not appropriate models for SSRQs.
To make a meaningful comparison between a model that invokes a distinct jet component and
the model attributing X-ray luminosity to the corona,
$\gamma_\mathrm{uv}$ of Model~\RomanNumeralCaps 2 is fixed to 0.63 in Table~\ref{tab:pars}.

\begin{table*}
\centering
    \caption{The fitting results using Model~\RomanNumeralCaps 3 with all parameters allowed to vary.}
\label{tab:free}
\begin{threeparttable}[b]
\begin{tabular}{lccccc}
\hline
\hline
    Sample & \multicolumn{5}{c}{\RomanNumeralCaps 3: $\log L_\mathrm{2keV}=\log \Big(AL_\mathrm{5GHz}^{\gamma_\mathrm{radio}^\prime}L_\mathrm{2500\angstrom}^{\gamma_\mathrm{uv}}+BL_\mathrm{5GHz}^{\gamma_\mathrm{radio}}\Big)$} \\
    \cmidrule(lr){2-6}
    & $A$& $B$& $\gamma_\mathrm{uv}$ & $\gamma_\mathrm{radio}$ & $\gamma_\mathrm{radio}^\prime$\\
\hline
    All RLQs &$0.75_{-0.20}^{+0.01}$&$0.03_{-0.01}^{+0.19}$&$0.50_{-0.01}^{+0.15}$&$0.92_{-0.32}^{+0.32}$&$0.17_{-0.04}^{+0.04}$\\
    FSRQs&$0.77_{-0.14}^{+0.03}$&$0.03_{-0.01}^{+0.12}$&$0.51_{-0.03}^{+0.09}$&$1.26_{-0.34}^{+0.34}$&$0.18_{-0.08}^{+0.03}$\\
SSRQs&$0.32_{-0.07}^{+0.13}$&$0.35_{-0.11}^{+0.06}$&$0.89_{-0.24}^{+0.18}$&$0.30_{-0.06}^{+0.07}$&$0.12_{-0.08}^{+0.08}$\\
\hline
\end{tabular}
\end{threeparttable}
\end{table*}

\begin{table}
\centering
    \caption{Model fitting results for SSRQs using Model~\RomanNumeralCaps 2 with all parameters set free to vary.}
\label{tab:free2}
\begin{threeparttable}[b]
\begin{tabular}{lcccc}
\hline
\hline
    Sample &\multicolumn{4}{c}{\RomanNumeralCaps 2: $\log L_\mathrm{2keV}=\log \Big(AL_\mathrm{2500\angstrom}^{\gamma_\mathrm{uv}}+BL_\mathrm{5GHz}^{\gamma_\mathrm{radio}}\Big) $} \\
    \cmidrule(lr){2-5}
    & $A$& $B$& $\gamma_\mathrm{uv}$ & $\gamma_\mathrm{radio}$\\
\hline
    SSRQs& $0.23_{-0.04}^{+0.07}$&$0.39_{-0.06}^{+0.03}$&$1.03_{-0.15}^{+0.09}$&$0.36_{-0.03}^{+0.05}$  \\
\hline
\end{tabular}
\end{threeparttable}
\end{table}

\bsp    
\label{lastpage}
\end{document}